# Reconstructing the lost eclipse events of the Saros spiral applying the Draconic gearing on the Antikythera Mechanism - The impact of the gearing errors on the pointers' position

(Manuscript + Supplementary material)

**Aristeidis Voulgaris[1], Christophoros Mouratidis[2], Andreas Vossinakis[3]**

[1]City of Thessaloniki, Directorate Culture and Tourism, Thessaloniki, GR-54625, Greece,
[2]Merchant Marine Academy of Syros, GR-84100, Greece,
[3]Thessaloniki Astronomy Club, Thessaloniki, GR-54646, Greece

[1]Corresponding author <arisvoulgaris@gmail.com>

**Abstract**

*We present new observations concerning the procedure for the reconstruction of the lost eclipse events engraved in the Saros spiral cells of the Antikythera Mechanism. For the reconstructed eclipse events we applied the assumed, albeit missing, Draconic gearing of the Antikythera Mechanism, which was probably a part of the Mechanism's gearing, representing the fourth lunar motion, the Draconic cycle. The Draconic gearing is very critical for the eclipse prediction and defines whether an eclipse will occur. For our research we created a program which presents the phase of the four lunar cycles - and the position of the Draconic pointer relative to the ecliptic limits. After calibrating the software according to the preserved eclipse events, the lost eclipse events of the Saros spiral were calculated and discussed. The procedure for the calculation of the events' times by using solely the Mechanism is also presented. The eccentricity error of a gear which is preserved on the ancient prototype is discussed. An experimental setup facilitated the analysis of the mechanical characteristics of gears with triangular teeth and the errors. The experimental study of the gears' errors revealed the strong impact the Antikythera Mechanism pointers have on the results.*

## 1. Introduction

The Antikythera Mechanism was a unique geared device of the Hellenistic era and was constructed in order to make calculations/predictions of astronomical events of the near future. These mechanical calculations were made by a large number of engaged gears. The results were presented by a number of pointers, which rotated on their measurement scales. The results were read by the user, who just had to observe the relative position of each pointer rotating on its corresponding measuring scale which also had subdivisions.

The Mechanism could predict the astronomical events of solar and lunar eclipses (Freeth et al., 2006; 2008), which are still very important today (Pasachoff 2018). On the lower half of the Back plate of the Mechanism there was the Saros spiral in which the sequence of the eclipse events was engraved. Today the Saros spiral is preserved by about 30%, in 3 parts: Fragment A2, F and E.

The Saros spiral consisted by four full spiral turns, divided in 223 sectors/cells representing the 223 synodic months of a Saros period (Ptolemy 1898 refers to this cycle as ΠΕΡΙΟΔΙΚΟΣ/Periodikos, Toomer 1984; Voulgaris et al., 2021). After 223 synodic months, equal to $18^y\ 11^d\ 8^h$, the eclipse events sequence is repeated in the same order, with a delay of 8 hours. After three Saros periods, equal to $54^y\ 34^d$, named Exeligmos, the eclipse events are visible from about the same observing place, at about the same time (Neugebauer 1975; Oppolzer 1962). We say "about" rather than "exactly", because the equality in integer number of cycles is an approximation; actually 223 synodic months = 241.998703 draconic months = 238.99195 Anomalistic months for Era of 1 AD (see Supplementary material). This means that the repeated eclipse path per each Saros changes through time, as the Node slightly shifts by about 0.46° (≈ one lunar diameter) per Saros cycle. After 3 Exeligmos cycles (starting with New Moon at Node-A) the New Moon will be located about 4.3° far away from the Node-A, either to the North or to the South. The repeated *Saros* sequence finishes with a last partial solar eclipse (only visible from one of the Earth's poles) or a penumbral lunar eclipse. As the Earth and Moon perpetual rotate around the Sun and around each other, new Saros series begin starting from one of the poles with a partial eclipse.

## 2. The eclipse events on the Antikythera Mechanism Saros spiral

The solar and lunar eclipse events sequence of the Antikythera Mechanism was presented on the Saros spiral. On some of the cells, information related to the eclipse events (whether it was solar or lunar and the time it occurred) was engraved. The letter Η, ΗΛΙΟΣ (Helios-Sun) meant that a solar eclipse would occur, whereas the letter Σ, ΣΕΛΗΝΗ (Selene-Moon) meant that a lunar eclipse would occur (Freeth et al., 2006, 2008; Anastasiou et al., 2016). There are also preserved cells with both letters engraved, meaning a lunar and a solar eclipse will occur in this month.

Following the calibration of the Mechanism pointers, using the very important astronomical and religious date of 22/23 December 178BC as a starting point (Voulgaris et al., 2022i unpublished results, https://arxiv.org/abs/2203.15045), additional astronomical information could be extracted, i.e. the position of the Moon and the Sun in the sky/Zodiac constellation and the corresponding day of the Egyptian calendar. Moreover, if a Lunar eclipse occurred on the constellation of Leo or Virgo, then by checking whether the Games pointer aims at the LΑ-ΟΛΥΜΠΙΑ (Olympiad) quadrant, the user could be informed if the Olympiad would start during the Lunar eclipse (Freeth et al., 2008; Perrotet 2004; Vaughan 2002).

## 3. Lunar cycles represented on the Antikythera Mechanism gearing trains

The ancient Greek astronomers of the Hellenistic era (the Babylonians and Egyptians too) studied the motions of the Moon and Sun by applying their periodicities equations resulting from a long time of observations. The sync of the three lunar cycles, Synodic, Draconic and Anomalistic, into a period that contains an integer number of each creates a composite cycle named Saros. This contains 223 Synodic months, 239 Anomalistic months and 242 Draconic months (also 241.029 Sidereal months).

From the currently preserved fragments of the Mechanism it is concluded that the Mechanism gearings represent three out the four known lunar cycles that were known and extensively used by the astronomers of the Hellenistic Era. The three preserved lunar cycles of the Mechanism - Sidereal, Synodic and Anomalistic - can be detected on the gearing of the Mechanism:

a) The Sidereal cycle, one rotation of the Moon relative to the "fixed" stars, is represented as one full rotation of the Lunar disc-pointer around the Zodiac constellation month ring. For example, the lunar pointer starts at the zodiac Sign of Capricorn constellation and after one rotation aims again at the same zodiac Sign.

b) The Synodic cycle, the time it takes the Moon to return to the same lunar phase, is represented by re-alignment of the Lunar pointer with the Golden sphere-Sun pointer (from New Moon to the next New Moon). At the same time the little lunar sphere located on the Lunar disc of the Mechanism, presents the same phase (black hemisphere). The Synodic cycle is the main measuring unit of the Mechanism. The units of the Saros (also Exeligmos), Metonic and Games scales are based on the Synodic month. The Antikythera Mechanism was a time machine calculator based on the Luni(solar) calendar, using units of the synodic month.

c) The Anomalistic cycle of the Moon presents the periodically variable angular velocity of the Moon as it travels in the sky. According to Geminus' description in *Introduction to the Phenomena*, chap. 18 *About Exeligmos*, the Moon crosses the zodiac sky in a variable angular velocity: initially it moves with its minimum angular velocity (i.e. the Moon located at Apogee) and through time increases its velocity. Afterwards, it gradually decreases its maximum angular velocity (the Moon located at Perigee) down to the initial smallest value. This cycle called Anomalistic cycle and starts when the Moon is located at Apogee position (Voulgaris et al., 2022ii, unpublished results https://arxiv.org/abs/2203.15045).

The Anomalistic cycle is also represented on the Antikythera Mechanism gearing: the ancient manufacturer introduced the unique idea of the conversion of a constant angular velocity into a periodically variable angular velocity introducing the *pin&slot* gearing design:
Two gears in a common, but eccentric modified axis so that their centers do not coincide (Freeth et al., 2006; Voulgaris et al., 2018b and 2019a). This system of the two gears k1/k2 and their k-axis is on-board of the large gear e3. The final angular velocity of this unique design system represents the Anomalistic lunar cycle of the Antikythera Mechanism.

The fourth lunar cycle, the Draconic cycle, also known as nodal or nodical month, is based on the periodic transits of the Moon through the Ecliptic. The name "*Draconic*" or "*Draconitic*" relates to the Dragon (ΔΡΑΚΩΝ in Greek) which eats the Sun or the Moon, as believed in the Middle Ages (see Kircher 1646, page 548). The ancient Greek astronomers used the phrase ΑΠΟΚΑΤΑΣΤΑΣΙΣ ΚΑΤΑ (ΕΚΛΕΙΠΤΙΚΟΝ) ΠΛΑΤΟΣ (restitution in Ecliptic latitude) or, alternatively the phrase ΠΛΑΤΙΚΑΙ ΑΠΟΚΑΤΑΣΤΑΣΕΙΣ (Jones 1990). The lunar orbit crosses periodically, the Ecliptic in two points, named ΑΝΑΒΙΒΑΖΩΝ ΣΥΝΔΕΣΜΟΣ/Ascending Node and ΚΑΤΑΒΙΒΑΖΩΝ ΣΥΝΔΕΣΜΟΣ/Descending Node. The Greek word ΕΚΛΕΙΠΤΙΚΗ (Ecliptic) derives from the word ΕΚΛΕΙΨΗ (Eclipse), which makes sense, since on this zone took place the solar and lunar eclipses. The Ecliptic is referred to in the Antikythera Mechanism inscription as ΕΓΛΕΙΠΤΙΚΗ (see Bitsakis and Jones 2016).

The relation between the Synodic phase and the Draconic phase determines the occurrence of an eclipse: when a New Moon coincides with the beginning or the middle of the Draconic cycle, a solar eclipse will occur. When a Full Moon coincides with the beginning or the middle of the Draconic cycle, a lunar eclipse will occur. This, very important lunar cycle, was known and used by the astronomers of the Hellenistic era.

The Antikythera Mechanism was a device which could predict the eclipse events. The ancient manufacturer dedicated a large number of gears and a large part of his creation to the eclipse event calculations. The lower half part of the Back plate and two pointers are used for the eclipse event presentation.

Today, the important Draconic cycle/Draconic gearing is not represented on the Antikythera Mechanism or is not preserved. As the four lunar cycles were extensively used during the Hellenistic Astronomy, the assumption of the Draconic gearing on the AM is not far-fetched.

In addition, by engaging the preserved Fragment D-gear r1 with the gear a1 (engaged to the b1 gear) it could become part of the Draconic gearing of the AM (Voulgaris et al., 2022i unpublished results, https://arxiv.org/abs/2104.06181). If the Antikythera Mechanism included the Draconic gearing, then the eclipse sequence information engraved on the Saros cells could be calculated without any additional external information and the eclipse events sequence would be an actual prediction by a pure mechanical procedure.
We believe that the Draconic gearing of the Mechanism was most probably included in the Mechanism as an important and crucial part, as it directly related to the eclipse prediction.

In this work we consider the Draconic gearing as included in the Mechanism.
Based on this hypothesis we will show that the specific eclipse event sequence was a result of the Draconic gearing existence on the Mechanism and the eclipse sequence was calculated through a mechanical procedure.
We will also show that the omitted eclipses (Freeth 2014; Carman and Evans 2014; Anastasiou et al., 2016; Freeth 2019; Iversen and Jones 2019) resulted from the mechanical errors of the gears, especially the error of eccentricity.

The Draconic cycle/gearing presents the position of the Moon relative to the Ecliptic. The Draconic pointer (the output of the Draconic gearing) was rotated around the Draconic scale. The Draconic scale consisted of a circle, representing the lunar orbit, and an engraved line-diameter, parallel to the plane which is defined by the Zodiac month ring located on the Front dial. This line represented the Line of Nodes, the Nodes marking the ends of this line. Out of this line, the Draconic pointer showed that the Moon is above or below the Ecliptic plane, where no eclipses could occur. Two arcs, each containing a Node, define the ecliptic limits (see section 4.1)
The inclusion of the Draconic gearing, scale and pointer in the Antikythera Mechanism allowed real eclipse prediction via a pure mechanical procedure: By checking the relative positions of the two pointers, Lunar pointer aiming at the Golden Sphere, i.e. New Moon (or, in opposite position, Full Moon), and the Draconic pointer aiming (or not) at/close to the Ecliptic plane, the ancient Manufacturer could know if a solar (or lunar) eclipse will be occur. The Moon phase and position in relation to the Ecliptic was the key for the eclipse events engravings on the Saros cells. This way, the ancient Manufacturer did not need any additional information in order to calculate the eclipse events sequence; he only needed the initial starting date, which defined the specific position of the Mechanism's pointers, i.e. the positions of the Sun in the zodiac sky, the Moon relative to the Sun, the anomalistic phase (position of the pin inside the slot) and the Moon position relative to a Node.

### 4.1 Ecliptic Limits and eclipses

Due to the sizes of the Earth's and the Moon's shadows, for an eclipse to occur the geometrical centers of the Sun, the Earth and the Moon need not lie on a straight line.
The ecliptic limits define the maximum angular distance (measured from the Earth) by which the Moon can deviate from the straight line connecting the Earth's center and a Node of the Moon's orbit still allowing for an eclipse occurrence. When the full disc of the New Moon/Full Moon is located between the Ecliptic Limits, either a total/ annular solar eclipse or a total lunar eclipse is bound to be visible somewhere on the Earth.
If the Earth had a bigger diameter, then the possibility of the lunar shadow projection on the Earth's surface (solar eclipse) would be greater and also its shadow would have been larger, increasing the probability and the duration of a lunar eclipse. On planet Saturn, solar and moon eclipses often occur somewhere on Saturn's surface (atmosphere).

The ecliptic limits have not fixed values because the Sun-Moon-Earth system is dynamically variable: the distances between them as well as the angular dimension of their shadows vary with time (Green 1985).

For a partial solar eclipse, the solar ecliptic limit has a maximum value of 18°.4; for a total solar eclipse it is 11°.8. For a partial lunar eclipse, the lunar ecliptic limit has a maximum value of 12°.2; for a total lunar eclipse, it is 5°.9.

Ptolemy in Almagest calculates and presents the ecliptic limits:

For any kind of a solar eclipse, the ecliptic limits are 20.68° to the north of a Node (measure to the moon's inclined circle), and 11.36° to the south.

For a lunar eclipse, the ecliptic limits are ±15.2° off a Node.

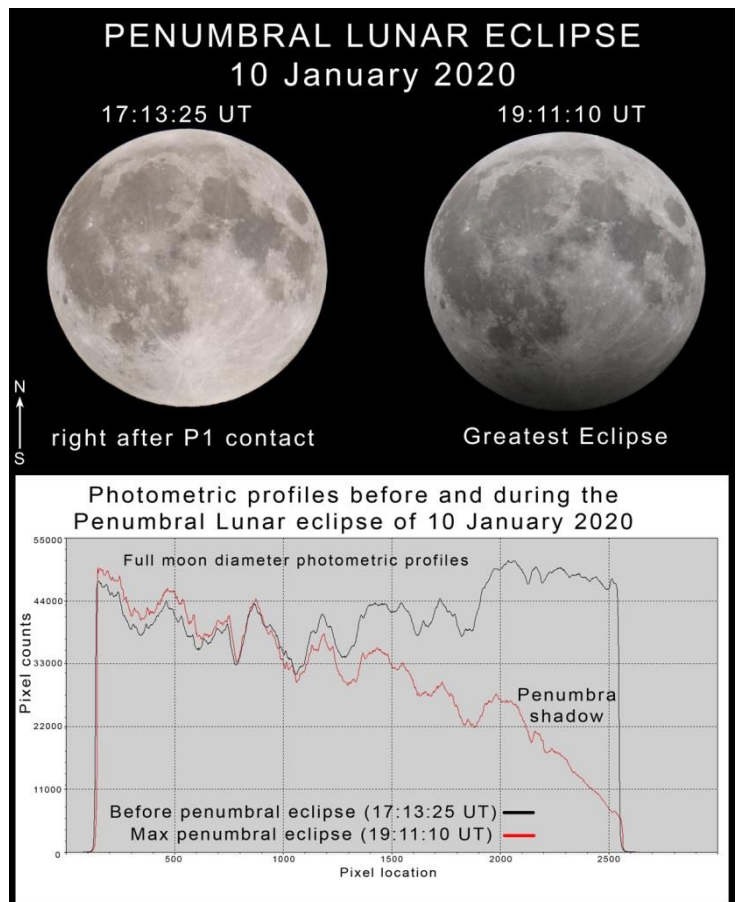

**Figure 1:** The Full Moon before (left) and during (right) the Penumbral eclipse of January 10, 2020, observed by the first author from Thessaloniki, Greece. For the capturing, an apochromatic refractor, 127mm diameter, 920mm focal length, was used. The exposure time was the same for both images. Note that the brightness of the upper parts of the lunar disc remained constant during the max of the penumbral eclipse. The Moon brightness at Greatest Eclipse is about 0.787 of the brightness of the Full Moon before P1 contact, yielding a brightness drop by 0.26 mag.

There are two characteristic patterns when three eclipses occur in two successive synodic months:
- If a total/annular solar eclipse occurs and just right on the Node, then one fortnight (half synodic month) before and one fortnight after this date a Lunar penumbral eclipse will occur (Espenak and Meeus 2008).

- If a total lunar eclipse occurs just right on the Node, then one fortnight before and after this date, two partial Solar Eclipses will be visible alternately from the Earth's poles (Espenak and Meeus 2008).

During a Penumbral Lunar Eclipse, the brightness of the Moon is slowly decreasing in a time span of about 2 hours, **Figure 1**. Even though the lunar disc enters on the penumbra shadow of the Earth, only a small part of the lunar disc is gradated and darkened. Most of the penumbral eclipses are not detectable by the eye because the darkening happens too slowly and gradually for the eye to notice. Even if this decreased brightness is detectable by some observer, they might as well assume that it is due to a thin cloud passing in front of the lunar disc. Therefore, the Penumbral lunar eclipse is difficult to be detected without any prior information regarding this event.

On the Saros spiral eclipse events, two successive events H-Σ in two successive cells or Σ-H in one cell, is preserved. As already mentioned, the penumbral lunar eclipses are difficult to detect and the partial solar eclipses only visible from the Earth's poles are also difficult to catch for an observer living around the Mediterranean Sea, because the percentage of covering is relatively small.
This is the reason why on the Antikythera Mechanism there should not be three eclipse events within two successive cells (synodic months), i.e. the eclipse event pattern
Cell-(X)=Σ+H and Cell-(X+1)=Σ (*Lunar-Solar-Lunar*) or
Cell-(Y)=H and Cell-(Y+1)=Σ+H (*Solar-Lunar-Solar*).

## 4.2 Reconstructing the lost eclipse events applying the Draconic gearing on the Antikythera Mechanism. The ecliptic limits of the Draconic scale.

In order to represent the phase correlation of the three lunar cycles (Synodic, Draconic, Anomalistic), we implemented a simulation created in *Python* programming language code. The simulation, which will be referred as *DracoNod* program, reproduces the position of the Moon phase in relation to the Apogee/Perigee, to the Nodes, (and to the constellations of the Ecliptic-Sidereal cycle which is not presented in this work) versus time, applying the equation 1 Saros cycle = 223 Synodic cycles = 242 Draconic cycles = 239 Anomalistic cycles (= 241.029 Sidereal cycles). For the calculation of the lunar phase anomaly we applied the approximation that derives from the *pin&slot* gear system as calculated in Voulgaris et al., 2018b, and not the robust one, that derives from Kepler's laws. This way we calculated the ecliptic limits as they would be derived from the ancient manufacturer using the Mechanism itself. The simulation was transformed into a graphic environment, where we used the *GeogeGra* software (https://www.geogebra.org) for the user's convenience. The lunar cycles are represented by three independent circles – the Synodic, the Draconic and the Anomalistic – with their characteristic points or arcs that represent: New Moon-Full Moon, Node A-Node B, and Apogee-Perigee respectively, **Figure 2**. For the angular velocities of the pointers on these circles, as mentioned before, we used the equivalence with the Saros circle and the *pin&slot* anomaly approximation.
As the initial starting position of the three lunar cycles the New Moon phase at Apogee and at Node-A was selected (and the Moon was set at the zodiac Sign of Capricorn, even though initially the sidereal cycle was not necessary for the lost eclipse events calculation).
In this position of the Moon, a solar eclipse occurred (Moon at Node-A), which was a long duration annular eclipse (Moon at Apogee).

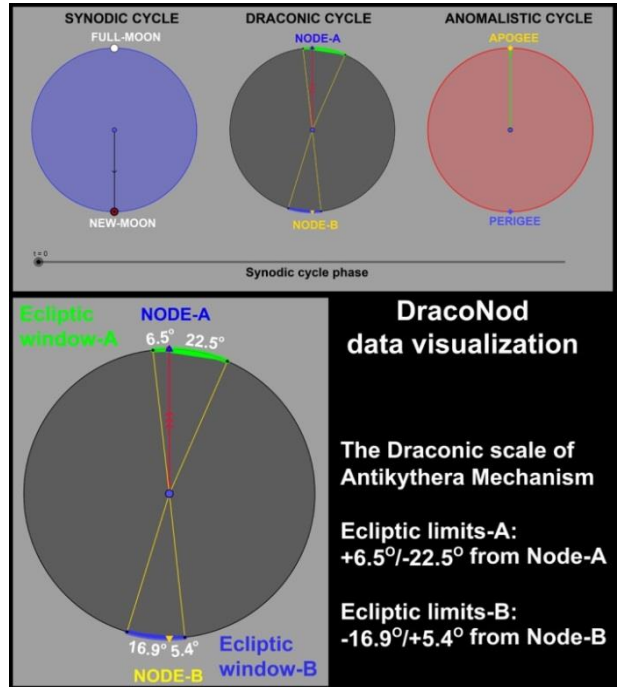

**Figure 2:** Data calculated by the program *DracoNod* were visualized using *GeoGebra* software. In this image the phases of the three lunar cycles during the beginning of Saros period/Saros spiral (end of cell-01) are presented: New Moon at Node-A and at Apogee. The two ecliptic windows are presented in green and blue arcs. The ecliptic limits are defined by the yellow lines. Each of the circle's orbital radiuses rotates with the period of the corresponding lunar cycle. An eclipse event occurs when the orbital radius of the Draconic cycle (red color) is located inside the ecliptic window during the phase of New Moon or Full Moon.

On the Draconic circle we firstly applied the ecliptic limits described in Ptolemy's Almagest:
20. 41° North of Nodes – 11.22° South of Nodes for a solar eclipse and
15.12° North/South of Nodes for a lunar eclipse.
Then, we ran the *DracoNod* and we recorded the solar and lunar eclipses on their corresponding cells.
By applying the limits referred by Ptolemy, the calculated eclipse events sequence was not in agreement with the preserved events of the Mechanism. For example, the *DracoNod* program calculated events Σ+H for cell-13 and no event for cell-12. Moreover, it calculated three eclipse events in two successive synodic months. This mismatch is a result of the wrong values of the ecliptic limits introduced to the program: Ptolemy's ecliptic limits were larger than the limits adopted by the ancient Manufacturer.
So, we re-calibrated the ecliptic limits of our program by giving shorter values, in order to achieve the best correlation to the Saros spiral preserved events.
According to this calibration, the proper ecliptic limits which present the best correlation with the preserved eclipse events are common for the solar and the lunar eclipses:
For the Ecliptic Window-A: –22.5° and +6.5° from Node-A (measured CCW) and for the Ecliptic Window-B: –16.9° and +5.4° from Node-B (measured CCW).
These non-symmetrical limit values might be considered as unusual and unjustified, but we present a satisfactory explanation in the next sections.
On **Table 1**, the missing eclipse events calculated by the program *DracoNod* and on **Table 2** the summary of the results are presented (see also Supplementary Material).

**Table 1:** Prediction of the lost eclipse events of the Saros spiral using the authors' program *DracoNod*. The new numbering of the Saros cells was applied according to Voulgaris et al., 2021. Out of the hooked A (Ⱥ) for A3 index number event (Iversen and Jones 2019), a second symbol ẞ was introduced as the index number (B3) for the last event(s) of the sequence, on cell 218. Summary, results and comments on Table 2 and in authors' Supplementary material.

| Event Number | Event index letter | New cell numbering, (*Voulgaris et al., 2021*) | Preserved Eclipse events of Saros cells | Reconstructed eclipse events sequence, applying the Draconic gearing. Calculated by *DracoNod* program | Comments/ Moon position relative to the Node/ecliptic limit |
|---|---|---|---|---|---|
| 1, (2) | [A1] | **Cell-1** |  | Sun<br><br>*Very long duration of the annular eclipse* | **Saros cycle begins. New Moon at *Apogee* and at *Node-A*.** *No lunar eclipse event in accordance to cell-112 (Sar period)* |
| 2, 3 | **B1** | **Cell-7** | **Moon, Sun** | **Moon, Sun** |  |
| 4 | **Γ1** | **Cell-12** | **Sun** | **Sun** | New Moon close to the ecliptic limit |
| 5 | Δ1 | Cell-13 |  | Moon |  |
| 6 | **E1** | **Cell-19** | **Moon** | **Moon** |  |
| 7 | **Z1** | **Cell-24** | **Sun** | **Sun** |  |
| 8 | **H1** | **Cell-25** | **Moon** | **Moon** | Full Moon at *Node-B* |
| 9 | Θ1 | Cell-30 |  | Sun |  |
| 10 | I1 | Cell-31 |  | Moon | Full Moon close to the ecliptic limit |
| 11 | K1 | Cell-36 |  | Sun |  |
| 12, 13 | Λ1 | Cell-42 |  | Moon Sun | New Moon close to Node-B |
| 14, 15 | M1 | Cell-48 |  | Moon Sun | New Moon close to Node-A |
| 16, 17 | N1 | Cell-54 |  | Moon Sun | New Moon on the ecliptic limit |
| 18 | Ξ1 | Cell-59 |  | Sun |  |
| 19 | [O1] | **Cell-60** | **Moon** | **Moon** |  |
| - |  | Cell-65 | Event not exists | Sun | New Moon just right on the ecliptic limit. Indeterminacy or eccentricity error |
| 20 | **Π1** | **Cell-66** | **Moon** | **Moon** | Full Moon close to the ecliptic limit |
| 21 | **P1** | **Cell-71** | **Sun** | **Sun** |  |
| 22 | [Σ1] | **Cell-72** |  | Moon | Full Moon close to Node-B |
| 23 | **T1** | **Cell-77** | **Sun** | **Sun** |  |
| 24 | **Y1** | **Cell-78** | **Moon** | **Moon** | Full Moon just right on the ecliptic limit |
| 25 | [Φ1] | Cell-83 |  | Sun |  |
| 26, 27 | [X1] | Cell-89 | | Moon Sun | New Moon close to Node-B |
| 28, 29 | [Ψ1] | Cell-95 | | Moon Sun | |
| 30 | [Ѡ1] | Cell-101 | | Moon | |
| 31 | [A2] | Cell-106 | | Sun | |
| 32 | [B2] | Cell-107 | | Moon | |
| - | | Cell-112 | Event not exists | Sun | New Moon approaches the ecliptic limit. Error of eccentricity or tooth un-uniformity of gear s2/ Draconic pointer. Same position of the pointer in the beginning of cell-01 |
| 33 | **Γ2** | **Cell-113 Middle of *Saros Cycle*, a new Sar period begins** | **Moon** | **Moon** *Very long duration of the total lunar eclipse* | **At the middle of Cell-113, Full Moon at *Perigee* and at *Node-A*.** The gears of the Draconic training (same tooth by tooth) repositioned as in the Saros beginning |
| 34 | **Δ2** | **Cell-118** | **Sun** | **Sun** | |
| 35 | **E2** | **Cell-119** | **Moon** | **Moon** | |
| 36, 37 | **Z2** | **Cell-124** | **Moon, Sun** | **Moon, Sun** | Full Moon on the ecliptic limit |
| 38, 39 | **H2** | **Cell-130** | **Moon, Sun** | **Sun** (prediction of one out of two preserved events) | Full Moon just out of the ecliptic limit. Error of eccentricity |
| 40, 41 | **Θ2** | **Cell-136** | **Moon, Sun** | **Moon, Sun** | *New Moon at Node-B* |
| 42, 43 | [I2] | Cell-142 | | Moon Sun | New Moon close to the ecliptic limit |
| 44 | [K2] | Cell-148 | | Moon | |
| 45 | [Λ2] | Cell-153 | | Sun | |
| 46 | [M2] | Cell-154 | | Moon | |
| 47 | [N2] | Cell-159 | | Sun | |
| 48 | [Ξ2] | Cell-160 | | Moon | Full Moon close to Node-A |
| 49 | [O2] | Cell-165 | | Sun | |
| - | | Cell-166 | Event not exists | Moon | Full Moon just on the ecliptic limit. Error of eccentricity or indeterminacy. |

| 50, 51 | **Π2** | **Cell-171** | **Moon, Sun** | **Moon, Sun** | |
| 52, 53 | **P2** | **Cell-177** | **Moon, Sun** | **Moon, Sun** | Full Moon just right on the ecliptic limit. New Moon at Node-A |
| 54, 55 | **Σ2** | **Cell-183** | **Moon Sun** | **Moon, Sun** | |
| 56 | **T2** | **Cell-189** | **Moon** | **Moon and Sun (prediction of one additional event)** | New Moon just on the ecliptic limit. Error of eccentricity or indeterminacy |
| 57 | [Y2] | Cell-195 | | Moon | |
| 58 | [Φ2] | Cell-200 | | Sun | |
| 59 | [X2] | Cell-201 | | Moon | Full Moon at Node-B |
| 60 | [Ψ2] | Cell-206 | | Sun | |
| 61 | Ω2 | Cell-207 | | Moon | |
| 62 | [2] (A3) | Cell-212 | | Sun | |
| 63, 64 | [ß] (B3) | Cell-218 | | Moon Sun | |

*DracoNod* predicted all the preserved solar eclipses plus three additional eclipses (Draconic pointer just on the limit) which are non-engraved events and present high indeterminacy. Was predicted all the preserved lunar eclipses out of one which is also presents high indeterminacy. Was also predicted one non-engraved eclipse which presents high indeterminacy. The predicted lunar eclipse on cell 01 is too doubtful if finally it was real engraved (comparing to Cell-112). See all the eclipse events visualization in authors' Supplementary material).

**Table 2.** Summary and comments of the predicted events via *DracoNod* program (see also Supplementary Material).

| *DracoNod* eclipse events prediction | |
|---|---|
| **Solar eclipse events on cell** | **Lunar eclipse events on cell** |
| 1, 7, 12, 24, 30, 36, 42, 48, 54, 59, 71, 77, 83, 89, 95, 106, 118, 124, 130, 136, 142, 153, 159, 165, 171, 177, 183, 200, 206, 212, 218 **(31 events H)** | 7, 13, 19, 25, 31, 42, 48, 54, 60, 66, 72, 78, 89, 95, 101, 107, 113, 119, 124, 130, 136, 142, 148, 154, 160, 171, 177, 183, 189, 195, 201, 207, 218 **(33 events Σ)** |
| Cell-65 non engraved-predicted Solar, Pointer at the limit (indeterminacy or gearing error) | Cell-130 engraved Lunar - not predicted Pointer just out of limit (gearing error) |
| Cell-112 non engraved-predicted Solar, Pointer close to the limit (gearing error) | Cell-166 predicted Lunar at limit (indeterminacy), but it should be not exist based on index numbering |
| Cell-189 non engraved-predicted Solar, Pointer at the limit (indeterminacy or gearing error) | Cell-224/Cell-01 predicted Lunar It could be exist or not exist (same with Solar of Cell-112) |

Taking into account that in a geared device there have to be mechanical errors, the omitted/additional eclipse events could be a result of mechanical errors in the Antikythera

Mechanism gearings, especially on the Draconic gears and pointer, or it could be an error of reading or a result of indeterminacy (discussed below).

## 5.1 Errors in time measurements using a geared measuring device

In his "*Introduction to the Phenomena*", chap. 18, About Exeligmos, Geminus 1880; 2002 describes the variable lunar angular velocity calculation using an angle measuring instrument and writes: …ΛΟΙΠΑ ΑΡΑ ΕΣΤΙ ΤΑ ΕΚΦΥΓΟΝΤΑ ΤΗΝ ΤΩΝ ΦΑΙΝΟΜΕΝΩΝ ΔΙΑ ΤΩΝ ΟΡΓΑΝΩΝ ΠΑΡΑΤΗΡΗΣΙΝ (…these angle measurements are lost as a result of the instrument's measuring errors).

When using a measuring instrument, measuring errors appear and affect every measurement. These errors are an inherent problem of the *instrument-user* system.

There are many types of measuring errors (Herrmann 1922; Muffly 1923). Some of the errors are owed to the instrument and some errors to the user of the instrument.

In this work we present and discuss the errors that can be directly related to the Antikythera Mechanism operation.

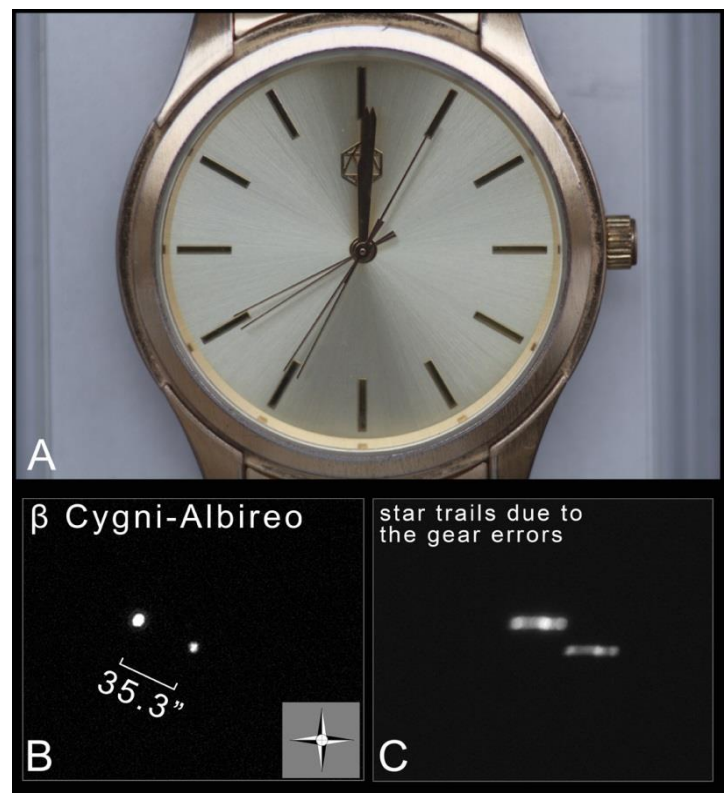

**Figure 3:** A) the pointer of the seconds aims at the dial subdivision (subdivision 1). In the opposite position, the pointer of the seconds should aim on the subdivision "7" for the $35^{th}$ sec and on "8" for the $40^{th}$ sec. Because of the pointer's (or dial's) eccentricity, the $35^{th}$ sec is located after the subdivision "7" and the $40^{th}$ sec is located after the subdivision "8". This effect is characteristic of the geared devices, as a result of a gear or axis or measuring scale eccentricity. If, now, we consider the dial line "7"=$35^{th}$ sec as the first ecliptic boundary limit of the Draconic scale, this means that the Draconic pointer aiming either out of the ecliptic limits, meaning that-no eclipse event occurs although it actually does, or within the ecliptic limits, meaning that an eclipse event occurs although it actually does not. Watch suffering from eccentricity, from first author's collection. B) A single photograph via a Maksutov telescope (diameter 150mm, Focal length 1500mm) of the double star β-Cygni, named Albireo. C) Multi-combined image of 201 single images of 1 sec exposure. As a result of the gears eccentricity of the equatorial mount, the two stars are continuously recorded in different position on the CCD camera and the stars' shift to east-west direction, is evident. Images by the first author.

The most common error owed to the user is the reading error. For example, the reading error with an old analog voltmeter/ampere meter is the parallax effect between the user's eye and the indication needle, which moves to indicate the measurement results. The reading error becomes extremely critical when the measurement results concern the *on/off* selection: For a measurement just right at the limit, the user must decide for a procedure that is "*on*" or "*off*" situation.

Measurements which are too close to the *on-off* limit, or just right *on* the limit, have high uncertainty: the user cannot judge which side of the limit the measuring needle lies on (resolution error) and the indeterminacy of the measurement rises (see Supplementary material).

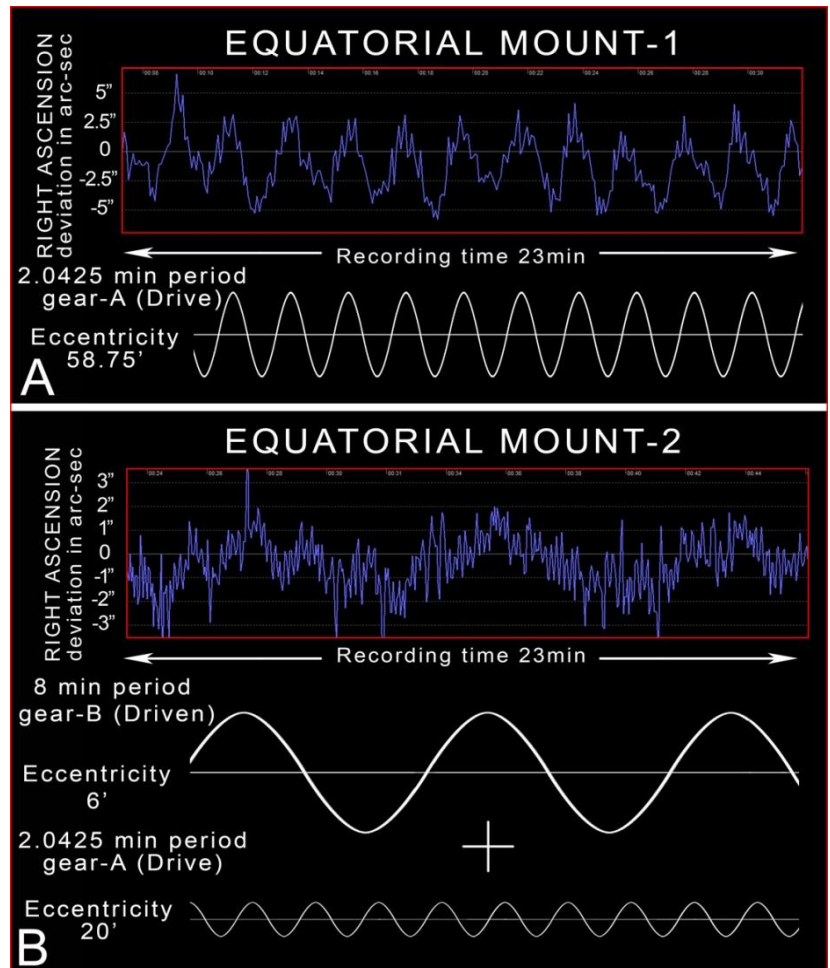

**Figure 4:** Periodic errors measurements (via a guiding software-*PHD2* Guiding Project and *PHD2 Log Viewer*) were detected in both of first author's equatorial mounts. The graphs represent the periodically varying position of a star (as it is focused on a CCD camera, via a telescope), through time (both graphs 23 min recording). Both mounts have the same design of the clock drive but different number of teeth on their drive/driven gears. The two graphs differ because the error of the eccentricity on the top graph concerns only one gear (drive gear-A with 9 teeth) and on the bottom graph the error of eccentricity appears in two gears (drive gear-A with 12 teeth and driven gear-B with 47 teeth). If the gears were perfect constructions, the graph should have appeared as a horizontal straight line coinciding with the x-axis, i.e. deviation "0" in Right Ascension, representing a constant angular velocity of the motor. The high frequency variations (9 picks per period of gear-A/Equatorial mount-1 and 12 picks per period of gear-A/Equatorial mount-2) is the variation in the transmission motion *tooth by tooth* from gear-A to gear-B. Even the two gears constructed with the involute teeth profile, the non-perfect transmission is visible on the graph (see also **Figure 7**).

As regards geared measuring instruments used for time measurement, such as clocks, equatorial mounts for telescopes, heliostats etc., there are mechanical errors that affect the results of an instrument's measurements and operation.

The most common mechanical errors of a geared device are the gear/axes eccentricities **Figure 3 and 4**, the axes/pointers eccentricities or precession, the gear teeth non-uniformity, and the imperfect transmission rotation between the gear teeth (the triangular shape of the gear teeth of the Antikythera Mechanism makes this problem even worse, see further below).

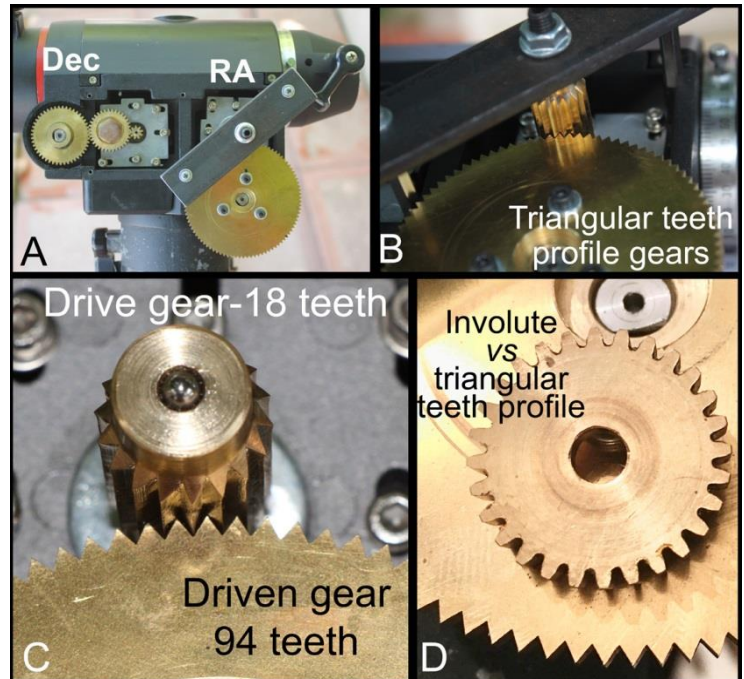

**Figure 5:** A) On the Equatorial mount-1, the two gears of the clock drive were replaced by gears with triangular tooth profile, as are the gears of the Antikythera Mechanism. The gears of the Declination axis (Dec) were not replaced. A bar with a screw (B) and a bearing ball (C) was placed in order to avoid the precession of the drive gear axis during its rotation. By applying a fast rotation of the input gear, the triangular teeth engagement produces a characteristic sound like a "*gun machine rattle*" (see authors' Video-1). D) The difference between the involute and the triangular teeth profile. Construction and images by the first author.

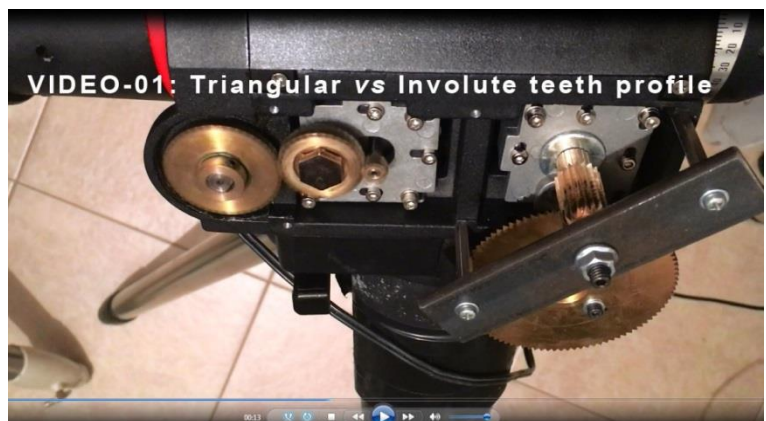

**Video-01:** Authors' video presents the rotation and the sound of gears with involute and triangular profile teeth, adapted on the Equatorial mount-1.

For time measuring by using geared devices, the error of eccentricity and the gear teeth non-uniformity, create mismatches between the true time and the calculated time:

The eccentricity is defined as the difference between the axis of rotation and the axis of symmetry. On a gear which suffers from eccentricity, the central hole of the gear does not coincide with the center of the circle defined by the teeth peaks, known as addendum circle, or with the center of the circle defined by the valleys, known as dedendum circle.

A gear (or an axis) eccentricity is an inherent and permanent mechanical problem of the equatorial mounts made for telescopes (also for heliostats/coelostats). These mounts are geared devices having a well-timed axis rotation. The equatorial mount compensates the Earth's rotation by only using one axis of rotation called *polar axis*. The polar axis rotates at the same (and constant) angular velocity as the celestial objects (sidereal for the stars, solar for the Sun etc.), about 1 turn/24h. The geared system (motor and gears) that rotates the polar axis is called clock drive.

As the equatorial mount compensates for the Earth's rotation, a celestial object, e.g. a star, when observed through the eyepiece of a telescope, should appear fixed on the center of field of view. Due to the eccentricity errors in the clock drive gears, the fixing of a star in the field of view is not perfect: The star periodically drifts back and forth (east-west) relative to a central point. This effect is continuously repeated during all of the time of observation **Figure 3 and 4**.

The motor of the clock drive rotates at a constant angular velocity (i.e. x-steps/time = constant). The eccentricity(ies) of the gears transform the constant velocity of the motor to a periodically variable velocity, resulting in the drifting of the star in the field of view.

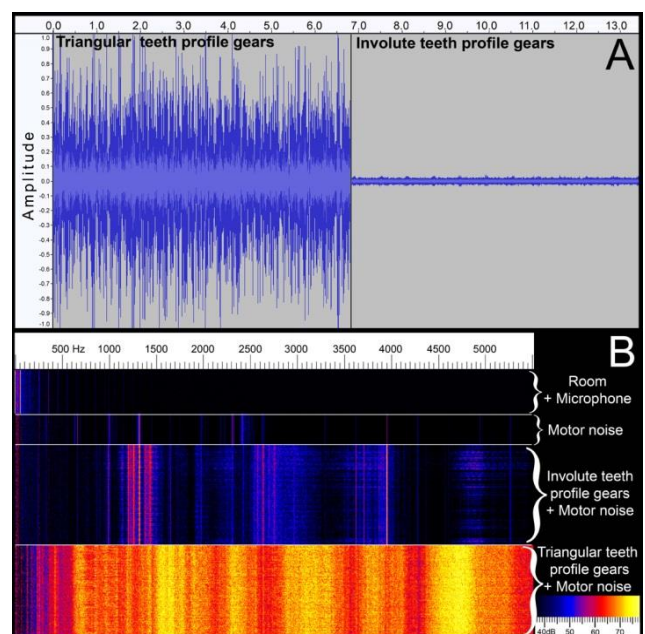

**Figure 6: A)** The sound amplitude difference between the operation of gears with triangular teeth profile (left graph) and the gears with the involute teeth profile (right graph). The sound graph of the gears with triangular teeth profile is equal to the sound graph produced by a "*gun machine rattle*". **B)** The sound spectrum analysis between 0-6000Hz. First spectral strip: the spectrum of noise from the recording room and the self-noise of the microphone (Average intensity ≈27dB). Second strip: the spectrum of the motor sound without engaged gears. Third strip: the sound spectrum during the rotation of the gears with involute teeth profile and the sound of the motor operation. The room and the microphone noise were subtracted (Average intensity ≈53dB). Fourth strip: the spectrum of the gears with triangular teeth profile and the sound of the motor operation. The spectral band concerns the full spectrum in average intensity ≈74dB and resulted from the strikes between the teeth of the engaged gears, because they are not in constant contact during their engagement (see also **Figure 7**). On the contrary, the sound spectrum produced by the gears with involute teeth has much lower intensity and presents broadband gaps.

There is no mechanical solution to fix the problem of eccentricity, but there is a solution via electronics and software: The PEC (*Periodic Error Correction*) procedure: via an educated algorithm, the motor is not rotated at constant velocity, but at variable (inversed) velocity, in order to compensate for the error of eccentricity (see Supplementary Material).

An additional error in the gearings is the transmission error in *tooth by tooth* engagement. The involute tooth profile offers the best uniformity of motion transmission between the engaged gears.

As the Earth rotates with a fixed period, the gears adaptation on an equatorial mount can be used for the quality evaluation of the gears' construction and their errors detection: In order to examine the motion transmission (*tooth by tooth*) of gears with triangular tooth profile, the two gears (with involute profile) of the clock drive/Equatorial mount-1 (A=9 teeth and B=47 teeth) have been replaced by two gears constructed with triangular tooth profile and having the same ratio of teeth numbers (drive gear *a*=18 teeth and driven gear *b*=94 teeth), **Figure 5** and **Video-01**. Both of gears were designed, constructed, adapted and tested by the first author.

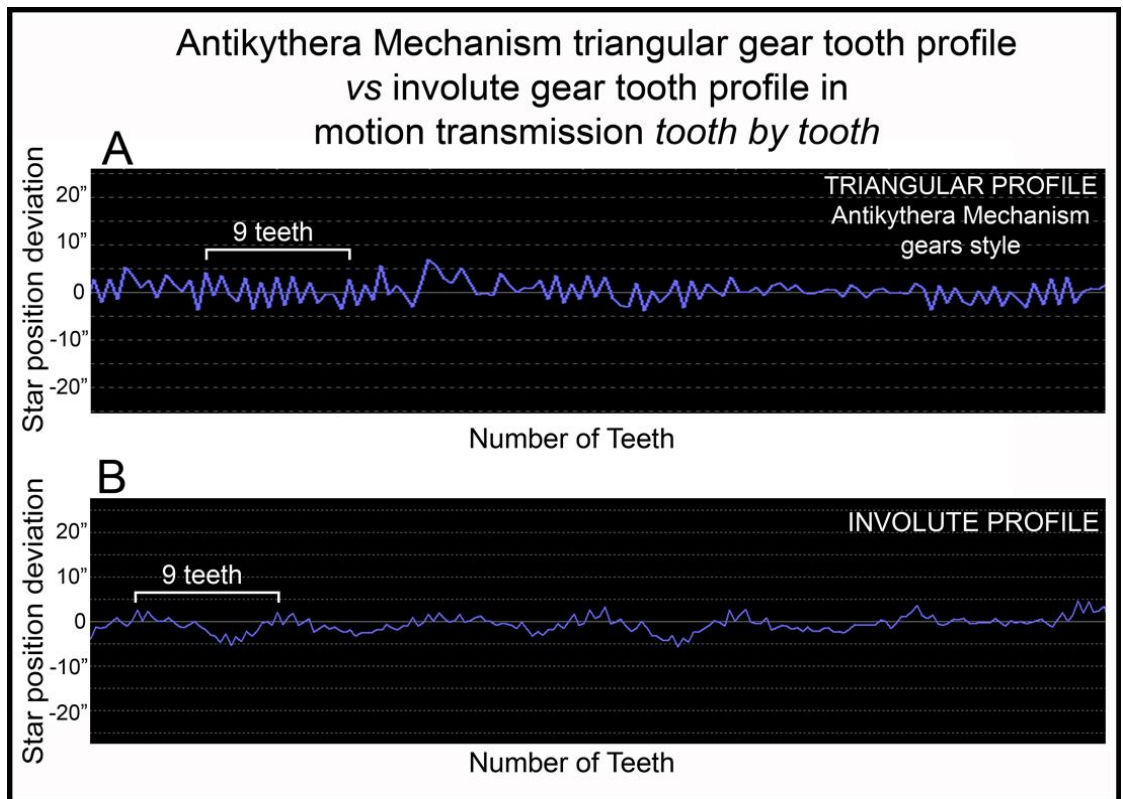

**Figure 7:** The graphs present the *star position deviation vs number of gear teeth*. **A)** The motion transmission by adaptation of gears with triangular teeth profile on the clock drive of the Equatorial mount-1. **B)** The motion transmission of the original gears with involute teeth profile. The motion transmission *tooth by tooth* in the gears with triangular tooth shape profile is much worse than the modern gears, about two times. If the motion transmission was perfect, the graph should be a smooth line without picks. Note that the modern gear-A (Drive) presents the error of eccentricity.

During the rotation of the gears we recorded the sound, by stabilizing a microphone on the Equatorial mount-1. The evaluation in the gears' errors and the transmission motion can be achieved applying the acoustic analysis (Korpel 1968; Gaylard et al., 1995; Van Riesen et al., 2006; Tavner 2008). The signals were digitized via an ADC sound card and afterwards were processed by the software *Audacity*, *Spectrum Lab* and *Radio Sky Pipe*. The amplitude and the sound spectrum graph analysis, show the strong noisy sound during the rotation of the gears with triangular teeth profile in comparison to the rotation of the gears with involute teeth profile **Figure 6**.

The graphs of **Figure 7** shows that the triangular tooth shape does not offer a smooth motion during the gears' rotation. The *tooth by tooth* motion appears more as intermittent motion.

## 5.2 The error of eccentricity in the Antikythera Mechanism gear(s)

By studying the AMRP radiography (Antikythera Mechanism Research Project) of Fragment A we found that the ancient Manufacturer fitted a thin copper sheet (thickness 0.1-0.2mm) into the central hole of the m2-gear: the Manufacturer probably made the central square hole of the gear by accident a bit larger (a mistake also made by the first author during the gear construction) and then he adapted the copper sheet in order to stabilize the gear on its shaft **Figure 8**. The ancient Manufacturer may also have detected a high eccentricity in this gear (the hole of this gear was not exactly centered to the gear's geometrical center) and he tried to eliminate it, by enlarging the hole and fitting this copper sheet into it. Both of these justifications of the copper sheet fitting, related to the existence of the eccentricity errors in some of the Mechanism's gears.

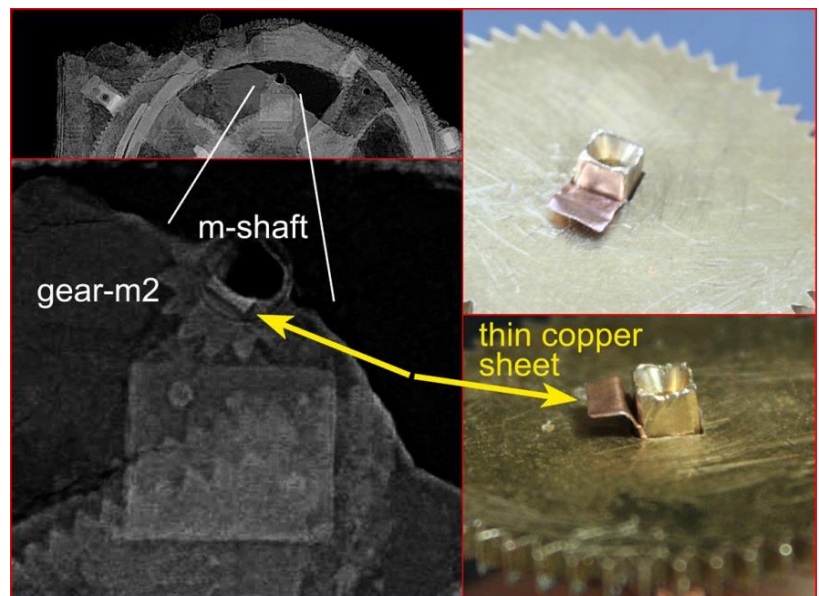

**Figure 8:** In the radiography of AMRP of Fragment A (processed by the authors), a thin part adapted on the square hole of gear m2 is visible. This thin copper sheet was in contact with the square shaft of m2 gear (which is poorly preserved). A reproduction of this observation is presented in the visual images by the first author. The copper sheet stabilizes (tightens) the gear to its axis.

As aforementioned, the eccentricity of the gears is an inherent mechanical error of the geared devices and it is mostly created during the construction of the gears.

The error of Eccentricity makes the ratio "*epicenter angle φ/teeth*", which otherwise would be constant, a variate:

When a shaft rotates at a constant angular velocity and the gear with eccentricity is adapted on the shaft, then the gear presents during its rotation a variable linear velocity (measured at the perimeter of the gear/teeth), because the radius of the gear differs. Therefore, the ratio "*tooth/time*" is not constant. This means that some of the teeth are moved faster and the teeth located opposite to them are moved slower.

A gear with radius $R_0$= 4.2mm and 15 teeth (m2-gear), having eccentricity *e*= 0.2mm (the copper sheet thickness), presents a max angle difference Δφ relative to the ideal gear:
tan(*φ*)= e/$R_0$ → tan(*φ*)= 0.2/4.2 → Δ*φ*= 2.79°, **Figure 9**.

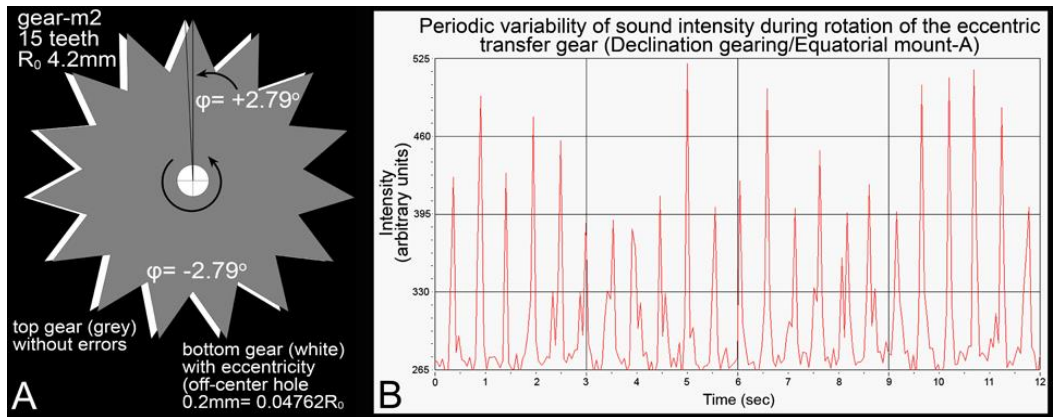

**Figure 9: A)** Two same gear scheme (m2 gear), in grey the (top) gear without errors and in white the (bottom) gear with eccentricity of 0.2mm. The centers of the gears coincide and the positioning difference of the gear teeth is visible. **B)** The periodic variability graph of the sound intensity (via *Radio Sky Pipe* software), during the operation of the Declination gearing (Equatorial mount-1). The intensity variability is due to the eccentricity of the transfer gear (see **Figure 5**).

## 5.3 The error of the gear eccentricity impacts the Antikythera Mechanism pointers' position

When gear A has an error of eccentricity and is engaged to gear B (with same teeth number and without errors), the periodically variable angular velocity of gear A is transmitted to gear B. If we stabilize a pointer on gear B, the pointer will be periodically rotated faster (for the first half period) and then slower (for the second half period), even though the input of the gearing is rotated at a constant angular velocity (see also Edmunds 2011).
Gear m2 is engaged to the (lost) gear n1 (53 teeth), which rotates the Metonic pointer (see gearing scheme Freeth et al., 2006 and Voulgaris et al., 2018b). The variable angular velocity of gear m2 (as result of the eccentricity) is transmitted to the gear n1, suppressed by the gearing reduction: $2.79° *(15/53) \to \Delta\varphi \approx 0.79°$ on the Metonic pointer. Taking into account that each Metonic cell corresponds to $\approx 7.66°$, the eccentricity phase of $\pm 0.79°$ is not critical.

Let us examine a different condition: If there was no error of eccentricity, the Lunar pointer would aims at the Golden sphere-Sun and at the same time the Draconic pointer would aim just right of the ecliptic limit - indicating a solar eclipse occurrence. But if the Draconic pointer/gear presents an eccentricity of $\pm 2°$, then during the phase of delay (-2°) the Draconic pointer aims 2° before the ecliptic limit, i.e. out of the ecliptic "window" - indicating no eclipse event occurrence (the inverse effect is also valid), **Figure 10**.

Even if no error of eccentricity existed in the Draconic gearing but an eccentricity of 0.2mm existed on gear b1 (130mm diameter and 225 teeth), then the angle difference $\Delta\theta$ would be $\tan(\theta) = 0.2/65 \to \Delta\theta = 0.176°$. The angle difference would be transmitted via the engaged gears of the Draconic gearing (multiplier gearing) to the Draconic pointer as:
$\Delta\theta*(b1/a1)*(r1/s1) = 0.176*(225/48)*(63/22) \approx \pm 2.36°$.
During one rotation of the b1, the Draconic pointer is rotated about 13.42 times.
For the first 13.42/2 = 6.71 rotations, the Draconic pointer would be rotated with gradually increasing angular velocity up to $\Delta\theta = +2.36°$, whereas for the next 6.71 rotations it would be rotated with decreasing velocity up to $\Delta\theta = -2.36°$ (while it should be rotating at a constant velocity).

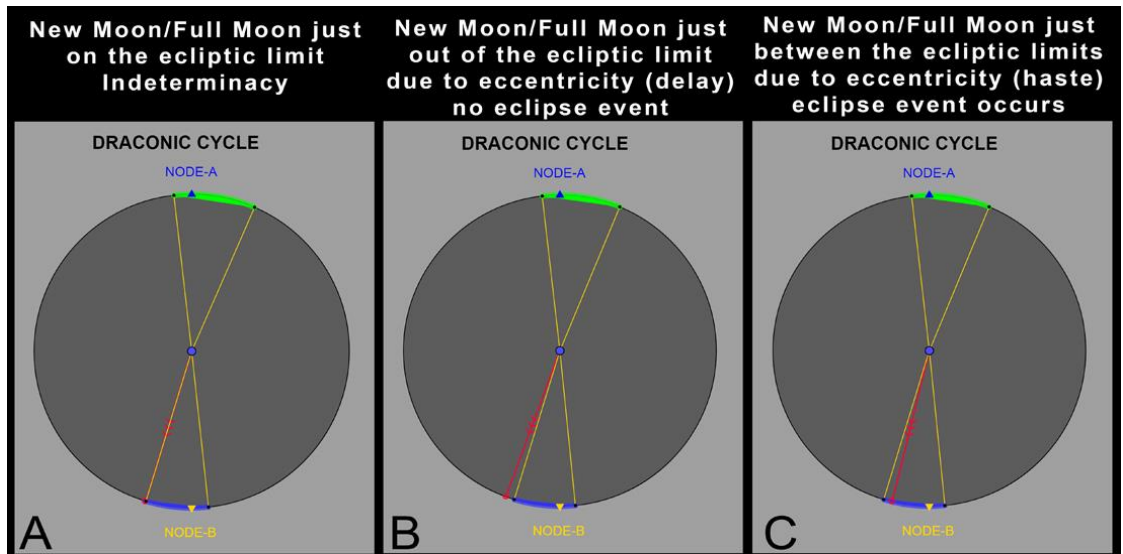

**Figure 10: A)** The actual calculated position (via *DracoNod*) of the Draconic pointer during the New Moon of Cell-65. The pointer aims just right on the ecliptic limit. This position presents a high indeterminacy. **B)** Via a gearing instrument, the Draconic pointer position is out of the ecliptic window (no eclipse event), due to the eccentricity error of -3° (phase of delay). **C)** The Draconic pointer position is inside the ecliptic window (eclipse event occurs), due to the eccentricity error of +3° (phase of haste). The gearing errors affect the position of the pointers and the final results.

Even if no error of eccentricity was existed in gear b1, but it did in gear b2, which is stabilized on gear b1, then the error of eccentricity would be transmitted to gear b1. An eccentricity of 0.1mm on gear b2 corresponds to an angle difference ≈ ±4.86° on the Draconic pointer.

Therefore, the error of eccentricity affects the positioning of the gears on which the pointers are adapted, i.e. the extracted results/calculations of the Antikythera Mechanism (Edmunds 2011).

The ratio "teeth/time" is also not constant if there is any dissimilarity on the gear teeth, i.e. the teeth of a gear have not the same dimensions or shape. This error is called pitch deviation.
For example, a (hypothetical) random accidental construction defect on a tooth of gear-b2, resulting in a shorter dimension by ≈15% corresponds to an epicenter angle of about 0.84° ≈1 subdivision on the Zodiac dial. Due to this error, the conjunction of the Lunar and the Golden sphere differs than to its original position, and changes the position of the Golden sphere-Sun on the Zodiac dial ring-Ecliptic.
These random errors also produce false results in a random distribution.

The omitted and the additional events were predicted by *DracoNod* they can be well justified as a result of the Draconic pointer eccentricity or the eccentricity of the (gearing/pointing) system of the Lunar Disc/Golden sphere-Sun, Figure 10.

In the *world of the hand-made geared devices* (Voulgaris et al., 2019a) using conventional machine tools of the Middle Ages, the Renaissance and the early Industrial Revolution, the gear errors (around 0.1mm-0.4mm), as a result of the lower precision of the measuring procedures and the machine tools, are much more probable than today. In our era, the dimensional errors of the commercial mechanical constructions are significantly smaller, around 0.05mm or less. Moreover, the use of high precision, special-cutting tools and of modern computerized machines (Computer Numerical Control-CNC) offers high repeatability in parts and shapes production, with dimensional errors around 0.02mm. These precisions are achieved by the use of optomechanical measuring devices (macro lenses and high resolution machine vision cameras).

For a high precision quality of the gear teeth shaping, the cutting tools and the material that is cut, are immersed in cooled oil, in order to avoid overheating of the material during the process, which would result in thermal expansion of it.

Today, the research of the Antikythera Mechanism, carried out using computers and simulations, has significantly progressed, yielding many answers and new discoveries. However, the mechanical errors of the Mechanism's gears are not taken into account (and it is difficult to be detected/measured, as some gears are missing and the preserved parts are deformed, shrunk and in different material and volume, Voulgaris et al., 2019b). Computer calculations and results presuppose that the AM was an ideal device operating perfectly, without mechanical errors or reading errors or errors of indeterminacy. This way, the AM research is done in a "*sterilized and perfect mechanical world*".

In the "*real material world*", mechanical errors exist and they have a big impact on the final results presented by the Antikythera Mechanism. Taking into account the random/unknown errors of the Mechanism gears, the original results could differ from the (perfect/ideal) results calculated by a computer and 3D simulations (see also Edmunds 2011).

## 5.4 Re-calculating the actual ecliptic limits of the Draconic scale taking into account the gearing errors

In section 4.2 we calibrated the ecliptic limits of the Draconic gearing considering that there are no mechanical errors in the Antikythera Mechanism gearing. The calibrated ecliptic limits resulted in a common ecliptic window for the solar and the lunar eclipses:

For the Ecliptic Window-A: –22.5° and +6.5° from Node-A
For the Ecliptic Window-B: –16.9° and +5.4° from Node-B.

The relative high value of 22.5° for the ecliptic limit of Node-A and the relative low value of 16.9° for the ecliptic limit of Node-B seem quite asymmetrical and are difficult to be justified in astronomical terms. However, their mean value is 19°.7.

Taking into account the existence of the errors of eccentricity in some of the Mechanism gears (or scales and pointers), these asymmetrical values can be a result of an eccentric gear or gears or an eccentricity Draconic scale or errors in gears b2/b1 which define the synodic cycle. It seems that a complex combination of such errors is more likely. But each gear with eccentricity error affects the Mechanism's results in a different degree.

Thus, our calculated (non-symmetrical/eccentric) ecliptic limits of Nodes A and B are the result of the gear(s) eccentricities, tooth un-uniformity and Draconic pointer/scale eccentricity.

Therefore, the specific eclipse events are the outcome either of perfect gears and an eccentric Draconic scale, or a faultless Draconic scale and imperfect gears (or both of them).

The approximate solution was found with the graphic design environment, the calculated error was ±0.5° and the ecliptic limits probably adopted from the ancient Manufacturer would be:

For the Ecliptic Window-A: –19.9° and +5.7° from Node-A
For the Ecliptic Window-B: –19.4° and +6.2° from Node-B.

The mean limit values for these Ecliptic Windows are 19.65° and 5.95°, very close to 20° and 6° (see Supplementary Material).

Unfortunately, the new limits cannot be used for the Antikythera Mechanism calculations, as the specific results (eclipse events) were calculated by the specific instrument which has specific mechanical errors (obviously, by applying the calculated symmetrical limits on the Mechanism gearings without errors, the eclipse event sequence would differ from the present Saros event sequence).

We don't know which of the gears have eccentricity errors, neither the degree of the eccentricity, nor the phase (direction axis) of the eccentricity, and we'll probably never find out, as some of the Mechanism gears are missing and the preserved ones have changed as regards their composition (material) and volume, plus they are deformed, broken and shrunk.

**6.1 The Antikythera Mechanism could also calculate the eclipse times**

On some of the preserved cells with eclipse events, the ancient Manufacturer engraved the times of the eclipse events. The times mentioned are between in 12 hours daytime and 12 hours nighttime.
A question arises: What was the source of the information that the ancient Manufacturer used to define the eclipse events sequence and times of their occurrence?
- A Babylonian papyrus with written events and times of occurrence?
- A mathematical process?
- A catalogue with written observed events? (Freeth et al., 2014; Carman and Evans 2014; Iversen and Jones 2019; Jones 2020).
In the present work we showed that the eclipse events sequence could be calculated by the Antikythera Mechanism applying the Draconic gearing and the omitted events can be well justified taking into account the existing mechanical errors.
Below we present the procedure of the eclipse event time calculation by using solely the Antikythera Mechanism and no external information.

On the Antikythera Mechanism, during a full rotation of the Lunar pointer (i.e. one Sidereal cycle), the Lunar pointer scans the Zodiac month ring of the Mechanism and its subdivisions. Geminus, in his work *Introduction to the Phenomena,* in Chap. 18 refers that
1 Exeligmos = 669 Synodic cycles = 19756 days = 723 zodiac cycles +32° = 723+(32°/360°)
= 723.08888 zodiac cycles (sidereal) (*eq. 1*).
19756$^d$/723.08888 sidereal cycles = 27.3216758$^d$/Sidereal cycle ≈**655.72 hours/Sidereal cycle.**
Let us divide the Zodiac month ring in 365 equal subdivisions-days (unit of time measuring), instead of 360 un-equal degrees (unit of angle measuring) see Voulgaris et al., 2018a. The Lunar pointer sweeps the 365.25 subdivisions (365+0.25 is achieved by the rotation of one subdivision/4 years) in one Sidereal cycle:
655.72 hours/365.25 subdivisions = **1.795263 hours/subdivision** (≈**1.8 hours/subdivision).**
Therefore, one (mean) synodic cycle of 29.5306427$^d$ (resulted from *eq. 1*) corresponds to ≈ 394.78 zodiac subdivisions = one full rotation (365.25)+29.53 subdivisions.

Geminus also writes: *Η ΣΕΛΗΝΗ ΑΝΩΜΑΛΩΣ ΦΑΙΝΕΤΑΙ ΔΙΑΠΟΡΕΥΟΜΕΝΗ ΤΟΝ ΖΩΔΙΑΚΟΝ ΚΥΚΛΟΝ* , *The Moon travels the zodiac circle at variable velocity*.
On the Antikythera Mechanism, during the re-aiming of the Lunar pointer to the Golden sphere (synodic cycle), the Lunar pointer scans a number of subdivisions. This number is not constant, because the synodic cycle duration varies as a result of the *pin&slot* variable angular velocity.
By applying the equations for the angular velocity produced by the *pin&slot* factor (without any external information or modern lunar theory), the synodic cycle duration varies between 29.31$^d$–29.81$^d$, thus the lunar pointer could scan 391.9–398.55 subdivisions in a synodic cycle.

For each synodic cycle, the ancient manufacturer measured the total number of the (Zodiac) subdivisions and then multiplied by 1.8 in order to transform them into ecliptic hours (on the Back Cover Inscription Part-II, the words *ΕΓΛΕΙΠΤΙΚΟΙ ΧΡΟΝΟΙ* (Ecliptic hours) are mentioned, Bitsakis and Jones 2016). Afterwards, using a timed map of the Ecliptic he transformed the ecliptic hours into hours of daytime/nighttime.

This way, the calculation of the eclipse event times, is achieved by only using the Antikythera Mechanism without any use of external information which is not directly related to the Mechanism.

Moreover, if he copied the time information from a papyrus or by applying mathematical calculation models, this could create a mismatch between the times of the papyrus and the times appearing on his creation on the Zodiac dial ring. As the Zodiac subdivisions/hours correspond to a specific position of the Zodiac sky at which the eclipse occurs, a mismatch could place the eclipse at a different position than where it appears on the Mechanism's Zodiac subdivisions.

## 6.2 Errors in the eclipse events times calculations, as a result of the gears/axes constructional imperfections

The calculation of the eclipse times demands high accuracy: The eclipse time calculation is subject to a bigger relative error than the eclipse events calculation:
- The eclipse events calculation is based on the synodic month.
- The time of an eclipse event is based on the duration of the synodic month calculated in hours, i.e. 24X29.53$^d$ = 708.7$^h$, meaning that the eclipse time calculation demands accuracy that is 2.85 orders of magnitude higher than the synodic cycle calculation.

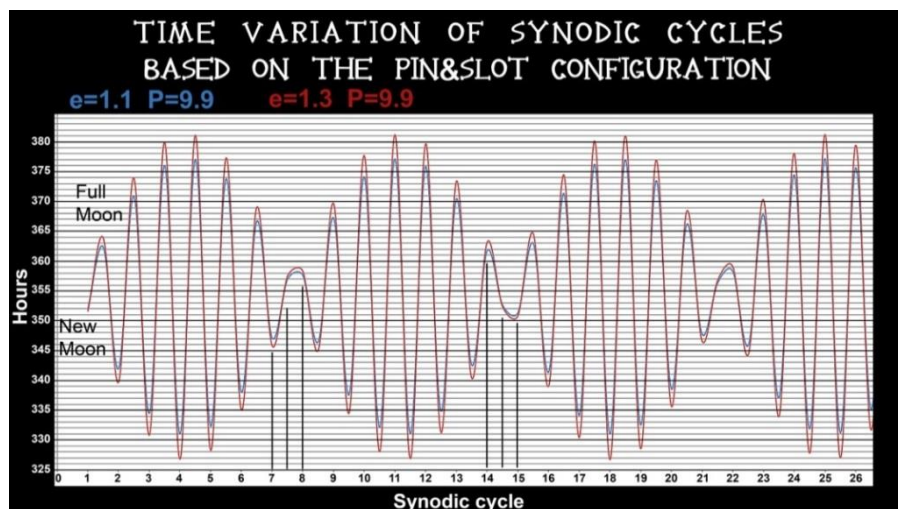

**Figure 11:** The graph(s) present the duration of each half synodic month (New Moon to Full Moon to New Moon), which was calculated based on the *pin&slot* configuration of the Antikythera Mechanism. X-axis: Synodic months-Saros cells (New Moon phase in Integer numbers, the Full Moon phase in integer +0.5. Y-axis: each half synodic month duration in hours. The mean synodic month is 708.7354 hours, and the mean half synodic month is 354.3677 hours. Blue graph: the results are based on current measurements of the pin position and the eccentricity of k-axis. Red graph: By increasing the value of k-axis eccentricity by =0.2mm, the half synodic month duration increases significantly (≈+8 hours).

Taking into account that in a geared machine there are constructional imperfections/mechanical errors, the time events calculation results would probably be affected by such errors.

On the Antikythera Mechanism, the variable duration of a synodic month is defined by the *pin&slot* operation. The dimensional characteristics of the *pin&slot*/eccentricity *e* of k-axis and the distance P of the pin from the gear axis (Voulgaris et al., 2018b), strongly affect the motion variability and the duration of each synodic month calculated by the Antikythera Mechanism gearing. The current values for the eccentricity of k-axis is e=1.1mm and for the pin distance P=9.9mm (measured on the parts in their present condition as the bronze material and volume has changed into Atacamite, Voulgaris et al., 2019b).

By slightly increasing the eccentricity *e* of k-axis by +0.2mm (= 1.3mm) the hours of the events could vary by about ±8 hours, **Figure 11**. Keeping the value for k-axis eccentricity at 1.1mm and increasing the pin distance at 10.1mm, the synodic month duration changes by about one hour.

**6.3 Sky screening during an eclipse event**

During an eclipse event, e.g. a solar eclipse, the lunar pointer aims at the Golden sphere-Sun and the solar pointer aims at a zodiac subdivision. This subdivision belongs to a specific zodiac constellation. Let us assume that the solar pointer-ΗΛΙΟΥ ΑΚΤΙΝ aims at the first zodiac subdivision of Leo in which the index letter Π is engraved. The index letter Π, probably corresponds to the Parapegma event **Π-ΛΕΩΝ ΑΡΧΕΤΑΙ ΕΠΙΤΕΛΛΕΙΝ** (Leo begins to rise), which is partially preserved on the PPI-2 (Bitsakis and Jones 2016).

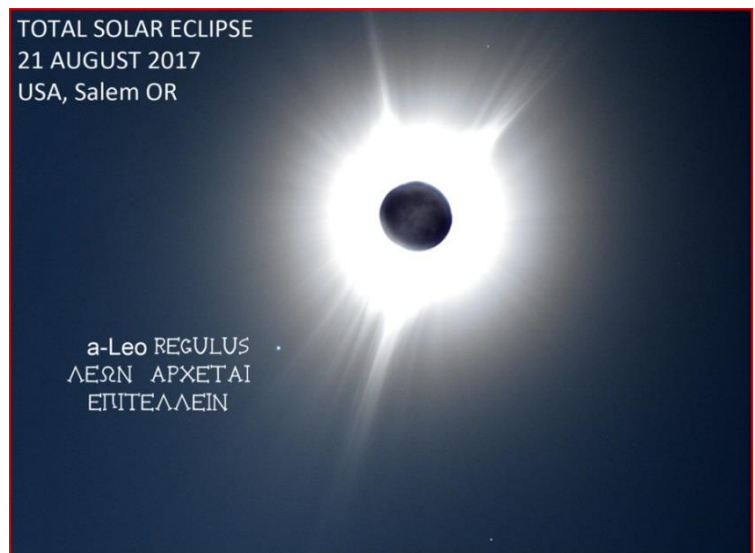

**Figure 12:** An overexposed processed wide angle photograph of the total solar eclipse of August 21, 2017, shot from Salem, Oregon,USA. Prominent are the bright inner corona (saturated), the extended outer corona with streamers, the polar plumes, the New Moon surface details and a number of stars. The brighter star is a-Leo-Regulus, which was visible to the naked eye during totality. The day of the eclipse, the Sun and Regulus rose at about the same time. Image by the first author.

The brightest star of Leo is called ΒΑΣΙΛΙΣΚΟΣ (Regulus). Therefore, during the specific solar eclipse, the Sun, the Moon and Regulus are in conjunction. If the solar eclipse is total, then Regulus is visible to the naked eye during totality **Figure 12**, as also mentioned by *Diodorus of Sicily* in his *Library of History* (XX: 5–6), describing the total solar eclipse of Agathocles, which occurred on August 15, 310 BC: *On the next day there occurred such an* *eclipse of the Sun that utter darkness set in and the stars were seen everywhere* (Stephenson et al., 2020).

This way the Antikythera Mechanism provides information regarding the background sky at which an event will be occurred.

## 7. Conclusions

Based on the results of this work, the suggested by the authors Draconic gearing appears less hypothetical and more likely to have really been a part of the Antikythera Mechanism, as it gives satisfactory results and answers to a number of important questions, that related to the specific eclipse events sequence and their times of occurrence. It also removes the question "*why did the ancient Manufacturer include in his creation only the three out of the four lunar cycles, rejecting the fourth, very important lunar cycle, well known and used during the Hellenistic era?*" It seems that the ancient Manufacturer was an experienced astronomer and constructor and knew for sure that the Draconic cycle is the key to real eclipse prediction. We strongly believe that the ancient Manufacturer, who lived before 178 BC (Voulgaris et al., 2022ii, unpublished results), constructed the Mechanism in order to find out the forthcoming eclipse events and the hours of their occurrence. By introducing the four lunar cycles he could predict the eclipse events sequence and the relative data based entirely on a mechanical procedure without any external information.

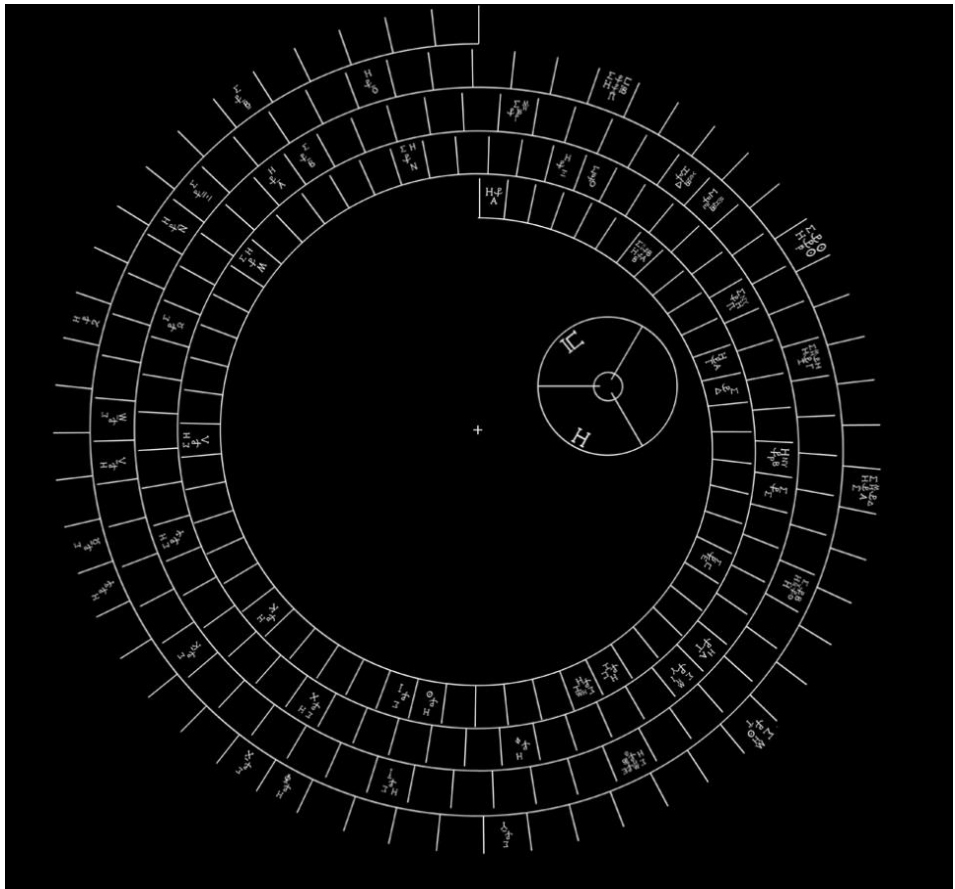

**Figure 13:** The reconstructed eclipse events of the Saros spiral using the *DracoNod* program and applying the Draconic gearing. The predicted events compromised according to the preserved events and the index numbering. The 64/(65) events in total are distributed in 50 cells.

When he finished the assembly of his creation, the ancient Manufacturer set the Mechanism pointers on their initial positions according to the starting date of December 23, 178 BC (Voulgaris et al., 2022ii unpublished results https://arxiv.org/abs/2203.15045) and at that time the 223 cells of the Saros spiral were blank cells - no engraved information.

Then, he started to operate his construction: when the Lunar pointer aimed at the Golden Sphere-Sun/New Moon (or in opposite position Full Moon) and the Draconic pointer was located somewhere between the ecliptic window, a solar eclipse would occur.

Afterwards, the ancient Manufacturer engraved the specific event and the time of its occurrence on the cell at which the Saros pointer aimed, **Figure 13**.

It seems that the ancient Manufacturer had the remarkable idea to construct a time-machine, an astronomical event predictor based solely on the mechanical procedure involving the engaged gears. This must have been a great achievement for that time!

His creation was a real predicting device of the future astronomical events. But the constructional imperfections/gearing errors created some deviations in the pointers' position by about 3°-4° (or 5°), affecting the final results.

In any measurement attempt, the use of a measuring instrument impacts more or less the outcome. Even the light emitted/reflected from an object, when is collected by an objective lens, is affected by the objective lens errors, which change the Reality, the Truth and the Originality of the object's information, known as *Module Transformation Function* (Hecht 2016; Vanidhis 2022). The smaller the impact of an instrument on the measurement, the better the proximity to the true value and the closer to representing Reality.

Today, we can measure dimensional errors that are impossible to be detected by the human eye. At the Hellenistic era, the only perceived dimensional/constructional errors were those that could be detected by naked eye, or by using some simple measuring tools (e.g. a compass), also having as a criterion the naked eye. As a result of the limited ability of error measuring, the mechanical calculations for the eclipse events prediction differed from current calculations that are based on the more complex calculations and the digital manipulation of the data, which is by about 4.5 orders of magnitude more accurate.

## Acknowledgement

*We are very grateful to prof. Xenophon Moussas (National and Kapodistrian University of Athens University) member of AMRP, who provided us with the X-ray Raw Volume data of the Mechanism fragments. Thanks are due to the National Archaeological Museum of Athens, Greece, for permitting us to photograph and study the Antikythera Mechanism fragments. We also thank G. Pistikoudis of "Astroevents in Greece" company for his hospitality and support in Trilofos of Thessaloniki, North Greece, in order to measure the periodic errors of our equatorial mounts and D. Varvarousis for his help for the data recording during the penumbral eclipse of 2020.*

\* \* \* \* \* \* \* \* \* \* \* \* \* \* \* \* \* \* \* \* \* \* \* \* \* \* \* \* \* \* \* \* \* \* \* \* \* \* \* \* \* \* \* \*

# SUPPLEMENTARY MATERIAL
# Data presentation, Comments, and Notes

**Aristeidis Voulgaris[1], Christophoros Mouratidis[2], Andreas Vossinakis[3]**

[1]City of Thessaloniki, Directorate Culture and Tourism, Thessaloniki, GR-54625, Greece,
[2]Merchant Marine Academy of Syros, GR-84100, Greece,
[3]Thessaloniki Astronomy Club, Thessaloniki, GR-54646, Greece

[1]Corresponding author arisvoulgaris@gmail.com

## 1. DracoNod program information

We present the integrated eclipse events sequence of the Saros spiral, including the lost eclipse events. The lost eclipse events reconstruction was achieved by the use of the authors' program named *DracoNod*, which presents the phase correlation of the three (four) lunar cycles Synodic, Draconic and Anomalistic and Sidereal), in the graphic environment. *DracoNod* program presents three cycles with their corresponding orbital radius and cycles subdivisions.

1st Circle: **Synodic cycle**, Period 1/223 of Saros. Two subdivisions: New Moon and Full Moon,
2nd Circle: **Draconic cycle**, Period 1/242 of Saros. Two subdivisions: Node-A and Node-B,
3rd Circle: **Anomalistic cycle**, Period 1/239 of Saros. Two subdivisions: Apogee and Perigee.

Synodic cycle index bar: for Number X = Cell (X+1) (New Moon phase @ 29th or 30th day of synodic month), for Number (X+0.5) = Cell (X+1.5) (Full Moon phase @15th day of synodic month). The epicenter angle of the ecliptic windows are depicted in yellow lines. The ecliptic windows depicted as arcs are presented in green and blue colors. The cycles' phase position is presented with the three independent orbital radii, black for Synodic cycle, red with three arrows for Draconic cycle and green for Anomalistic cycle.

When New Moon is located between the ecliptic limits (ecliptic window in green and in blue color) a solar eclipse occurs (symbol H-Helios-Sun on the cell-X). When Full Moon is located between the ecliptic limits, a lunar eclipse occurs (symbol Σ-Selene-Moon on the cell-X+0.5). When Full Moon and New Moon of the same synodic month is located between the ecliptic limits, then the symbols Σ+H exist on the same cell.
The events presented as **Cell number/Event index**. The preserved eclipse events are in green color background. The phrase "*No event*" is in red color background.

The program starts with New Moon at Node-A and at Apogee according to Geminus definition for Exeligmos/(Saros) (Voulgaris et al., unpublished results, https://arxiv.org/ftp/arxiv/papers/2203/2203.15045.pdf).

## 1.1 Antikythera Mechanism pointers' starting position

**Cell 01/A1 (end of Cell-01): Solar eclipse event (H). New moon at Node-A and at Apogee. Saros period begins.**

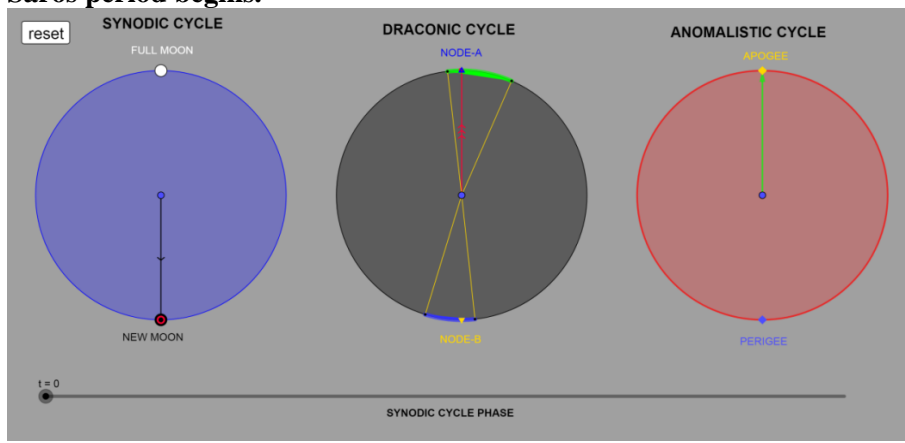

## 2. Indeterminacy error or gearing errors

Below, we present two characteristic events, which were predicted from *DracoNod*:
- On cell 65 no event occurs: The position of New Moon is just on of the ecliptic limit (a solar eclipse occurs or not?). The specific position of the Draconic pointer (red radius) presents a high indeterminacy. It is difficult for the user to decide if the Draconic pointer is located *on* or *out* of the ecliptic limit.

**Cell 65: New Moon just on the ecliptic limit. High indeterminacy.**

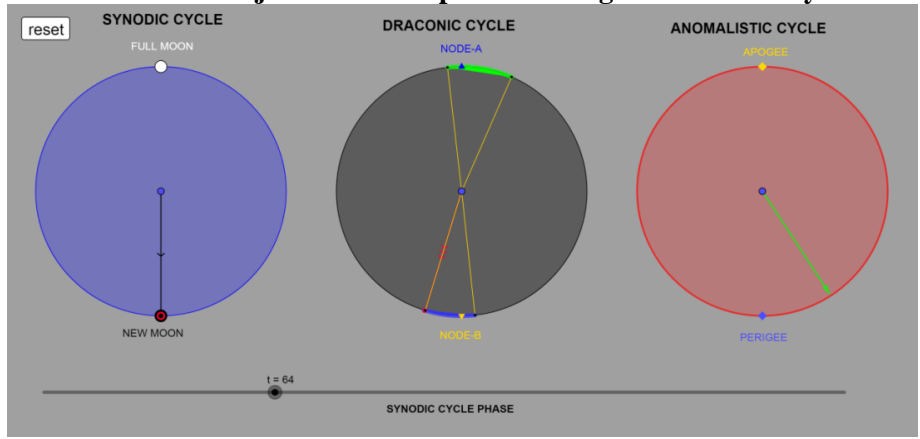

**Cell 166: Lunar eclipse event (Σ). Full Moon to close to ecliptic limit, but inside the ecliptic window - event at the limit.**

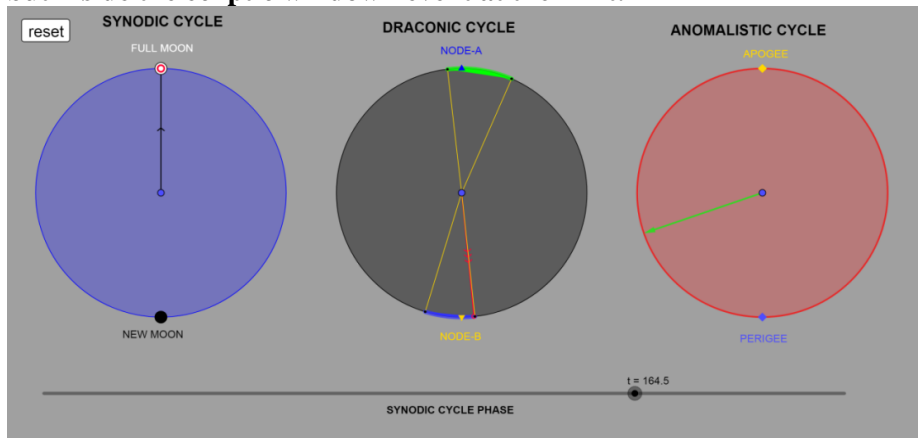

Additionally, the gear(s) error (eccentricity or teeth un-uniformity) affects the position of the Draconic pointer creating a deviation from the original position see section **7**. Events just on the ecliptic limit can be considered *inside* or *outside* the ecliptic window due to the gearing errors. On the preserved eclipse events sequence, three events should be existed. However, they are not engraved on the Saros spiral (Freeth 2014; Carman and Evans 2014; Anastasiou et al., 2016, Iversen and Jones 2019).

## 3. Ecliptic limits and Eclipses

Some characteristic positions of the Moon relative to a Node are generally discussed:
- If the New Moon is located exactly at the Node, then this solar eclipse will be total or annular and the shadow path will be projected around the Earth's equator.
- If the New Moon is located to the north and faraway from a Node, the eclipse will be visible from the northern parts of the Earth.
- If the New Moon is located just right to the southern ecliptic limit, the solar eclipse will be visible from the South Pole and will be a partial eclipse.
- If a Lunar eclipse occurs just right on the Node, it is a total Lunar eclipse.
- If the Full Moon is located faraway from a Node, then this eclipse is partial.
- If it is located too close to an ecliptic limit, then it is a Penumbral eclipse.

When the New Moon/Full Moon is located exactly at a Node, the Gamma of the eclipse will be too close to 0 and the eclipse shadow will be very central.

There are two characteristic patterns when three eclipses occur in two successive synodic months:
- If a total/annular solar eclipse occurs and just right on the Node (https://eclipse.gsfc.nasa.gov/SEplot/SEplot2001/SE2020Jun21A.GIF), then one fortnight (half synodic month) before (https://eclipse.gsfc.nasa.gov/LEplot/LEplot2001/LE2020Jun05N.pdf) and one fortnight after this date (https://eclipse.gsfc.nasa.gov/LEplot/LEplot2001/LE2020Jul05N.pdf) a Lunar penumbral eclipse will occur.
- If a total lunar eclipse occurs just right on the Node (https://eclipse.gsfc.nasa.gov/LEplot/LEplot2001/LE2018Jul27T.pdf), then one fortnight before (https://eclipse.gsfc.nasa.gov/SEplot/SEplot2001/SE2018Jul13P.GIF) and after (https://eclipse.gsfc.nasa.gov/SEplot/SEplot2001/SE2018Aug11P.GIF) this date, two partial Solar Eclipses will be visible alternately from the Earth's poles.

Two patterns of successive events are preserved on the Saros spiral:
Σ+H (in one cell), or
H+Σ in successive cells.
Not both of the eclipses can occur on/close to the Nodes. If one of the eclipses occurs on the Node (H: total/annular solar eclipse or Σ: total lunar), the second eclipse will occur faraway from the Node (Σ: partial lunar eclipse or H: solar eclipse visible in high latitudes).
The deviation (parallax) of an observing place from the Sun-New Moon line strongly affects the visibility of a total solar eclipse, which out of the totality path is seen as a partial eclipse. Usually, the width of the eclipse path for a large duration total solar eclipse (eclipse magnitude 1.0396) is about 260km and for a hybrid eclipse (coverage close to 100%, i.e. at the limit between total and annular eclipse) is about 2km-50km.

## 4. The errors of the *Instrument-User* system

https://www.watelectrical.com/different-types-of-errors-in-measurement-and-measurement-error-calculation/;
https://www.webassign.net/question_assets/unccolphysmechl1/measurements/manual.html;
https://www.elprocus.com/what-are-errors-in-measurement-types-of-errors-with-calculation/).
(https://www.wikihow.com/Read-a-Multimeter).

## 5. The Periodic Error Correction-PEC

The PEC (Periodic Error Correction) procedure: via an educated algorithm, the motor is not rotated at constant velocity, but at variable (inversed) velocity, in order to compensate for the error of eccentricity.
http://eq-mod.sourceforge.net/eqspeed.htm
https://astrojolo.com/gears/mount-periodic-error/.

# 6. Predicting of the eclipse events sequence using the *DracoNod* program

**1) Cell 01/A1: Solar eclipse event (H). New moon at Node-A and at Apogee. Saros period begins.**

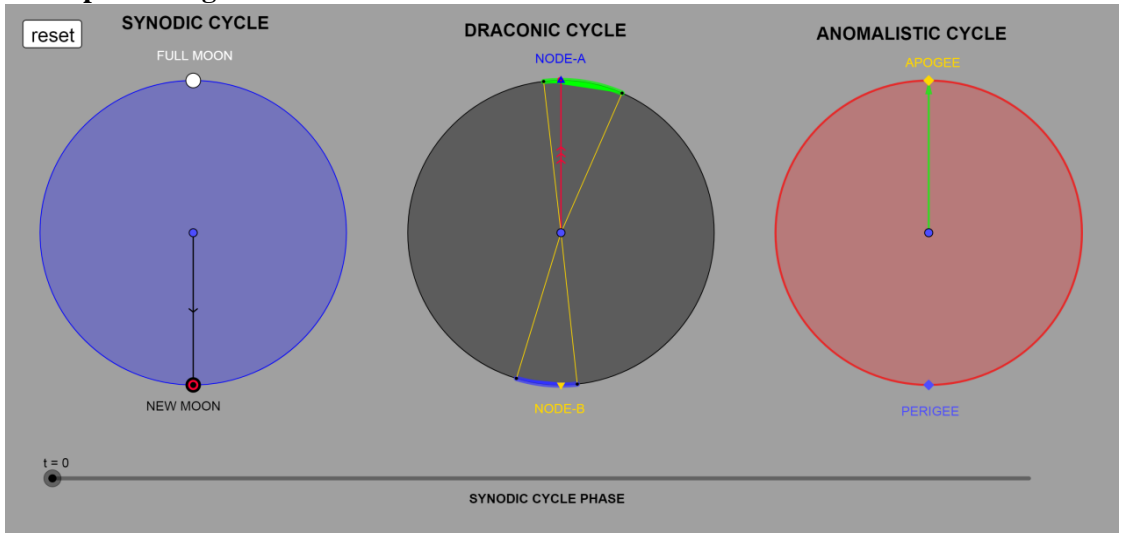

**2) Cell 07/B1: Lunar eclipse event (Σ).**

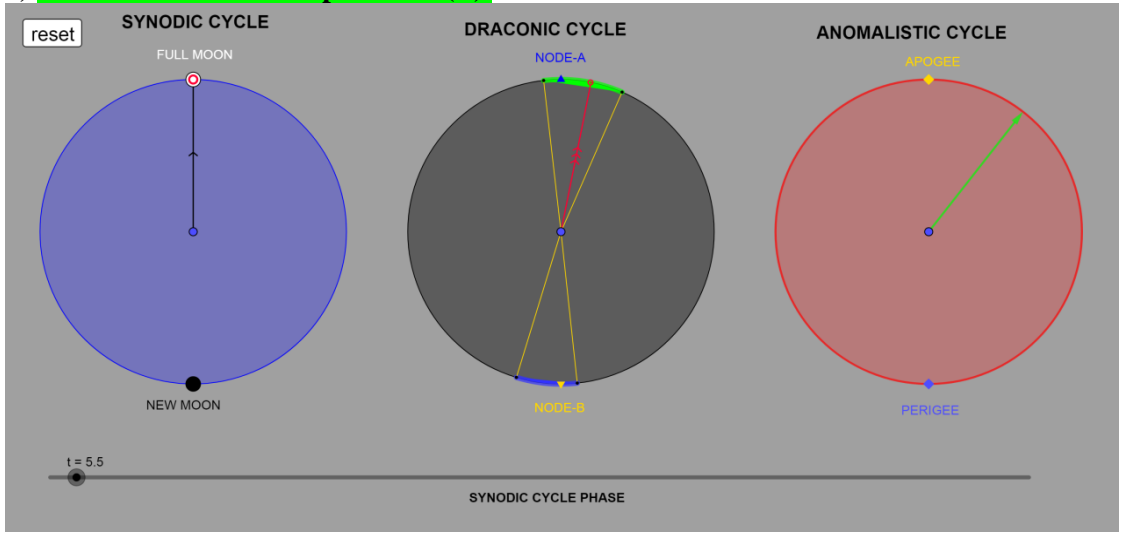

**3) Cell 07/B1: Solar eclipse event (H). New Moon close to the ecliptic limit.**

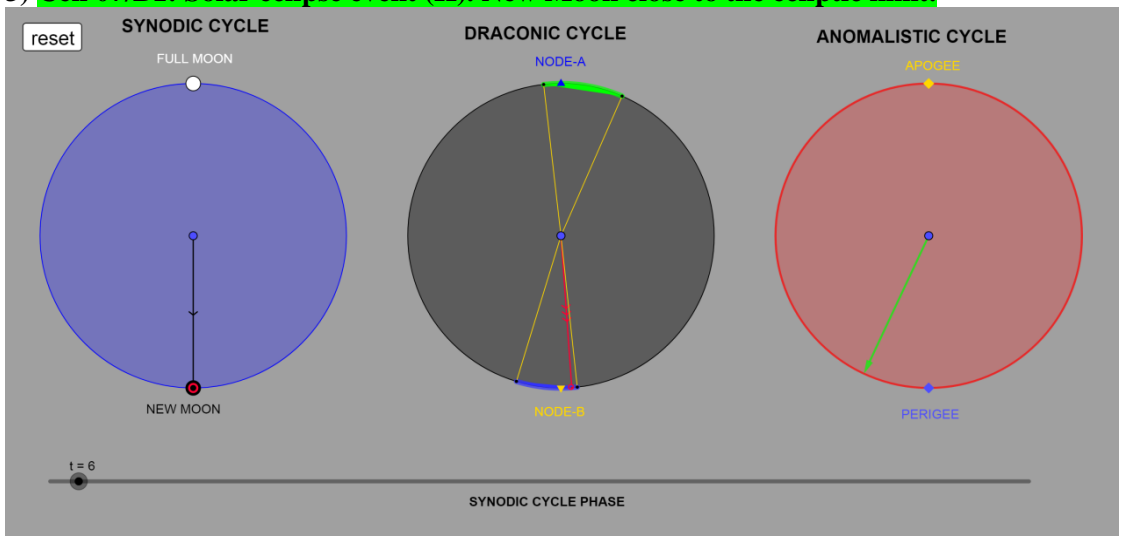

**4) Cell 12/Γ1: Solar eclipse event (Η). New Moon close to the ecliptic limit.**

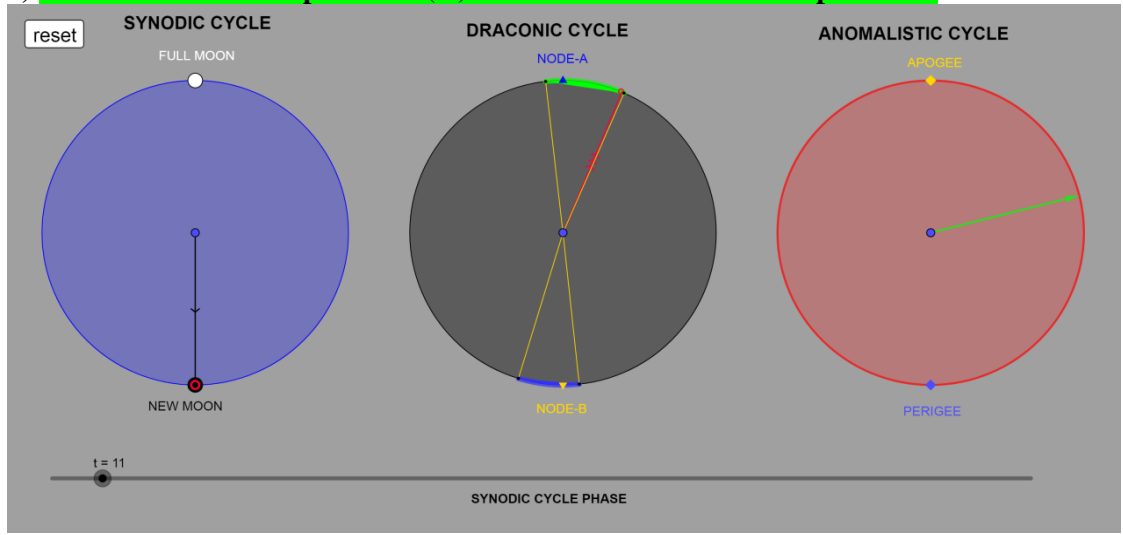

**5) Cell 13/Δ1: Lunar eclipse event (Σ).**

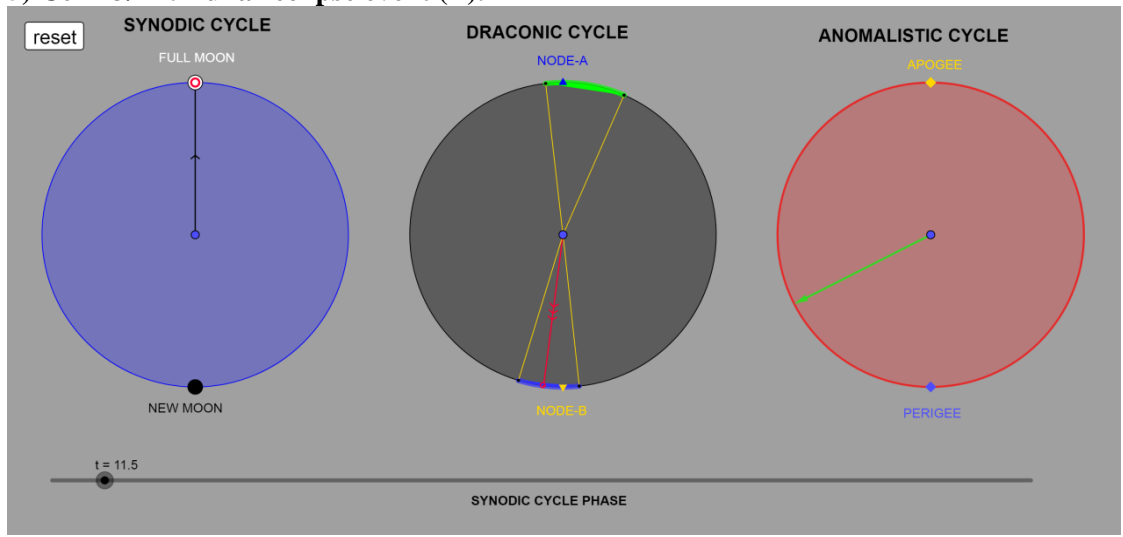

**6) Cell 19/Ε1: Lunar eclipse event (Σ). Full Moon close to Node-A.**

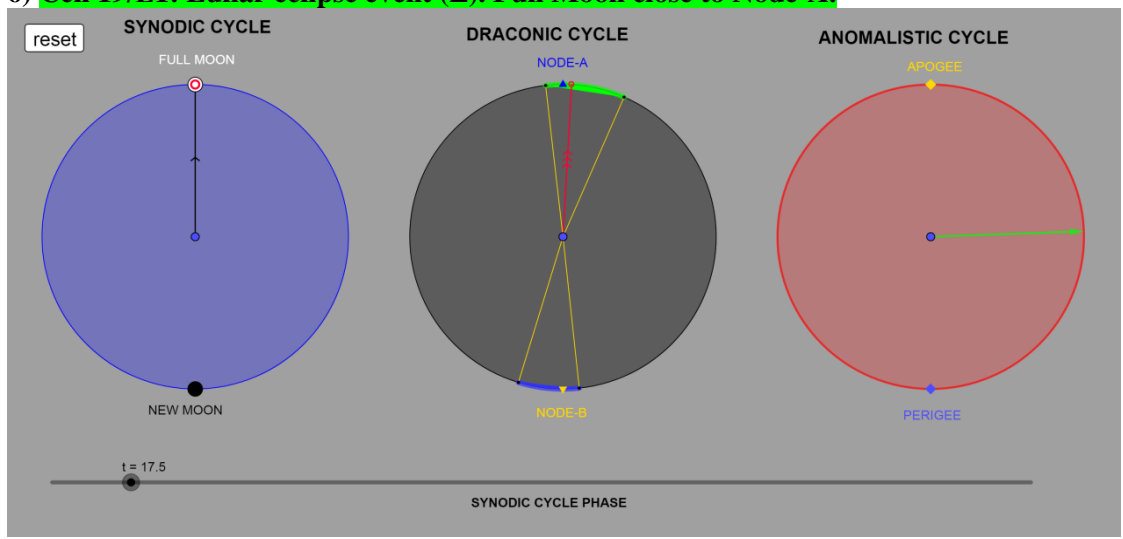

**7) Cell 24/Z1: Solar eclipse event (H).**

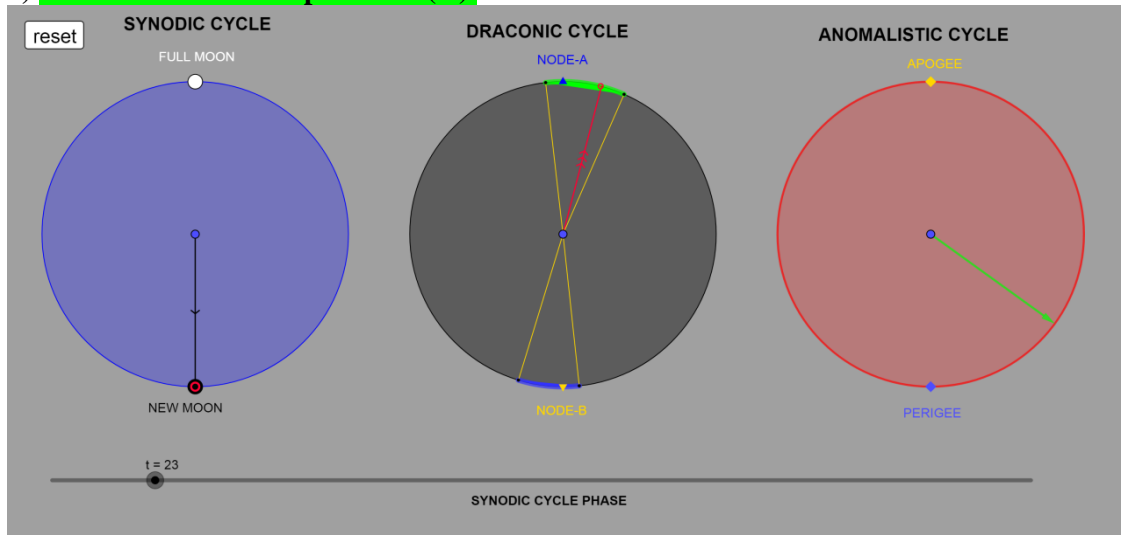

**8) Cell 25/H1: Lunar eclipse event (Σ). Full Moon at Node-B.**

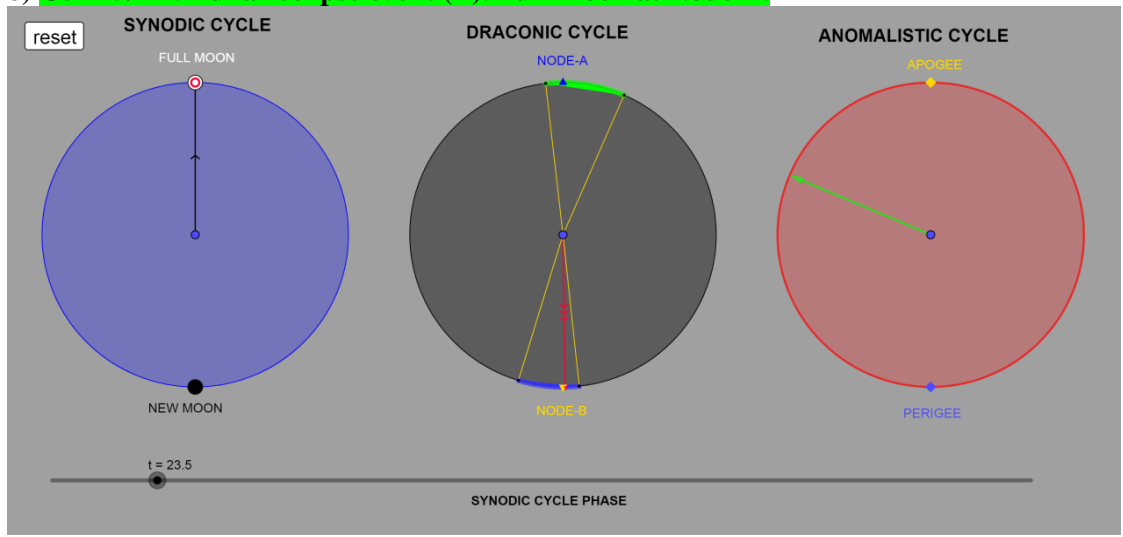

**9) Cell 30/Ө1: Solar eclipse event (H).**

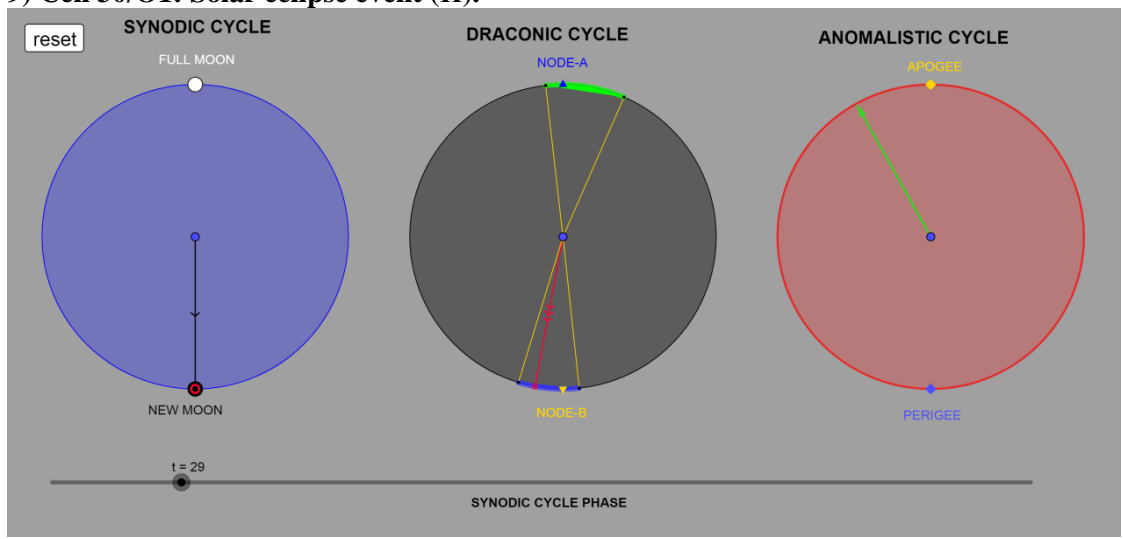

**10) Cell 31/I1: Lunar eclipse event (Σ). Full Moon close to ecliptic limit.**

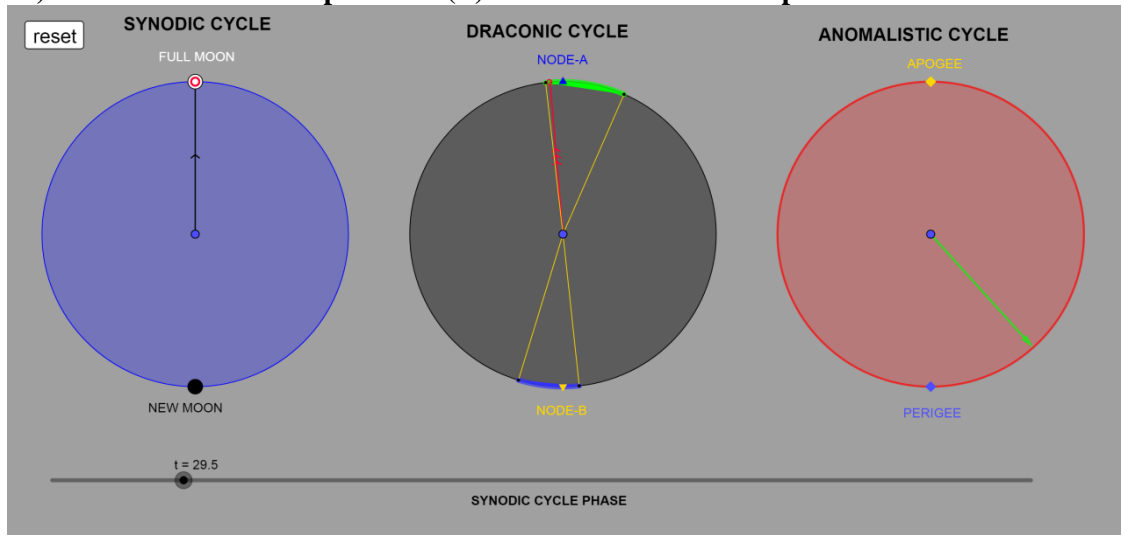

**11) Cell 36/K1: Solar eclipse event (H). New Moon close to Node-A and at Perigee.**

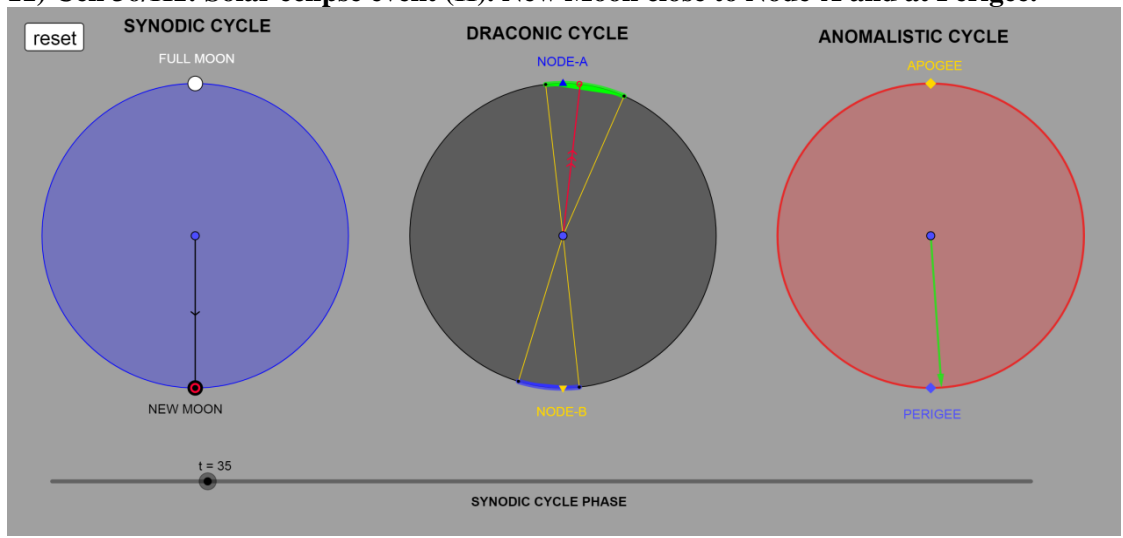

**12) Cell 42/Λ1: Lunar eclipse event (Σ).**

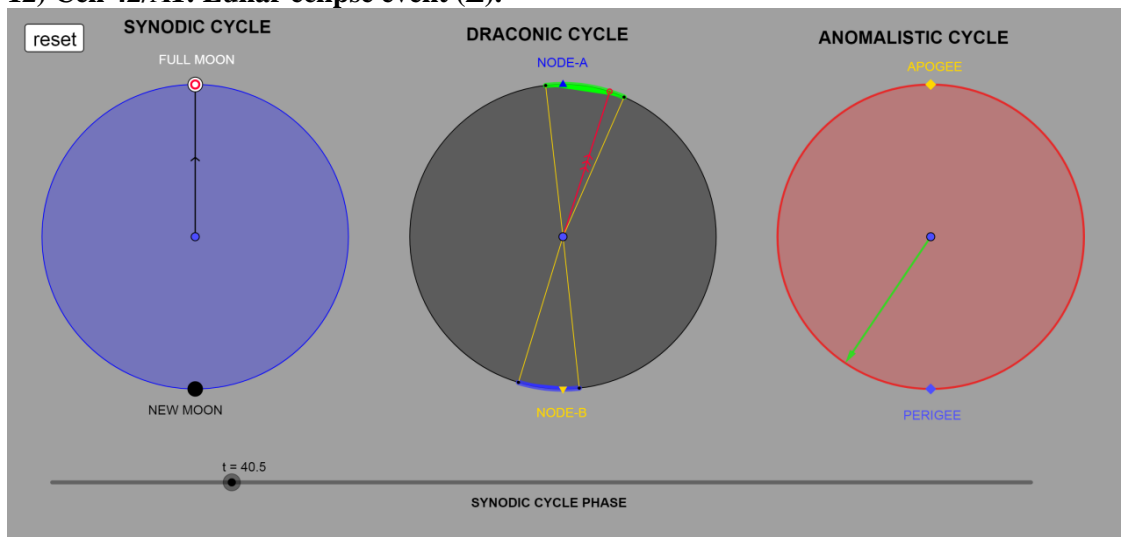

**13) Cell 42/Λ1: Solar eclipse event (H). New Moon close to Node-B**

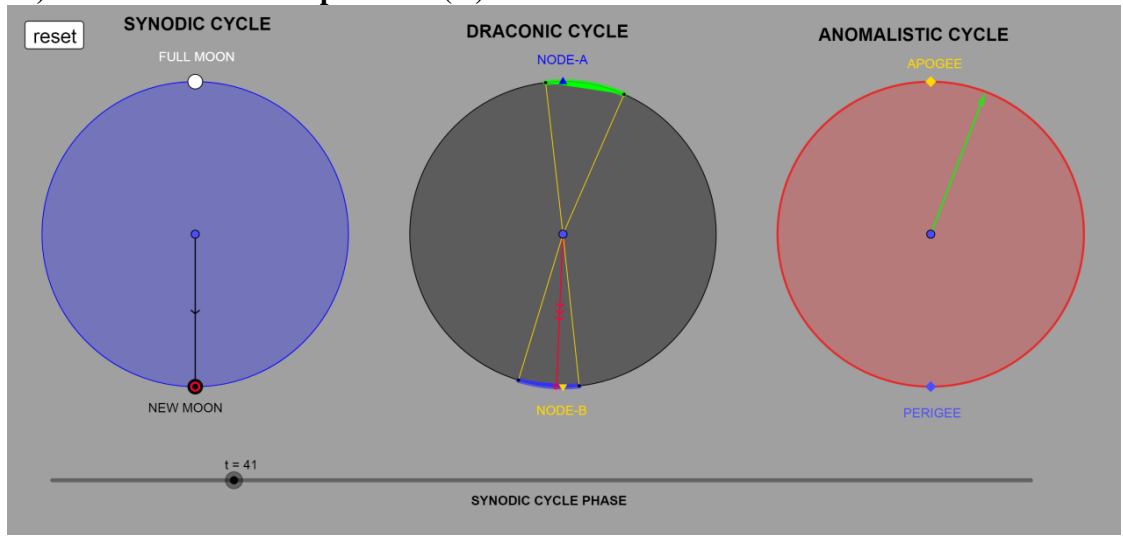

**14) Cell 48/M1: Lunar eclipse event (Σ). Full Moon close to ecliptic limit**

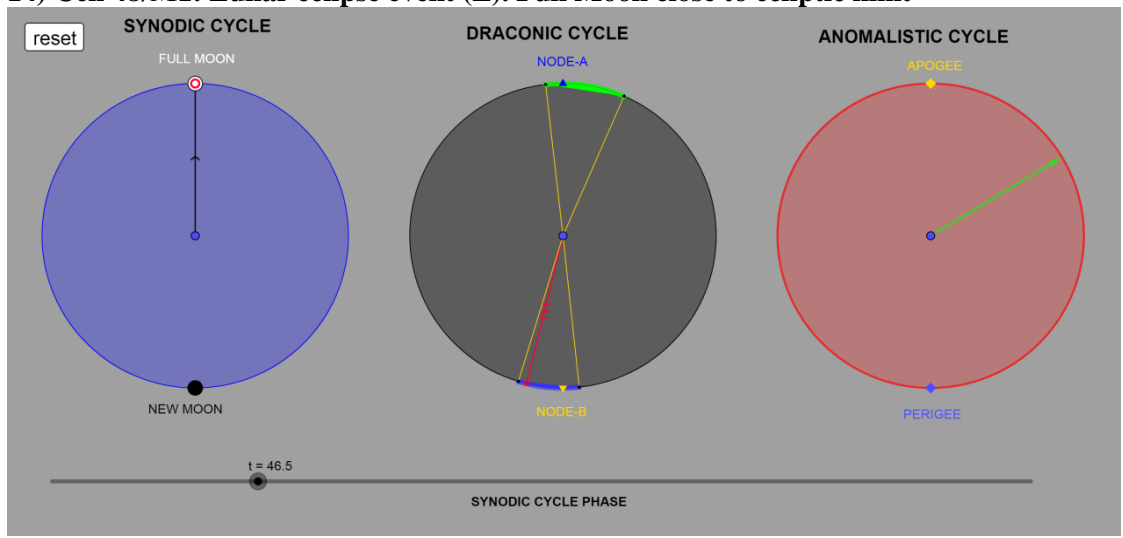

**15) Cell 48/M1: Solar eclipse event (H). New Moon close to Node-A**

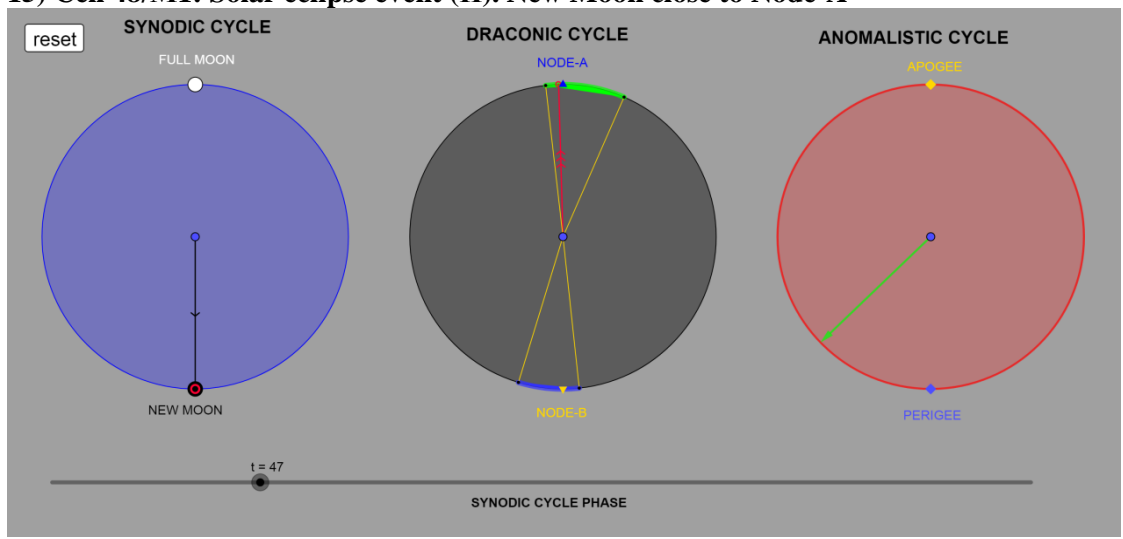

**16) Cell 54/N1: Lunar eclipse event (Σ).**

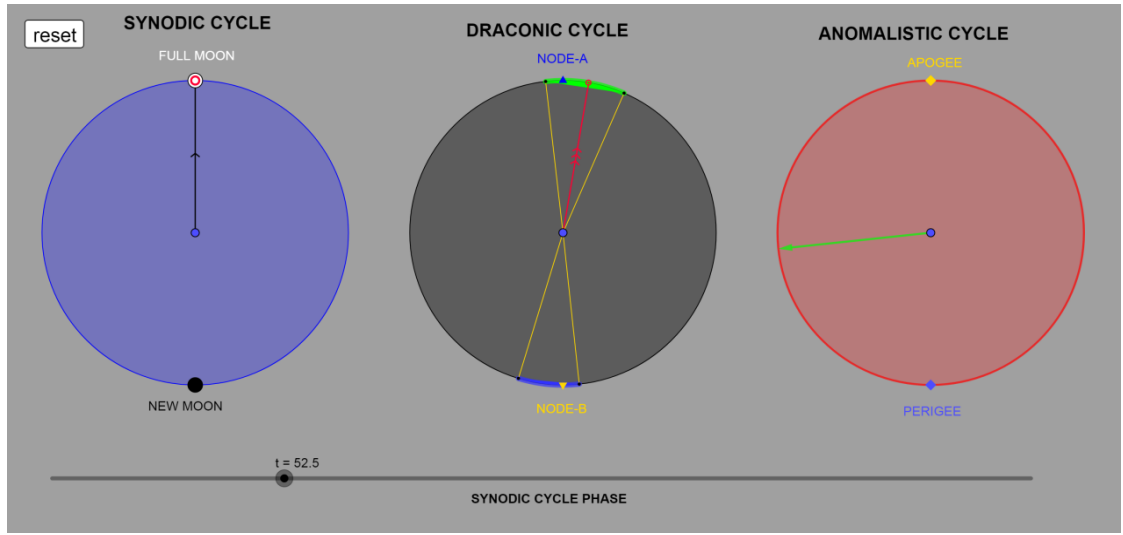

**17) Cell 54/N1: Solar eclipse event (H). New Moon on the ecliptic limit but inside the ecliptic window**

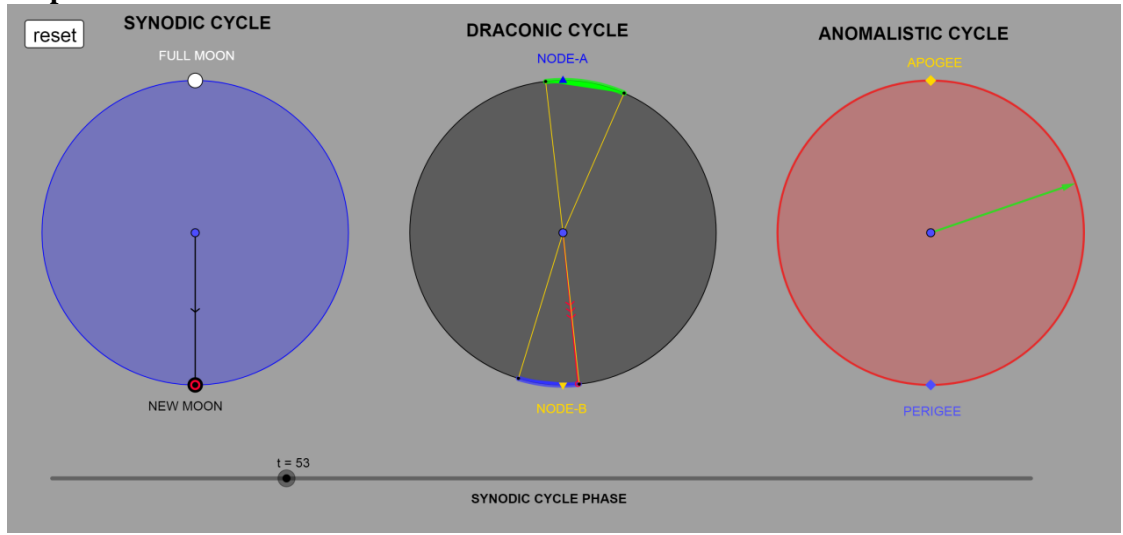

**18) Cell 59/Ξ1: Solar eclipse event (H). New Moon too close to the ecliptic limit**

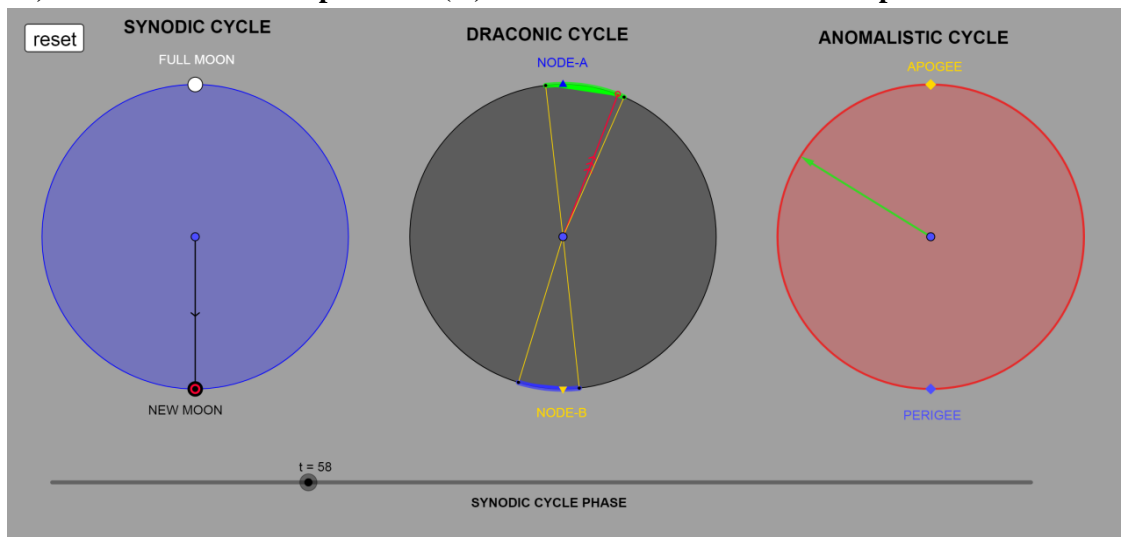

**19) Cell 60/O1: Lunar eclipse event (Σ). New Moon close to Node-B.**

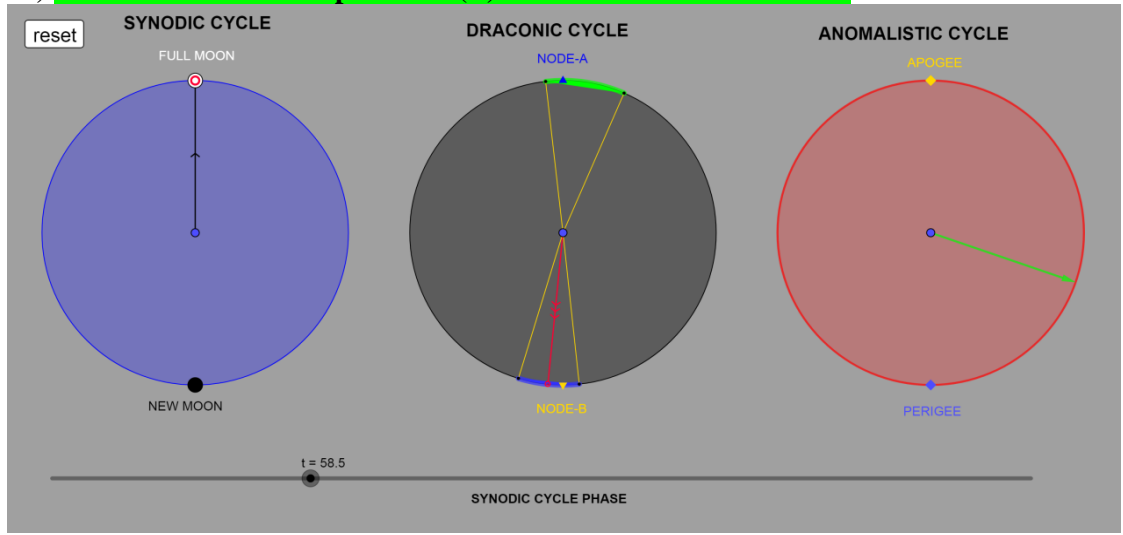

**I) Cell 65: Full Moon just right on the ecliptic limit. Based on the events' index numbering it should be no engraved event. Indeterminacy or eccentricity error.**

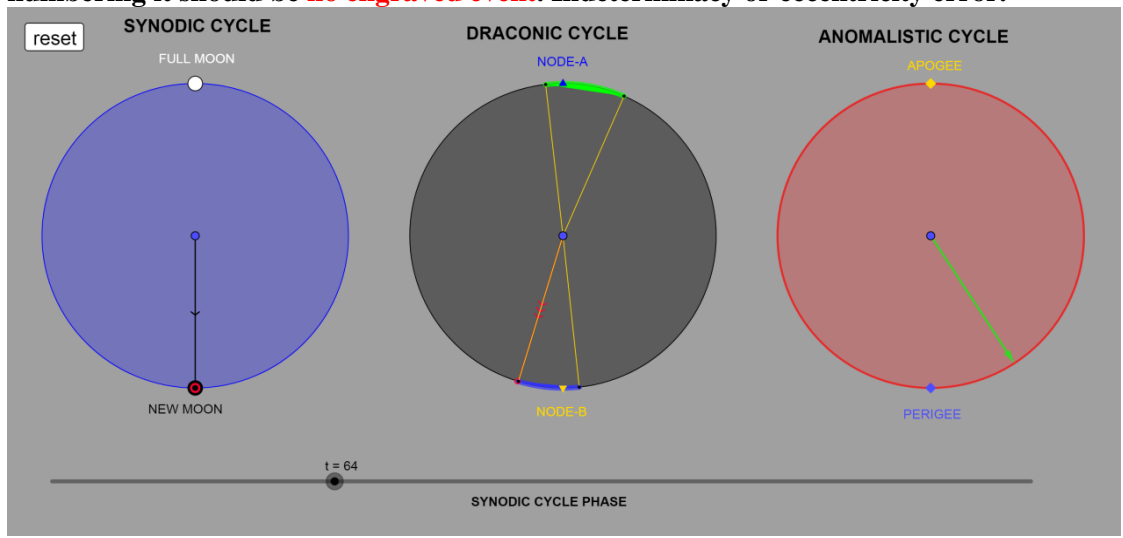

**20) Cell 66/II1: Lunar eclipse event (Σ). New Moon too close to Node-A.**

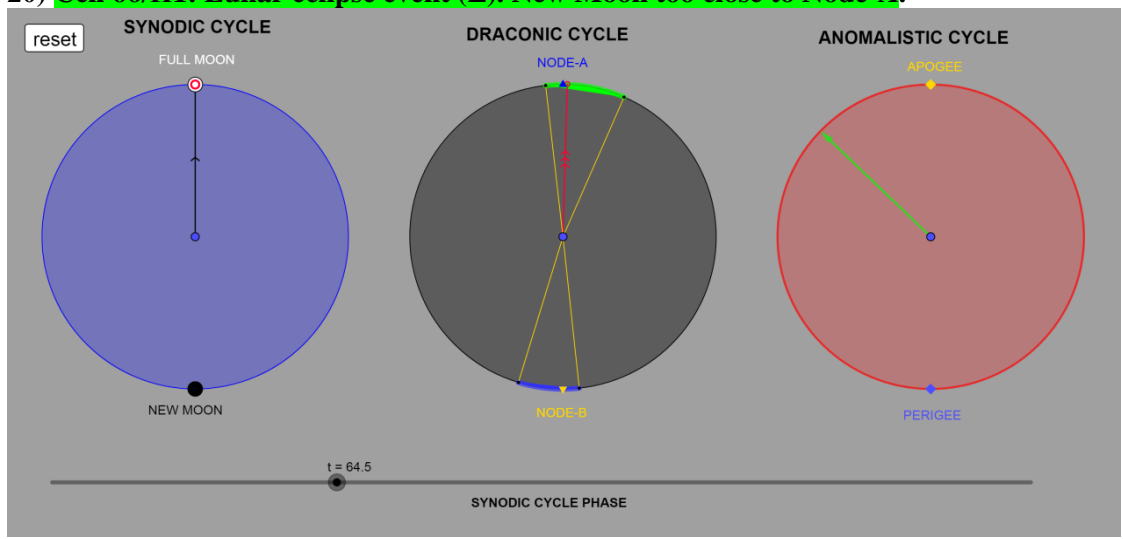

**21) Cell 71/P1: Solar eclipse event (H). Full Moon closes at Apogee.**

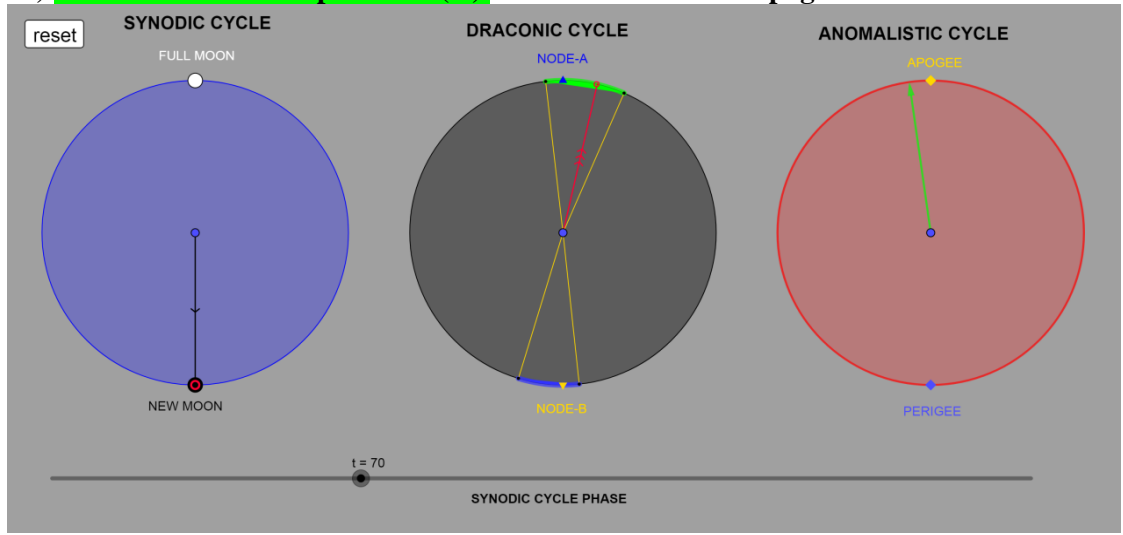

**22) Cell 72/Σ1: Lunar eclipse event (Σ). Full Moon closes to Node-B and at Perigee.**

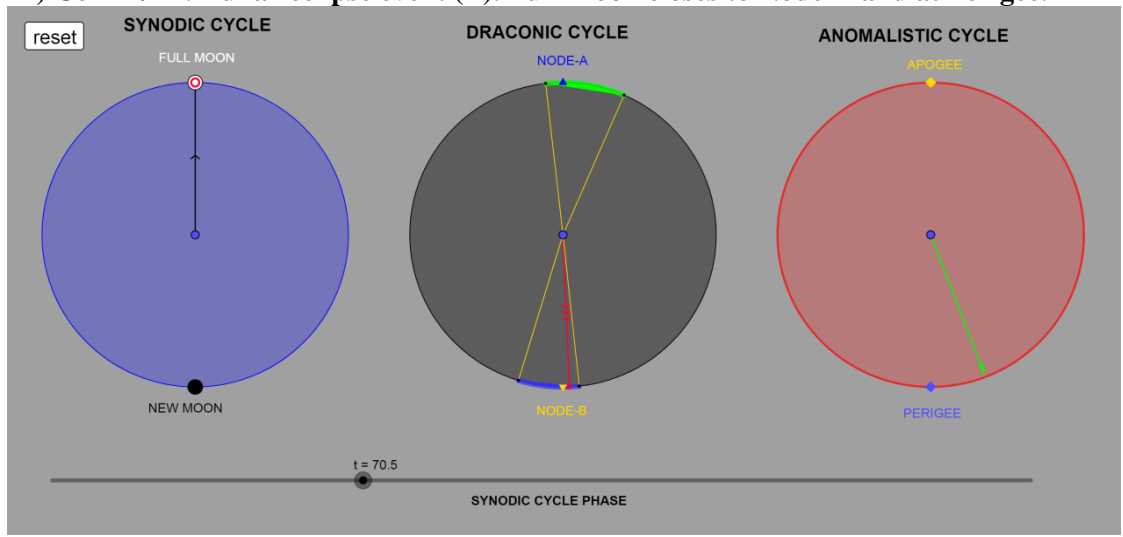

**23) Cell 77/T1: Solar eclipse event (H). Full Moon closes at Perigee.**

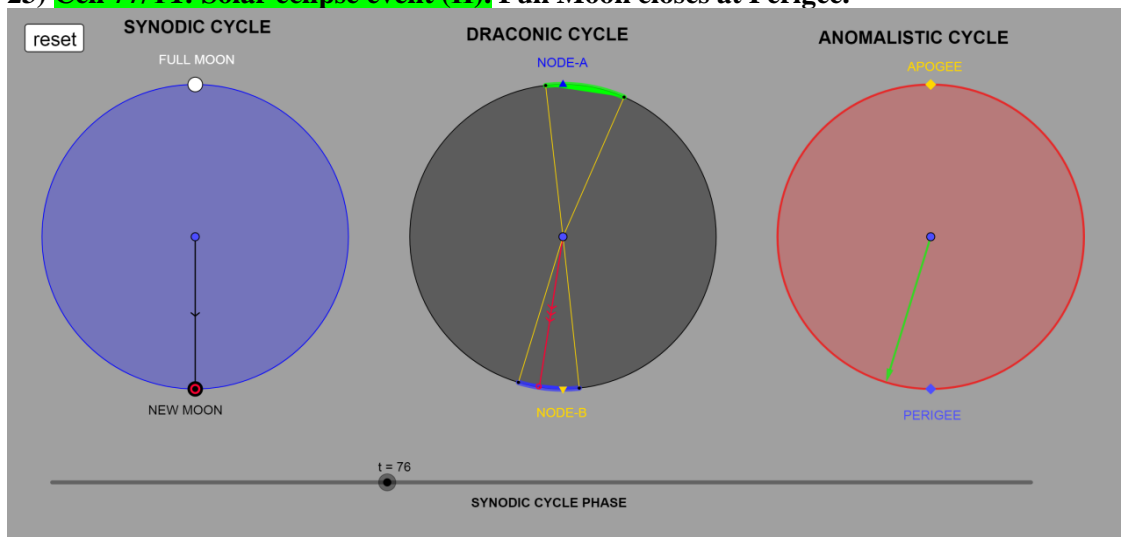

**24) Cell 78/Y1: Lunar eclipse event (Σ). Full Moon just on the ecliptic limit and at Apogee.**

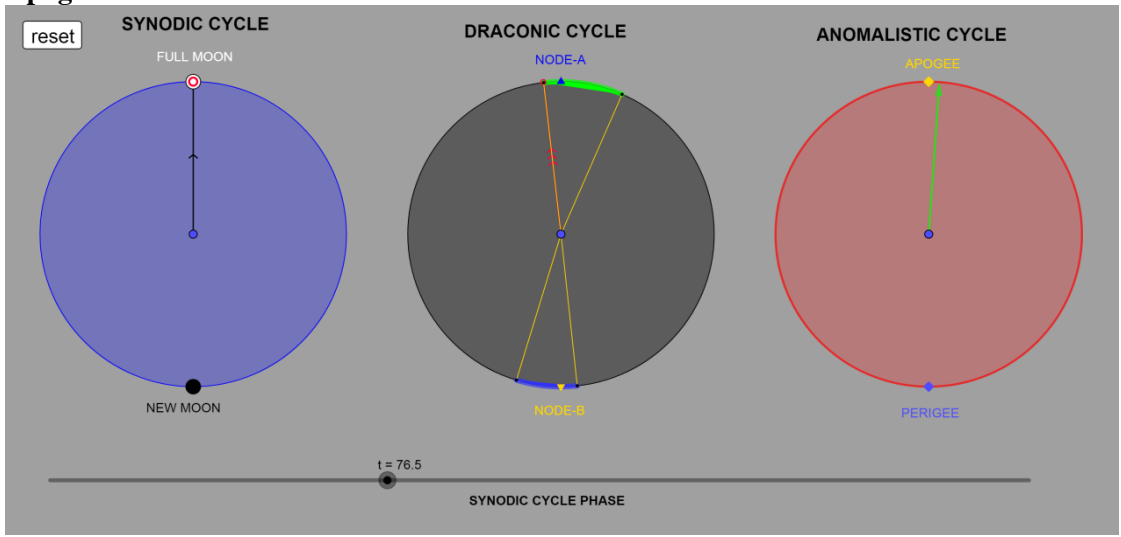

**25) Cell 83/Φ1: Solar eclipse event (H). New Moon closes to Node-A**

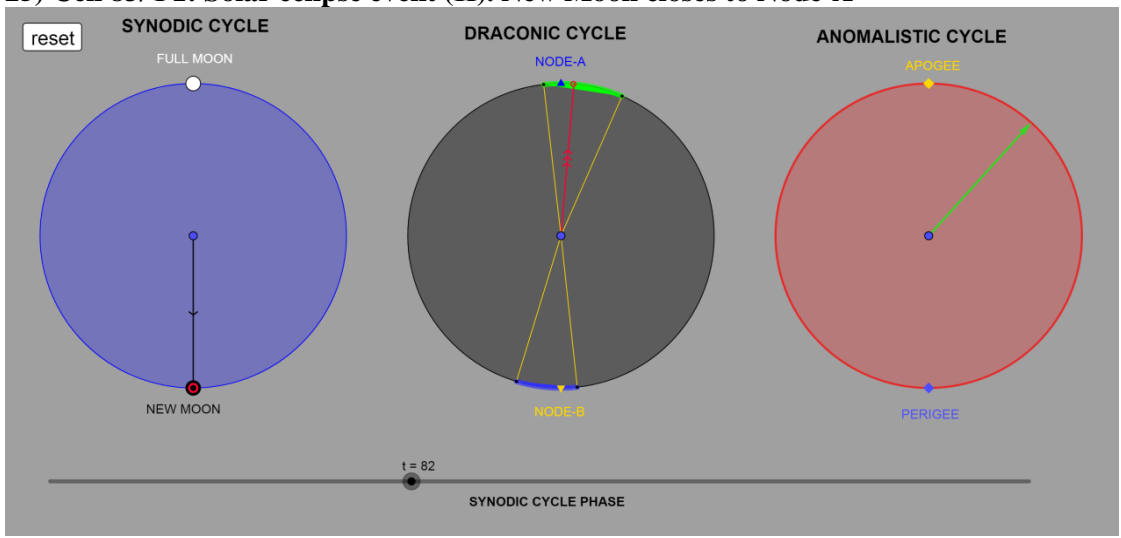

**26) Cell 89/X1: Lunar eclipse event (Σ).**

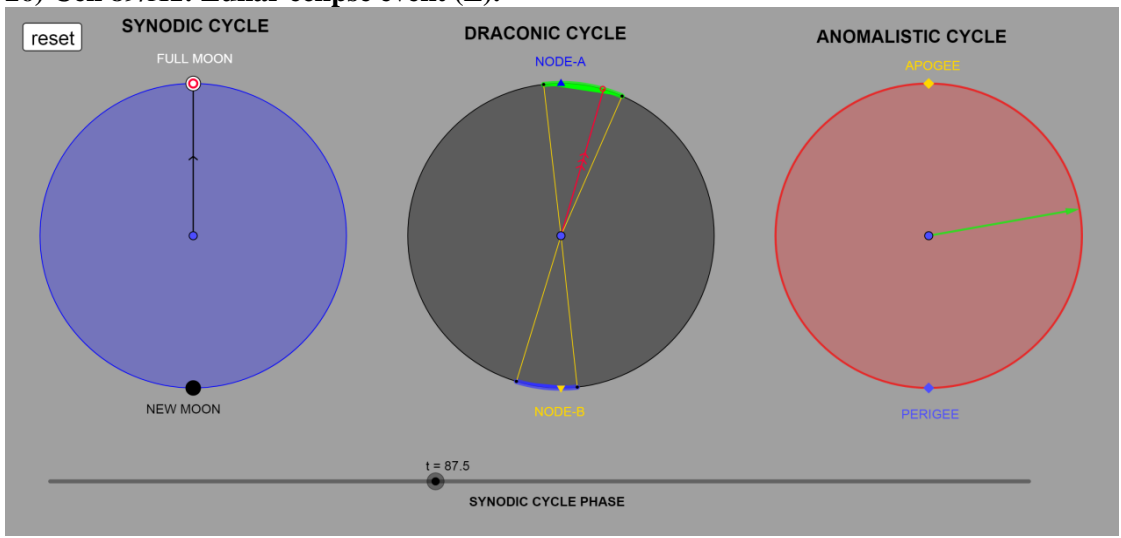

**27) Cell 89/X1: Solar eclipse event (H). New Moon on Node-B**

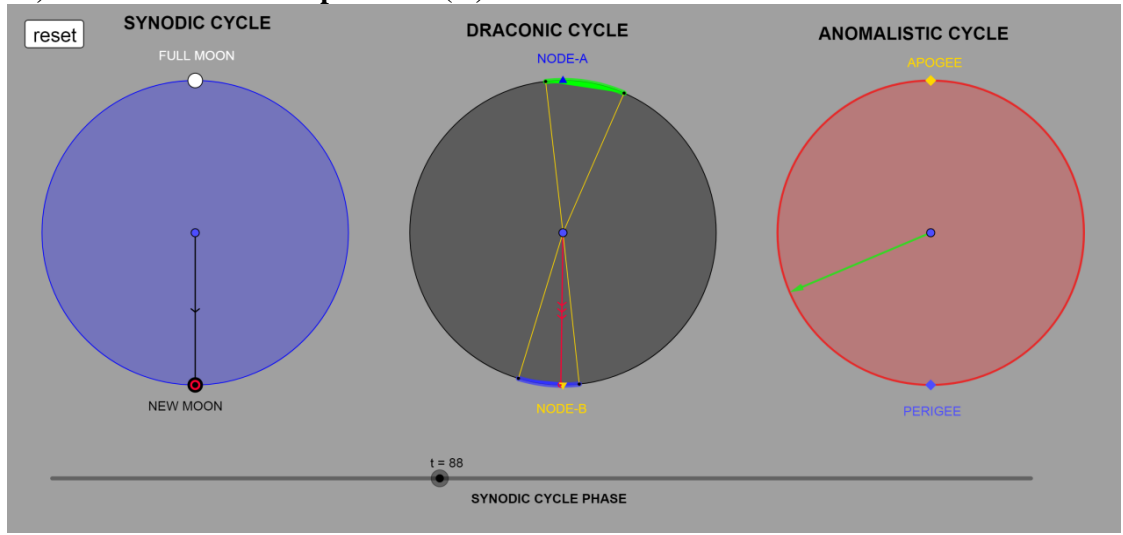

**28) Cell 95/Ψ1: Lunar eclipse event (Σ).**

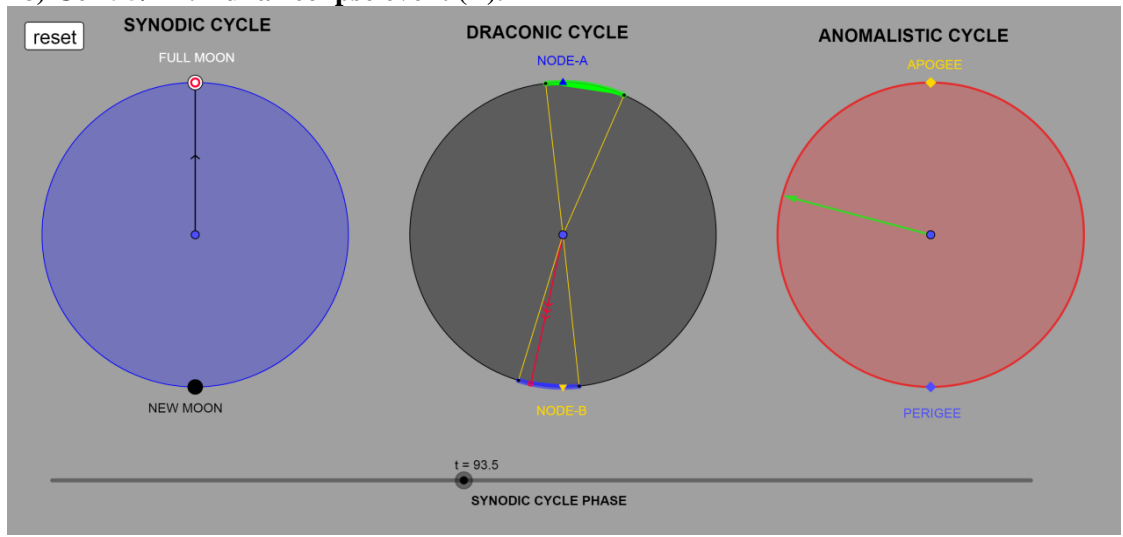

**29) Cell 95/Ψ1: Solar eclipse event (H). New Moon close to ecliptic limit.**

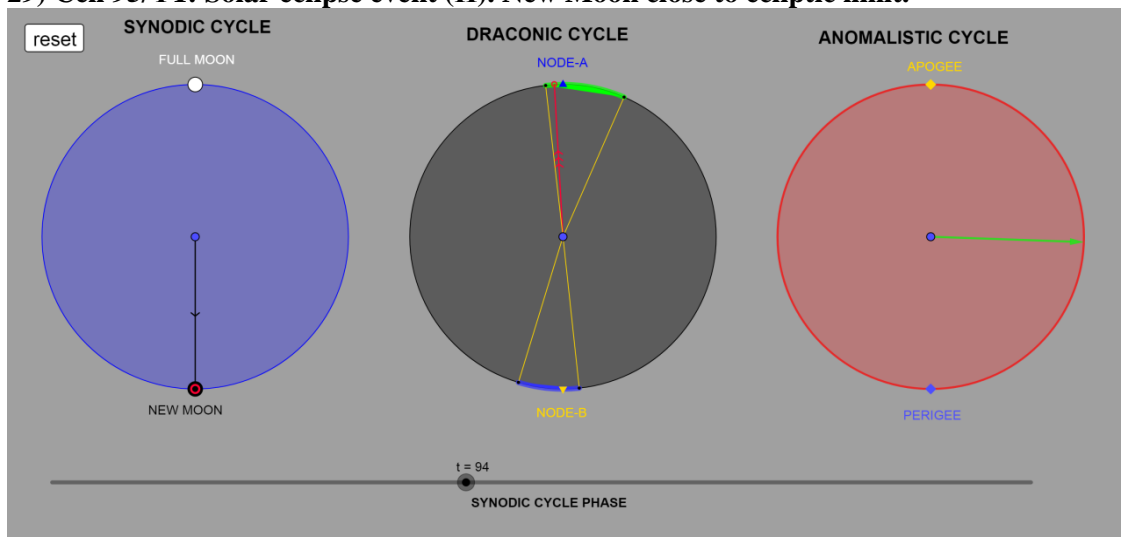

**30) Cell 101/GƆ1: Lunar eclipse event (Σ).**

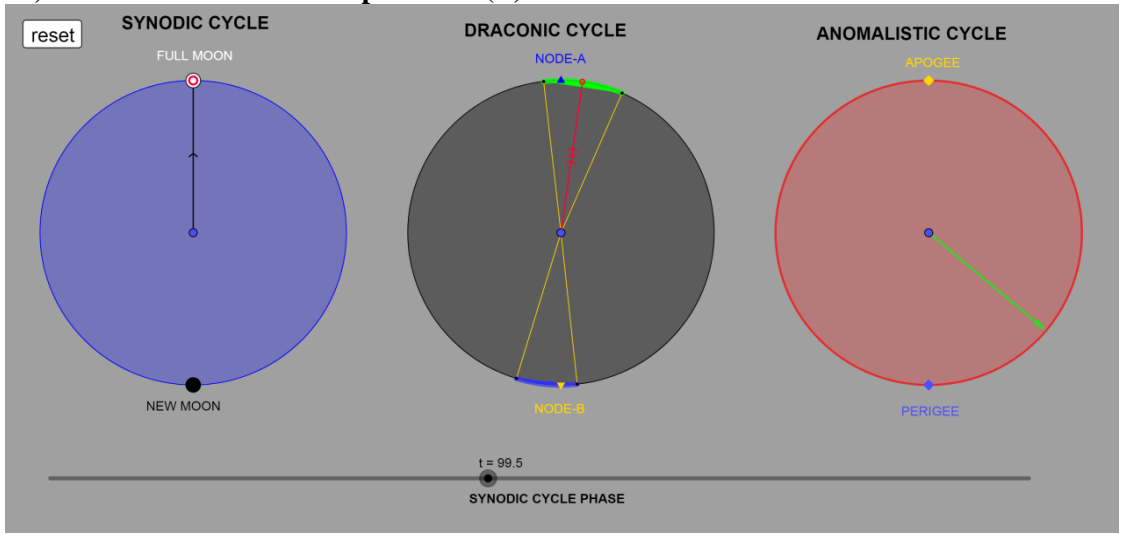

**31) Cell 106/A2: Solar eclipse event (H). New Moon close to the ecliptic limit and at Perigee.**

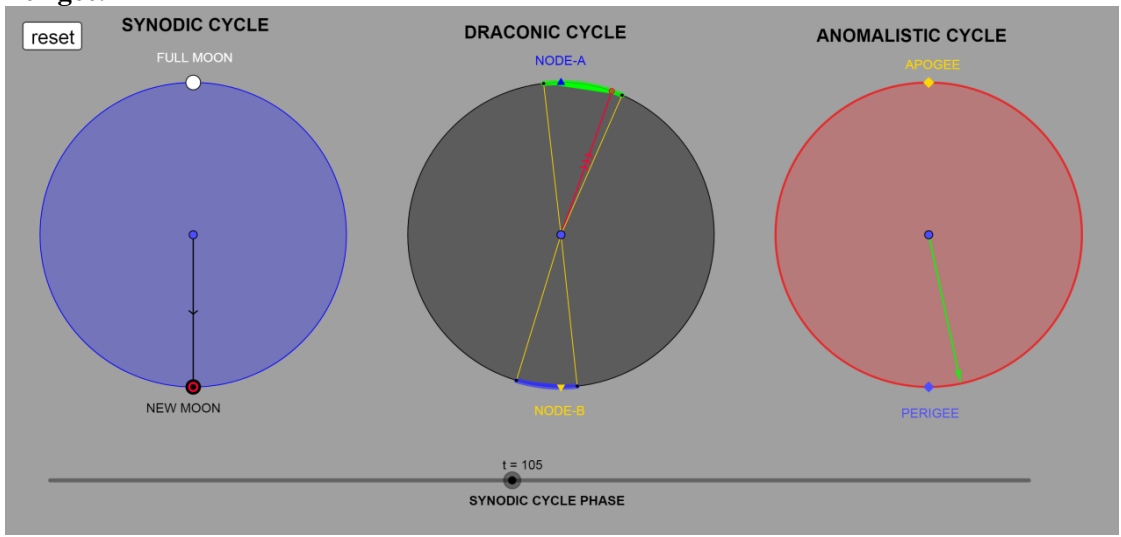

**32) Cell 107/B2: Lunar eclipse event (Σ). Full Moon close to Node-B**

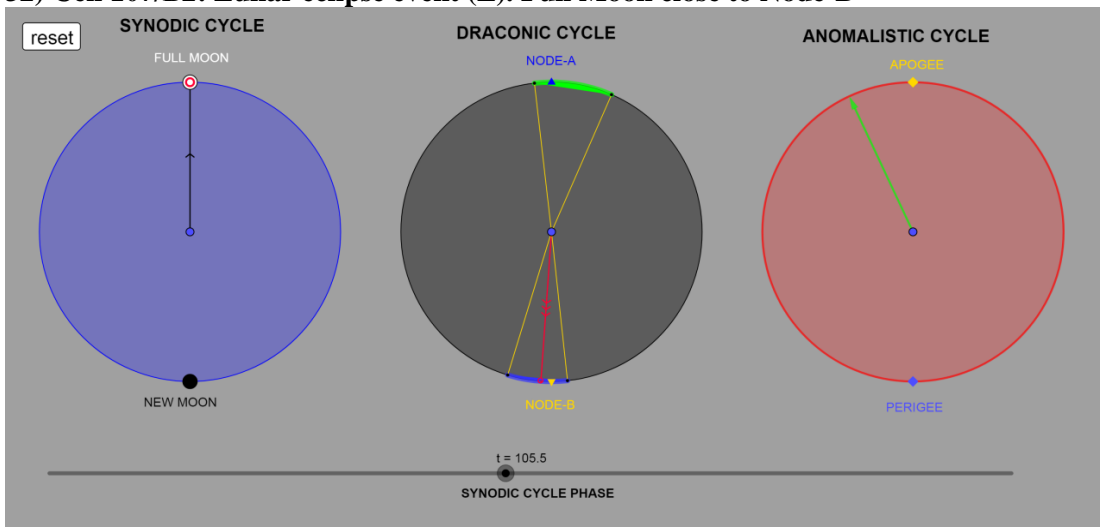

**II) Cell 112: Solar eclipse event (H).** Calculated by the program **but is not an engraved event,** according to the present sequence of the index letters. New Moon too close to the ecliptic limit. Indeterminacy or eccentricity error.

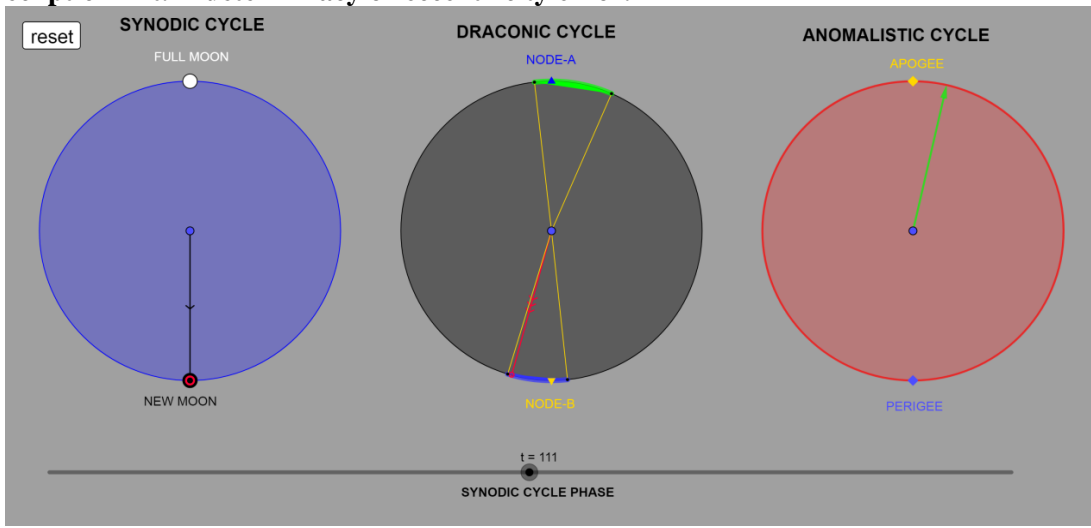

**33) Cell 113/Γ2: Lunar eclipse event (Σ). Full Moon at Node-A and at Perigee. A New Sar period begins.**

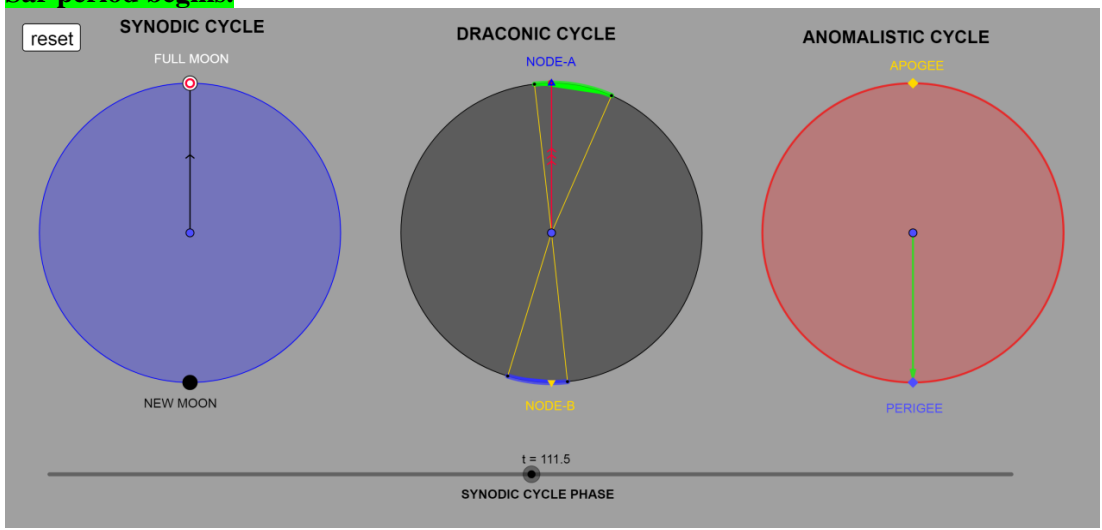

**34) Cell 118/Δ2: Solar eclipse event (H).**

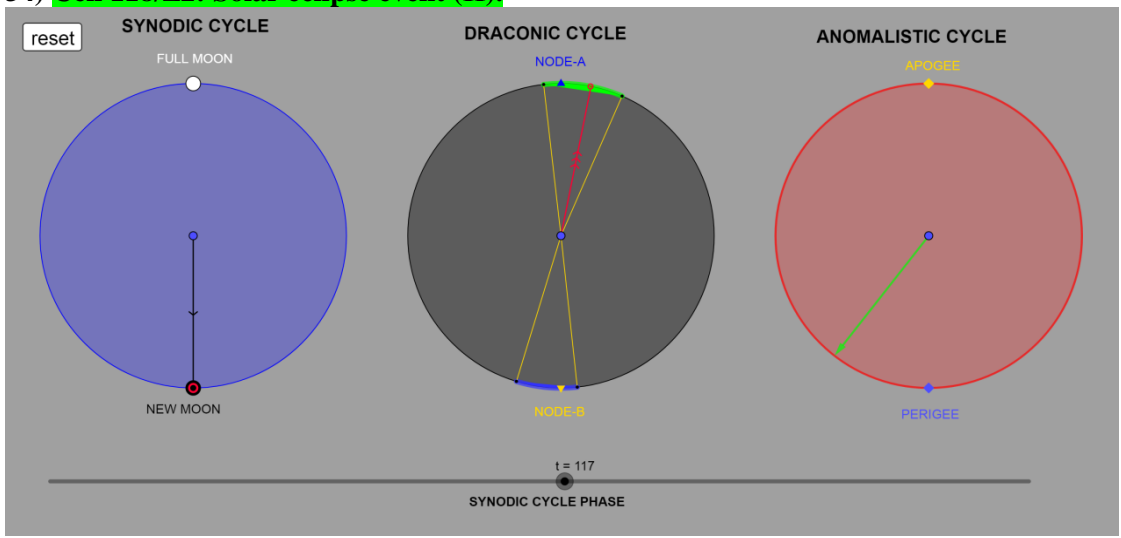

**35) Cell 119/E2: Lunar eclipse event (Σ). Full Moon too close to ecliptic limit.**

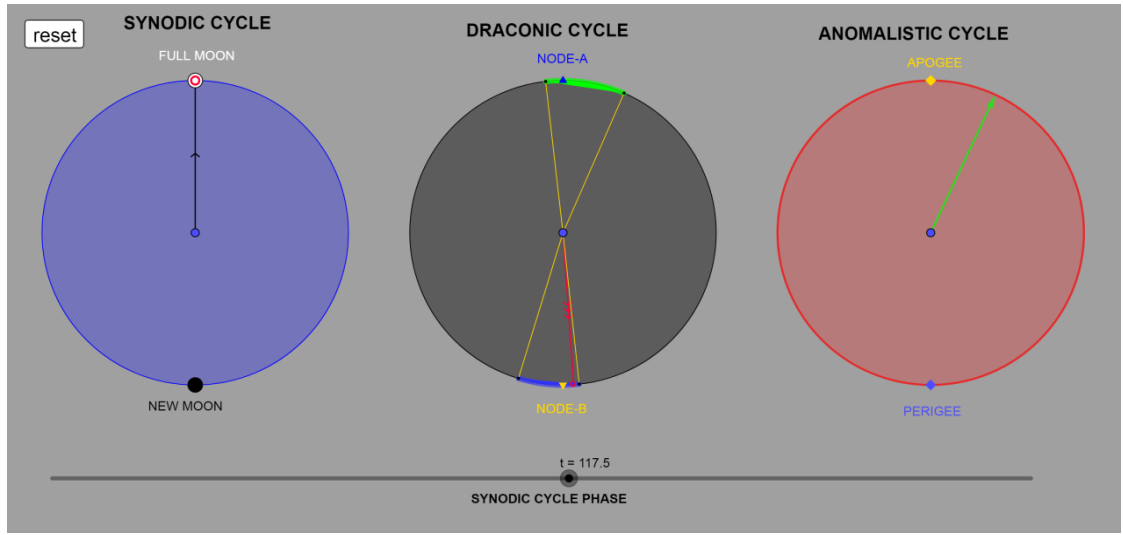

**36) Cell 124/Z2: Lunar eclipse event (Σ). Full Moon too close to ecliptic limit.**

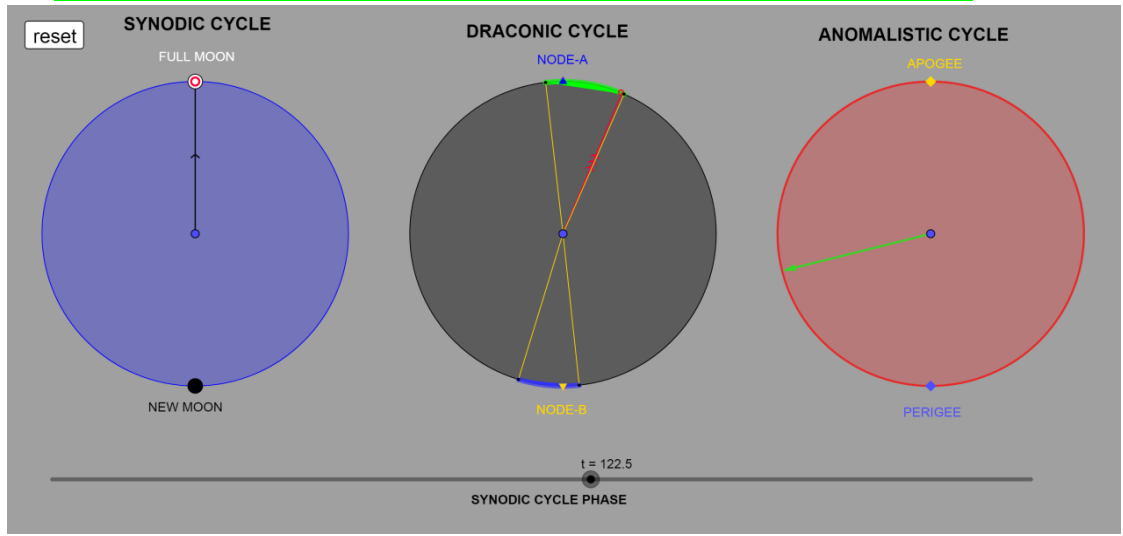

**37) Cell 124/Z2: Solar eclipse event (H).**

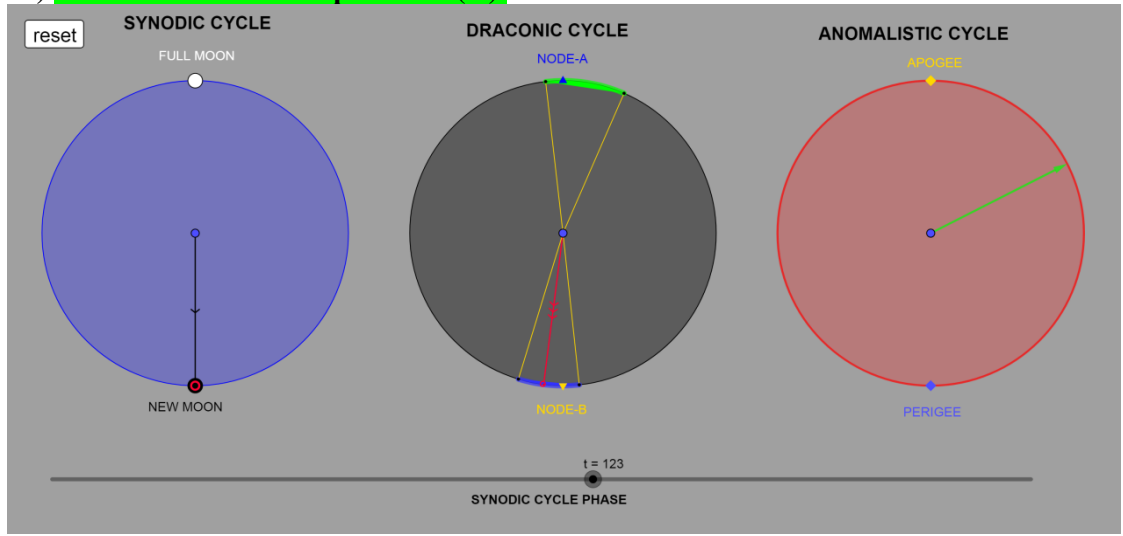

**38) Cell 130/H2: engraved event Lunar eclipse-Σ, but according to *DracoNod* program no eclipse event: the Full Moon is just out of the ecliptic limit. Probable error of eccentricity.**

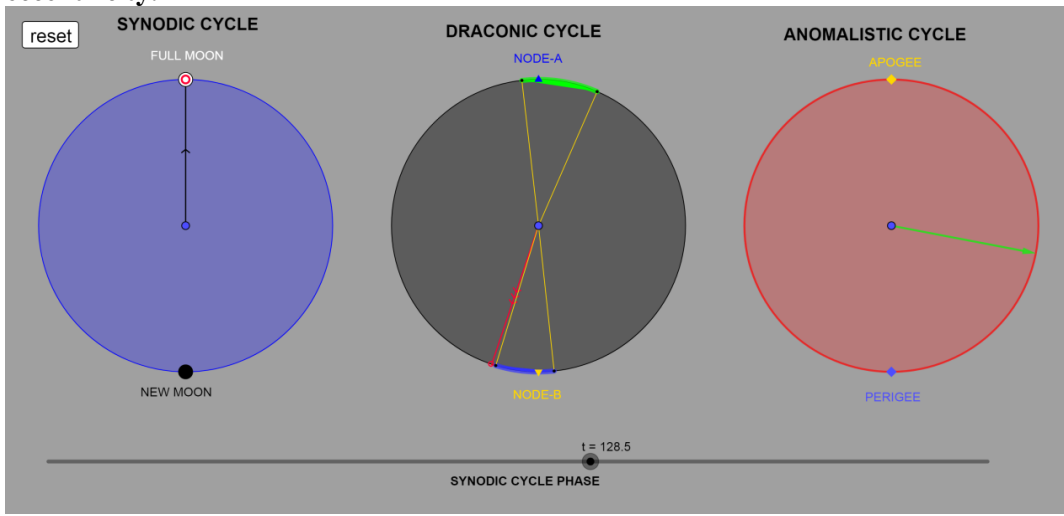

**39) Cell 130/H2: Solar eclipse event (H). New Moon closes at Node-A.**

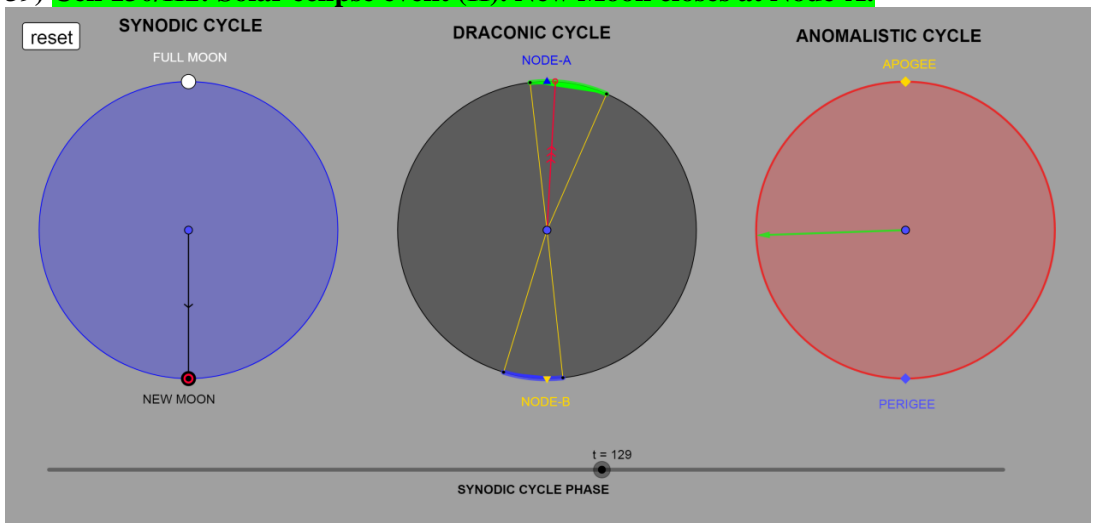

**40) Cell 136/Θ2: Lunar eclipse event (Σ).**

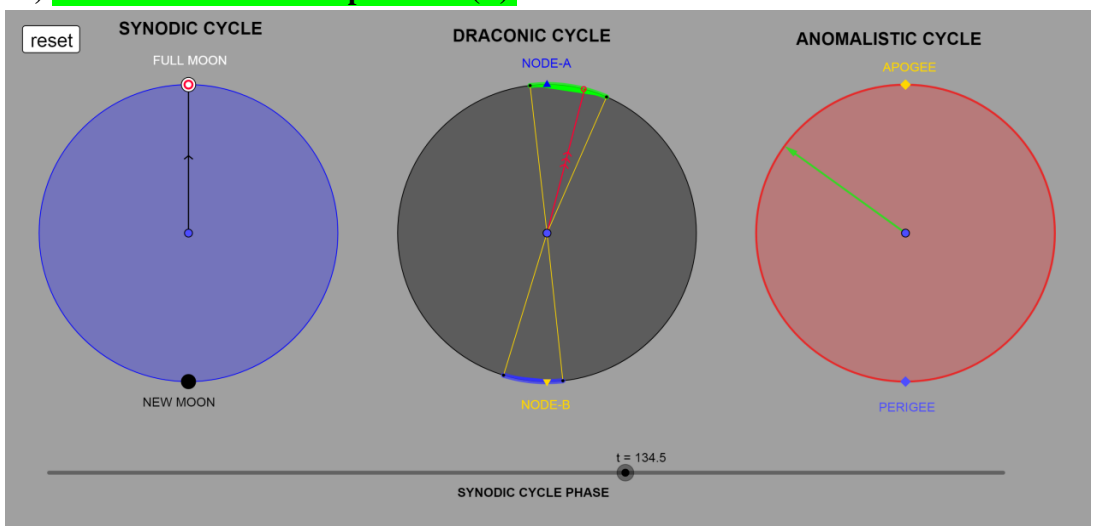

**41) Cell 136/Θ2: Solar eclipse event (H). New Moon at Node-B.**

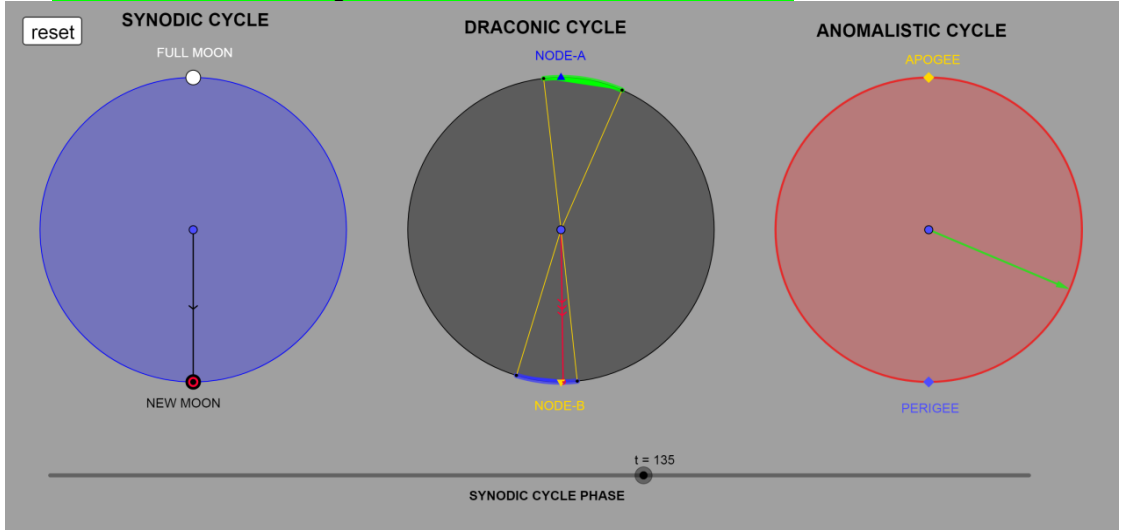

**42) Cell 142/I2: Lunar eclipse event (Σ).**

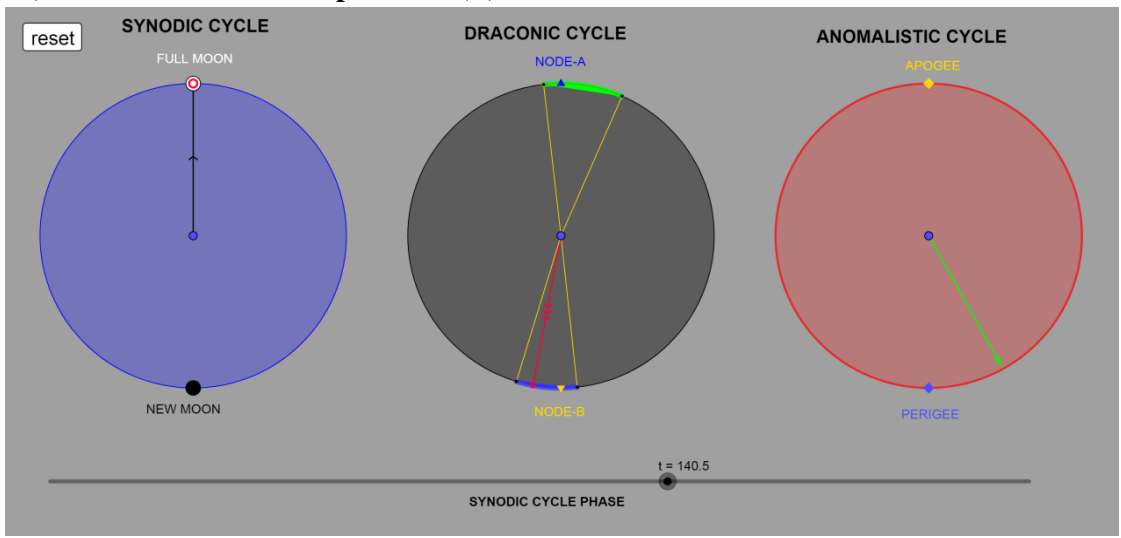

**43) Cell 142/I2: Solar eclipse event (H). New Moon too close to ecliptic limit.**

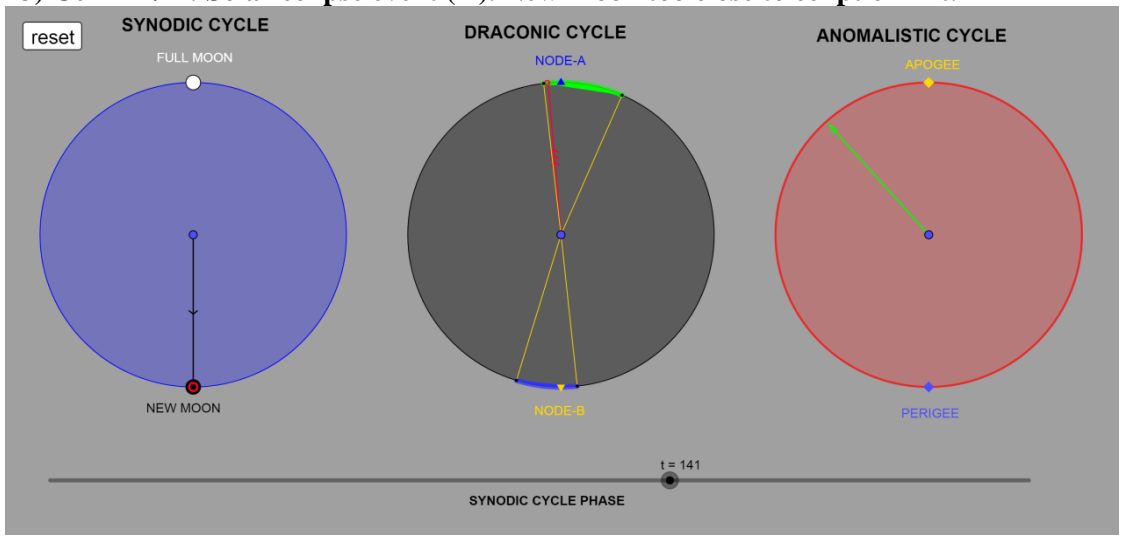

**44) Cell 148/K2: Lunar eclipse event (Σ). Full Moon close to Node-A and at Apogee.**

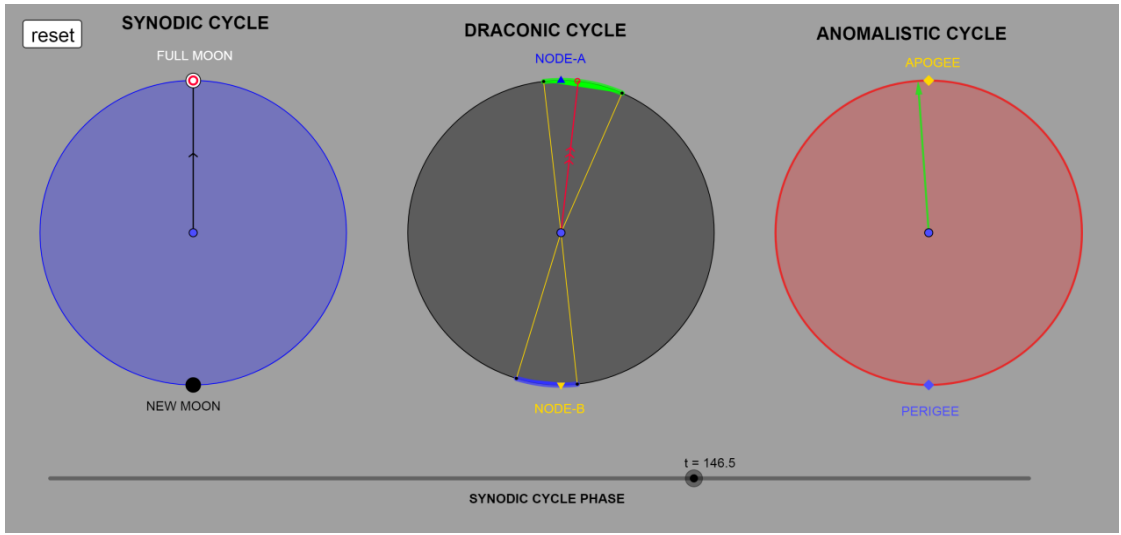

**45) Cell 153/Λ2: Solar eclipse event (H). New Moon closes to ecliptic limit.**

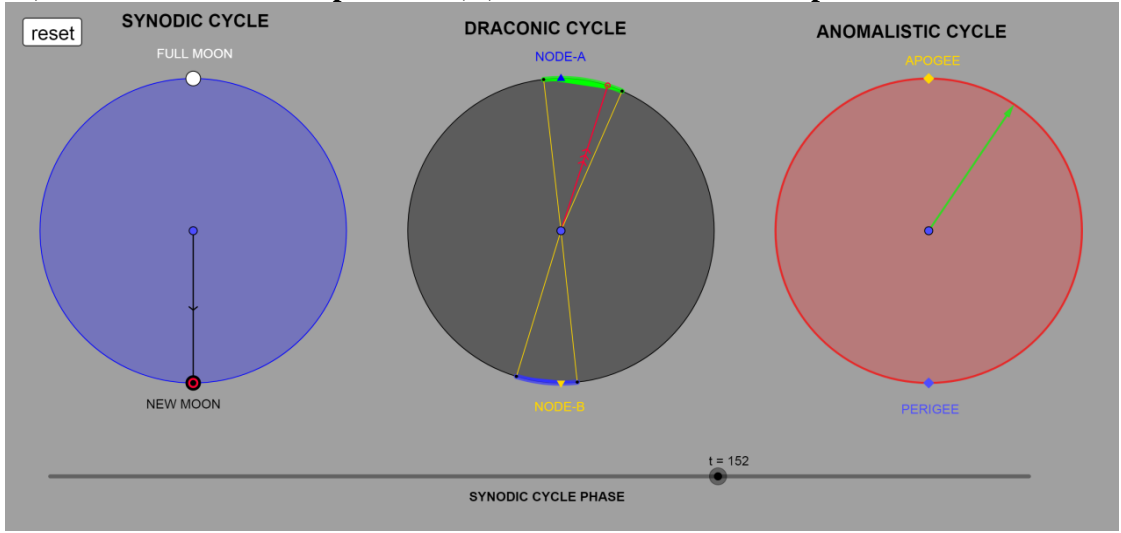

**46) Cell 154/M2: Lunar eclipse event (Σ). Full Moon on Node-B**

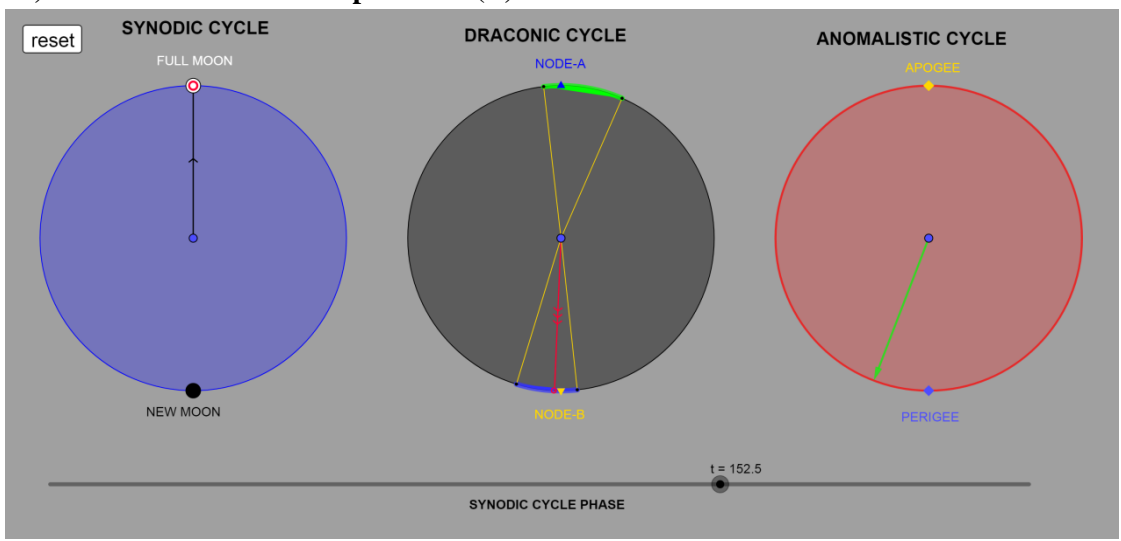

**47) Cell 159/N2: Solar eclipse event (H). New Moon too close to ecliptic limit.**

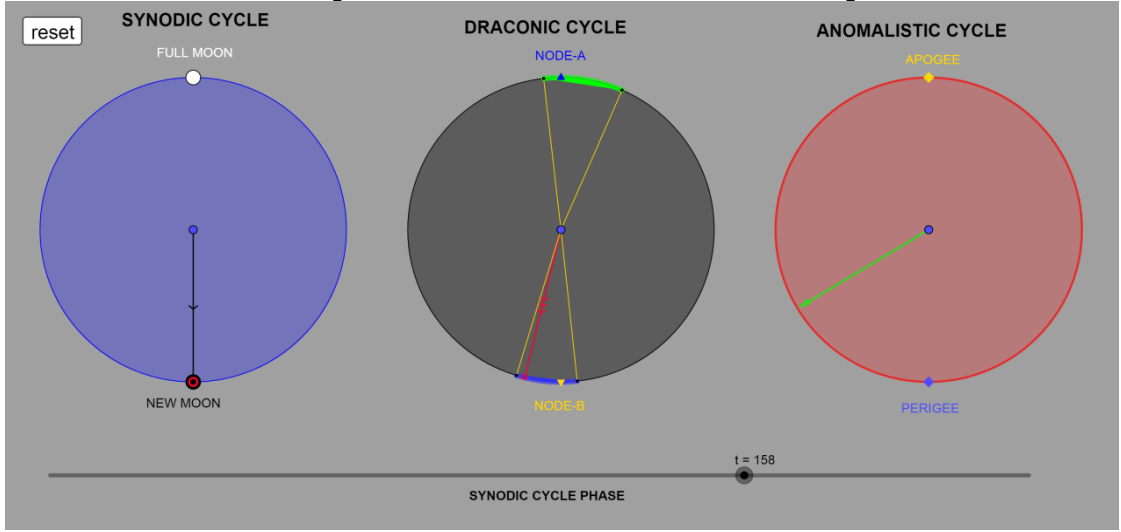

**48) Cell 160/Ξ2: Lunar eclipse event (Σ). Full Moon at Node-A.**

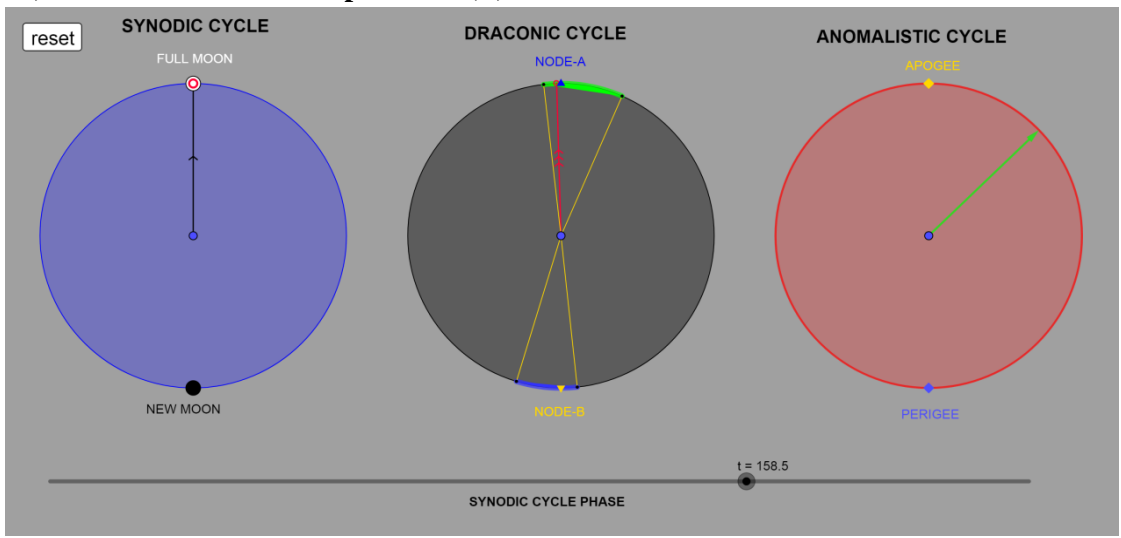

**49) Cell 165/O2: Solar eclipse event (H).**

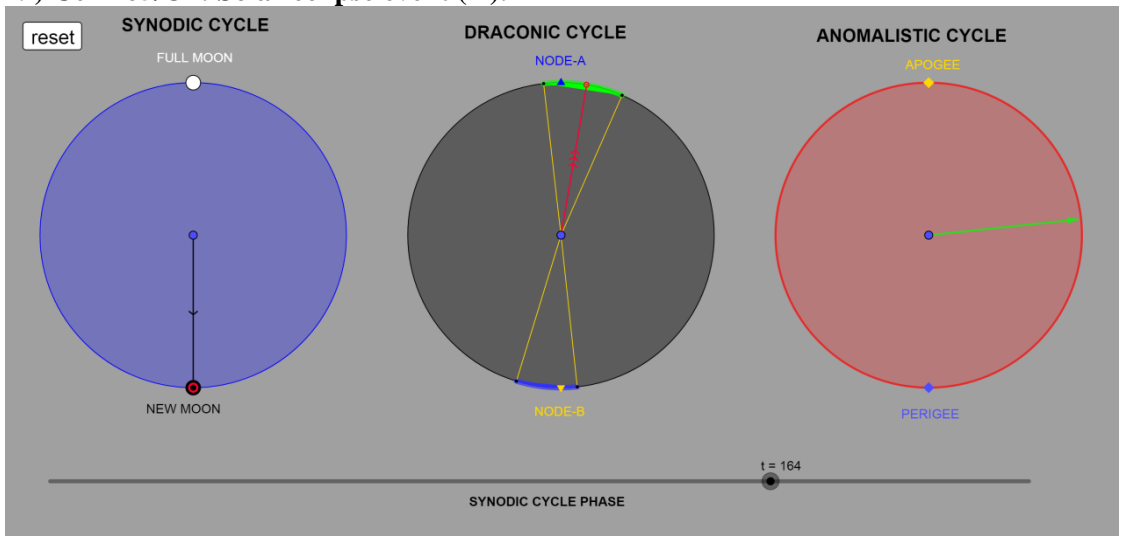

**III) Cell 166:** DracoNod program predicts a Lunar eclipse. Full Moon just right on the ecliptic limit. Based on the events' index numbering, between cells 137 and 170 they should be 7 cells with events. Cell 166 is the 8th cell with event, and it is the only one event presenting high indeterminacy. Cell 166 is considered as cell without event.

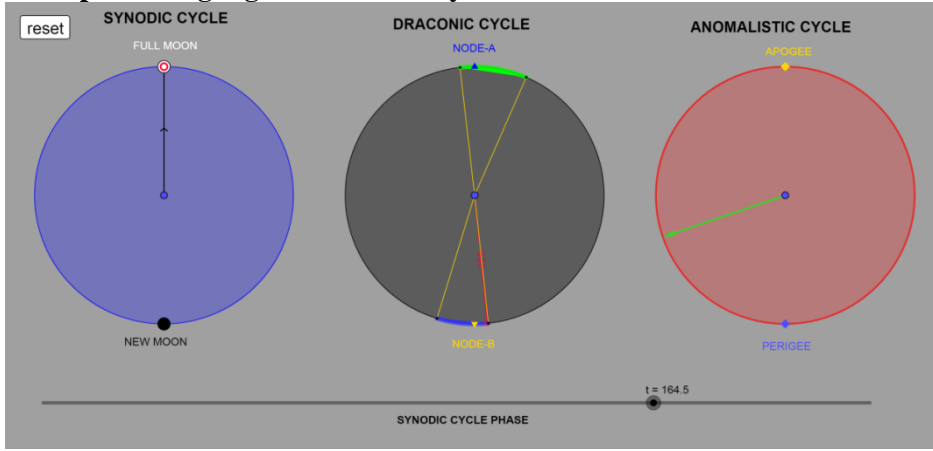

**50) Cell 171/Π2: Lunar eclipse event (Σ). Full Moon close to ecliptic limit.**

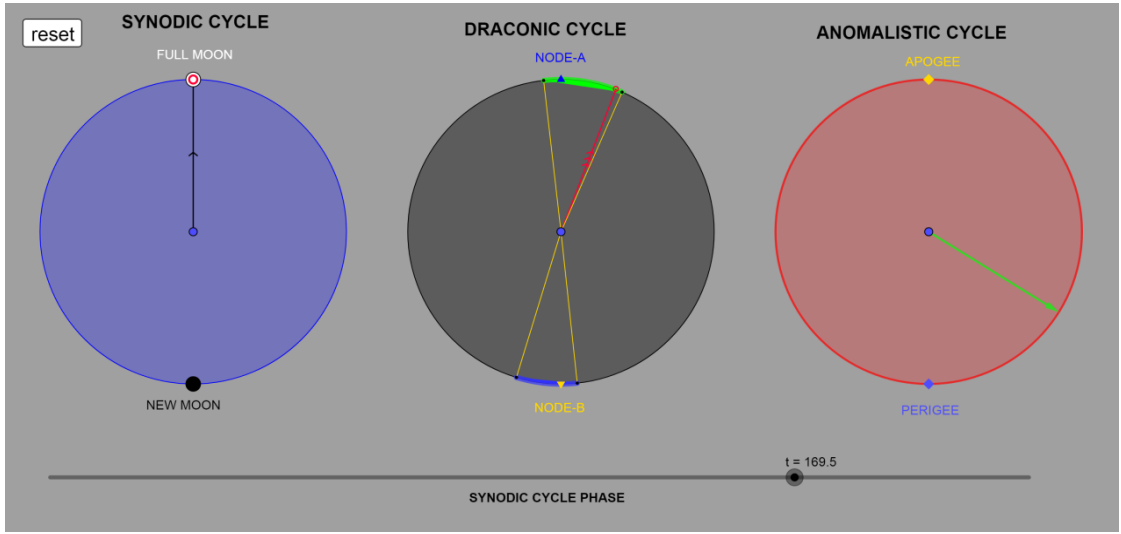

**51) Cell 171/Π2: Solar eclipse event (H). New Moon close to Node-B.**

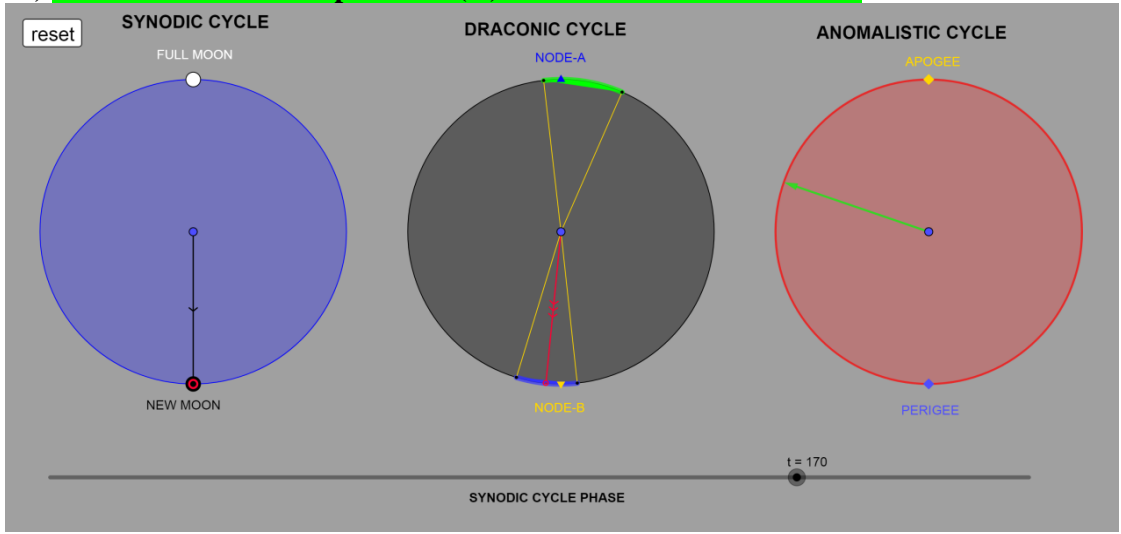

**52) Cell 177/P2: Lunar eclipse event (Σ). Full Moon just on ecliptic limit.**

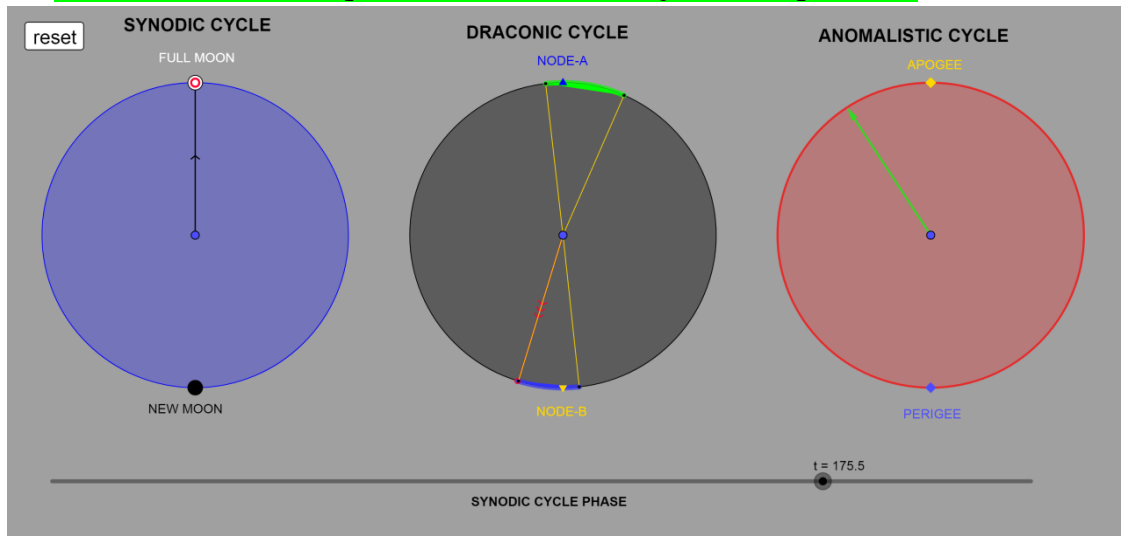

**53) Cell 177/P2: Solar eclipse event (H). New Moon at Node-A.**

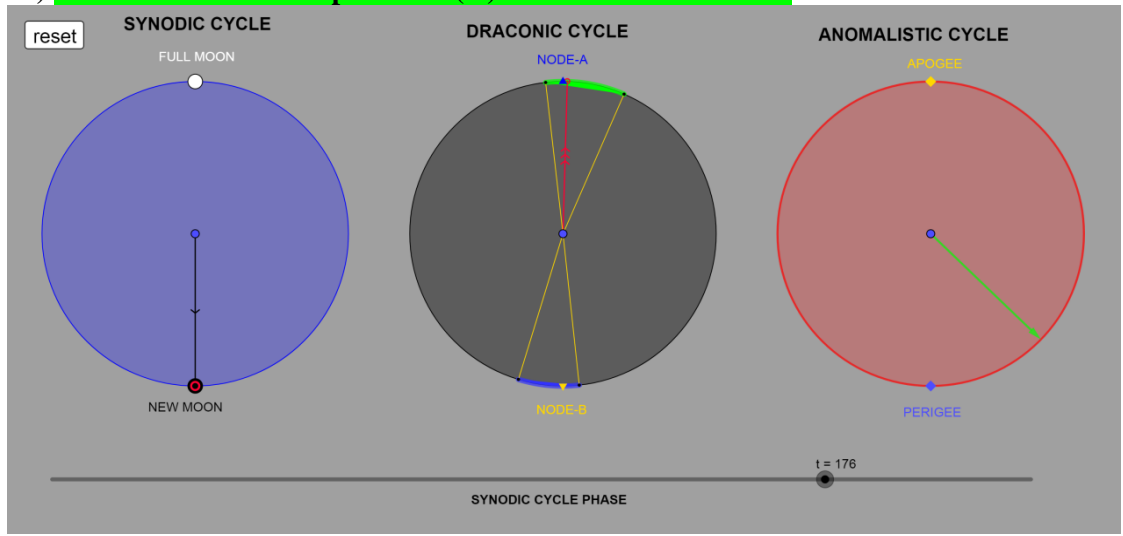

**54) Cell 183/Σ2: Lunar eclipse event (Σ). Full Moon closes at Perigee.**

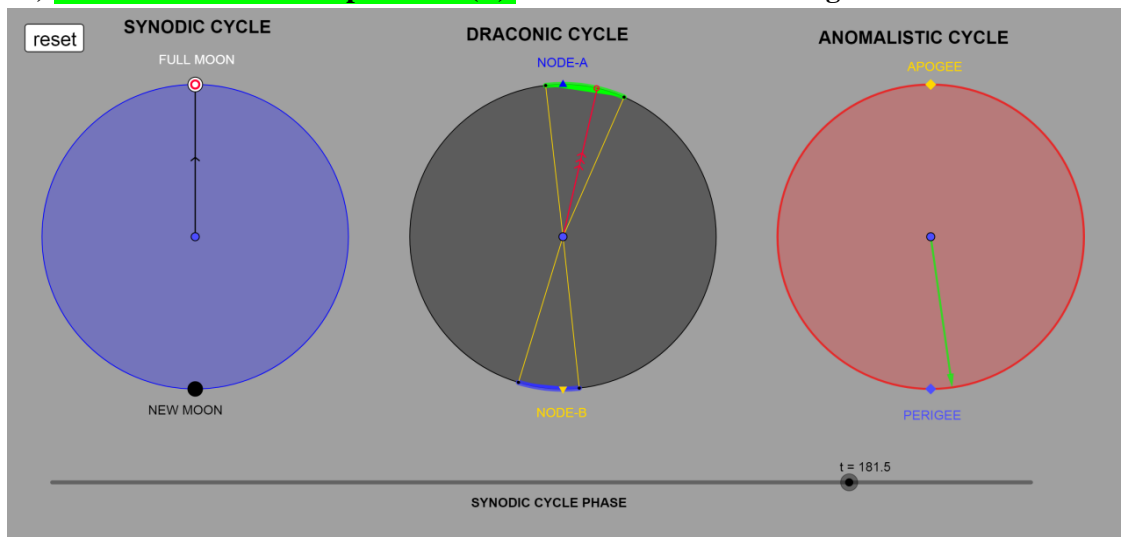

**55) Cell 183/Σ2: Solar eclipse event (H). New Moon closes at Node-B.**

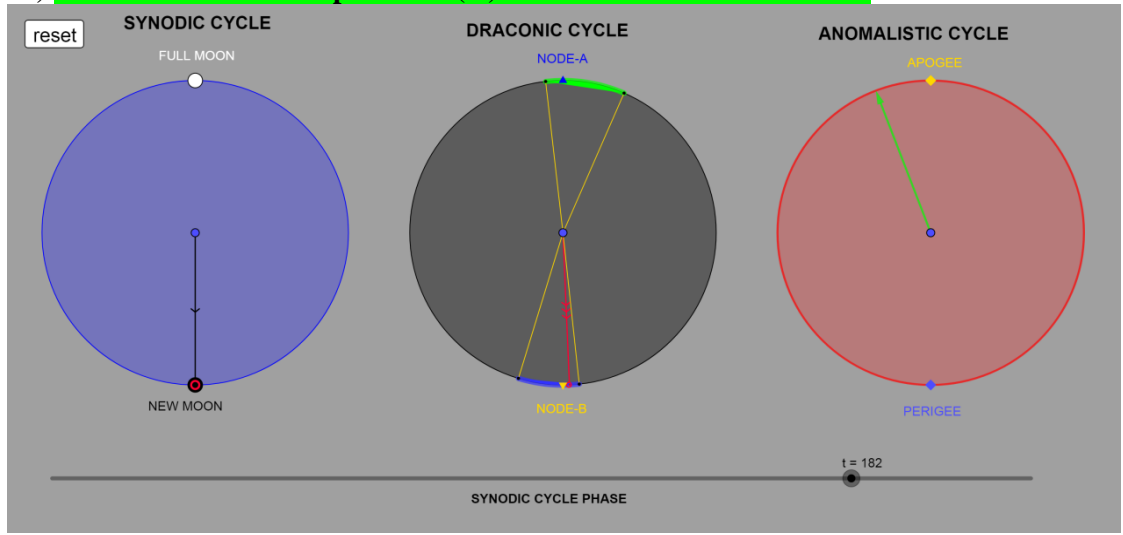

**56) Cell 189/T2: Lunar eclipse event (Σ). Full Moon closes at Apogee.**

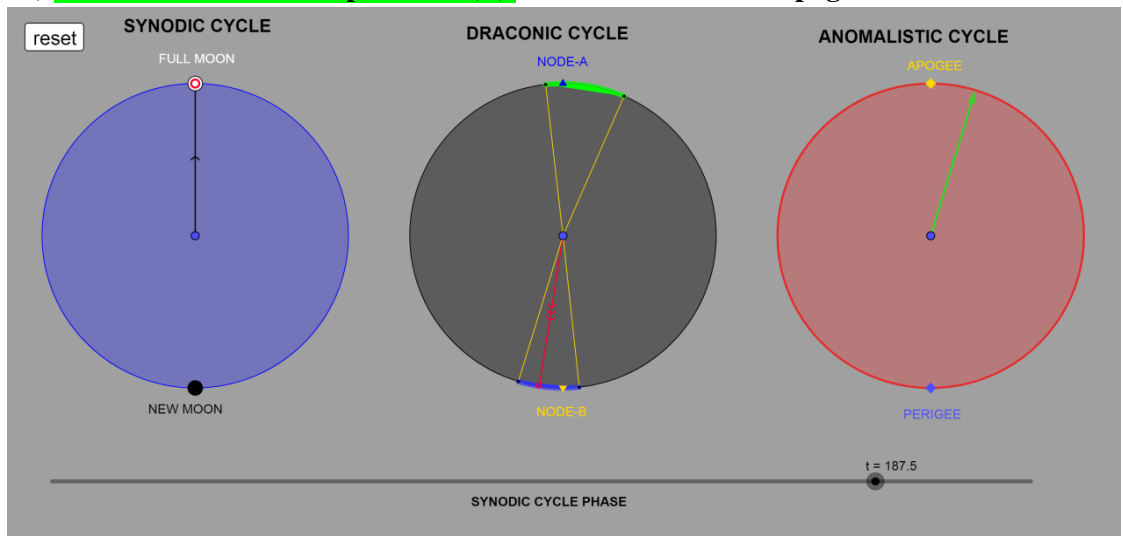

**IV) Cell 189: No engraved event. DracoNod program predicts a second event, a Solar eclipse. New Moon just right on the ecliptic limit. Indeterminacy or gearing errors.**

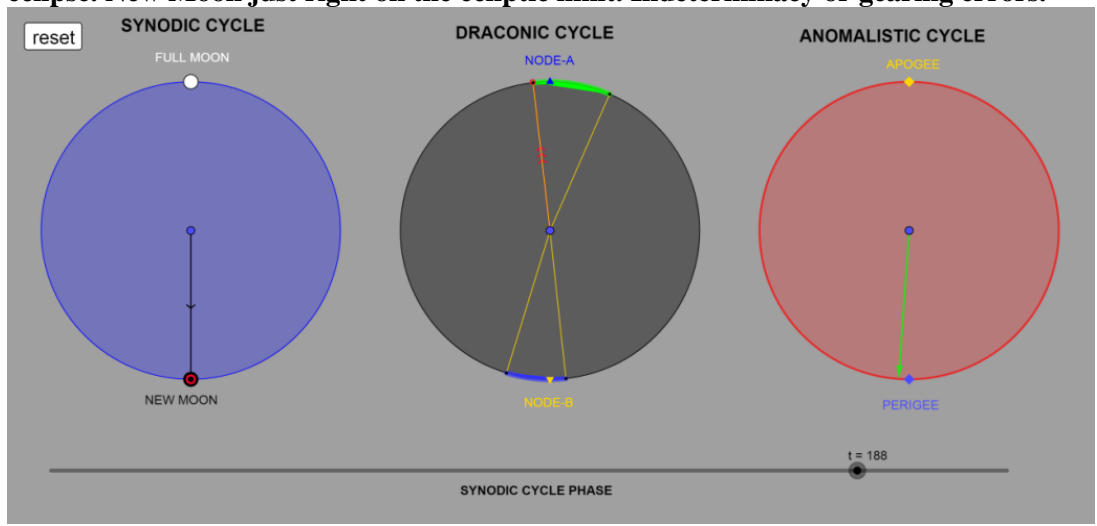

**57) Cell 195/Y2: Lunar eclipse event (Σ). Full Moon close to Node-A**

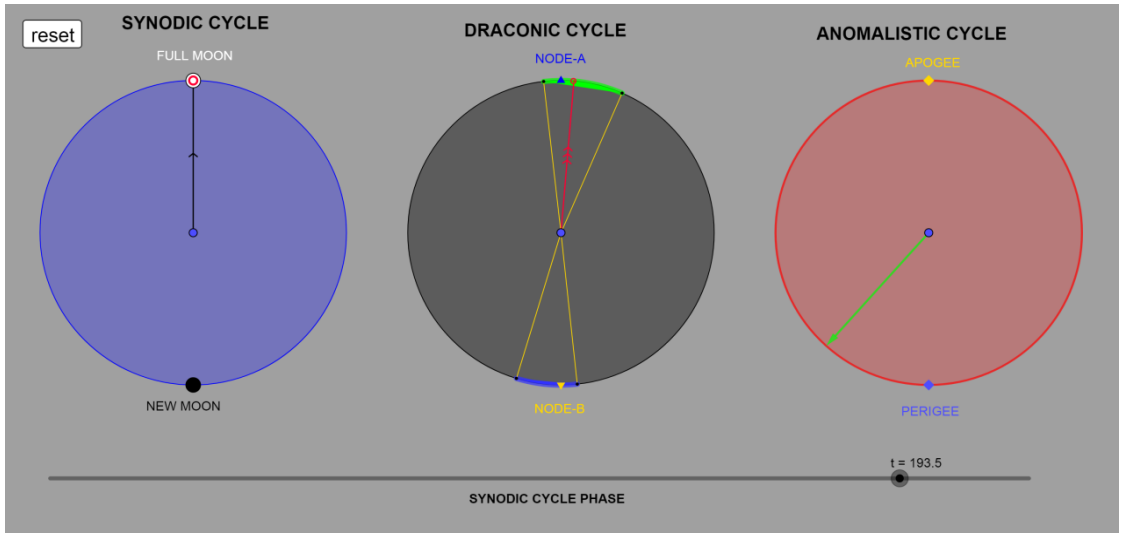

**58) Cell 200/Φ2: Solar eclipse event (H).**

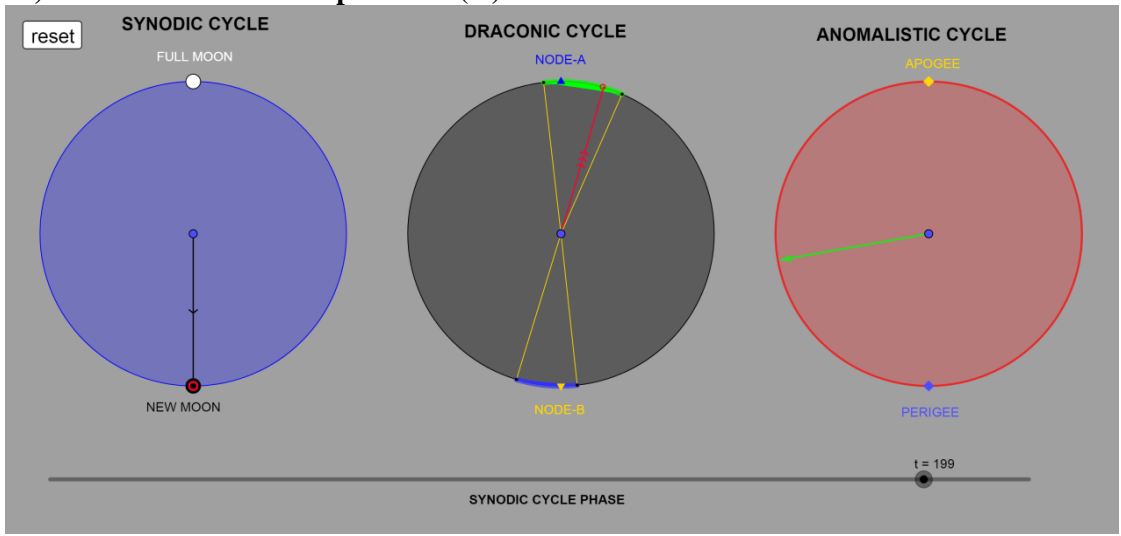

**59) Cell 201/X2: Lunar eclipse event (Σ). Full Moon on Node-B**

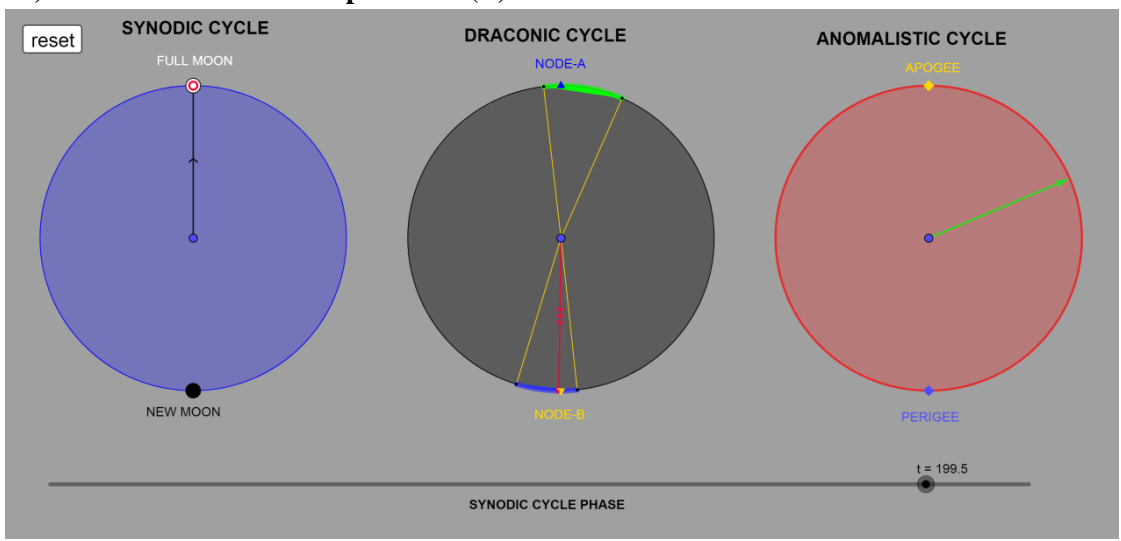

**60) Cell 206/Ψ2: Solar eclipse event (H). New Moon closes to ecliptic limit.**

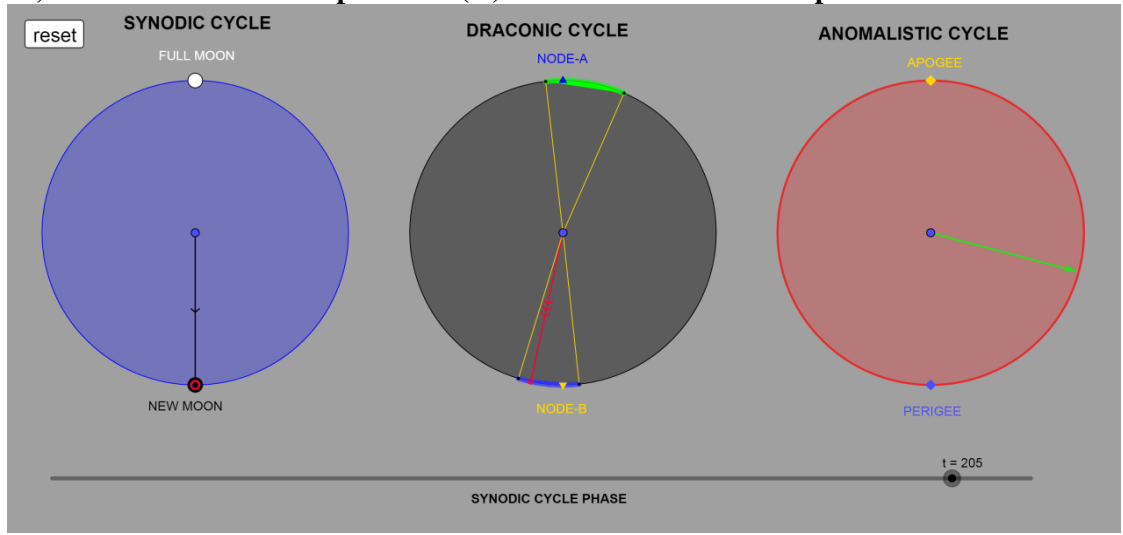

**61) Cell 207/♋2: Lunar eclipse event (Σ).**

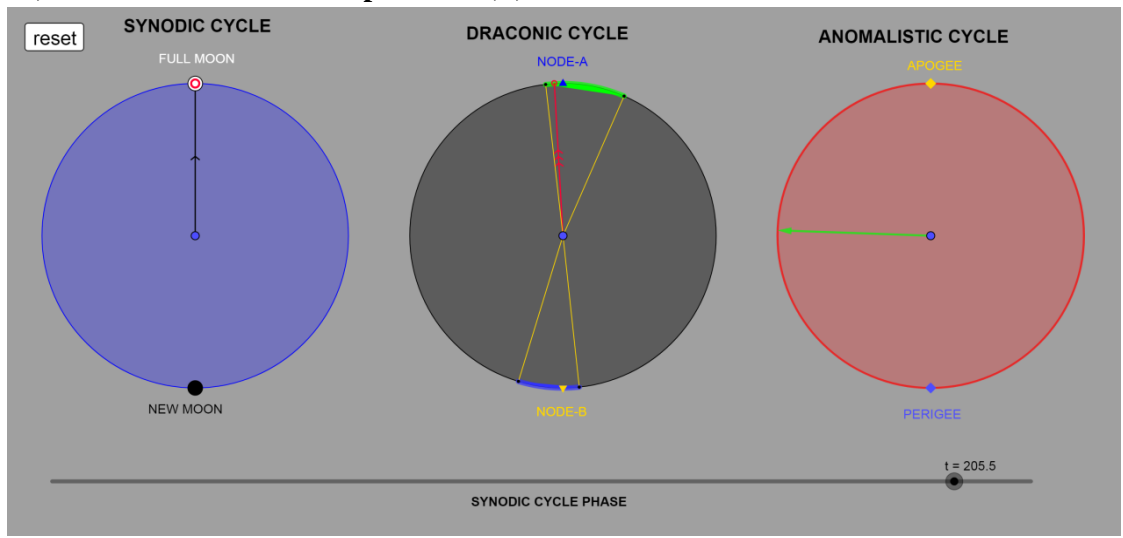

**62) Cell 212/2 (A3): Solar eclipse event (H).**

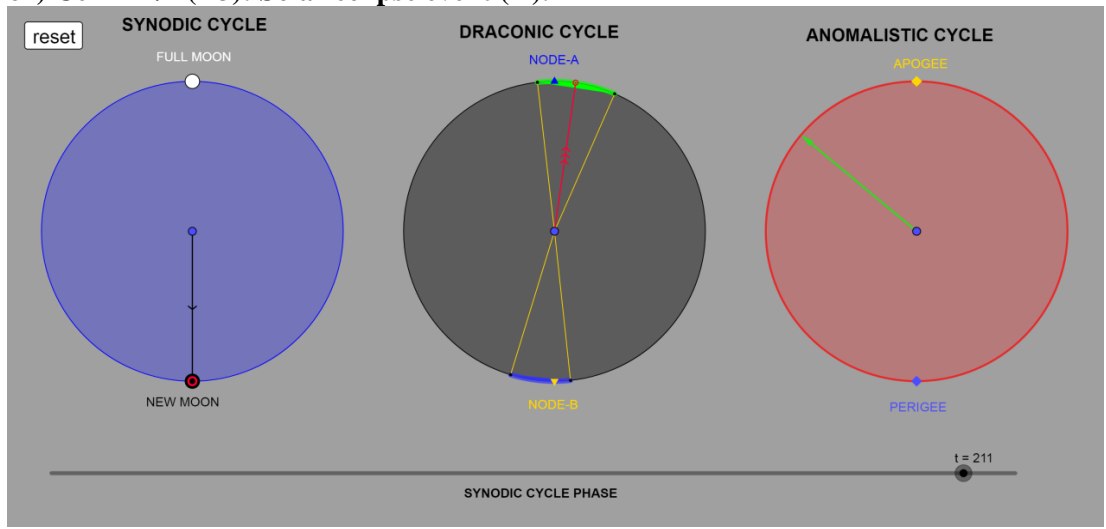

**63) Cell 218/ß (B3): Lunar eclipse event (Σ). Full Moon close to ecliptic limit and at Apogee.**

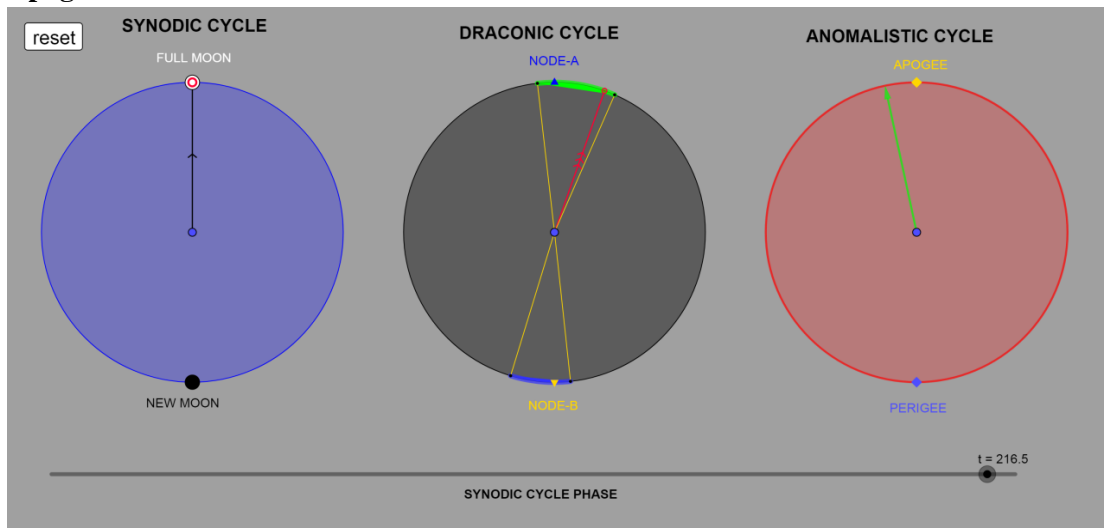

**64) Cell 218/ß (B3): Solar eclipse event (H). New Moon close to Node-B**

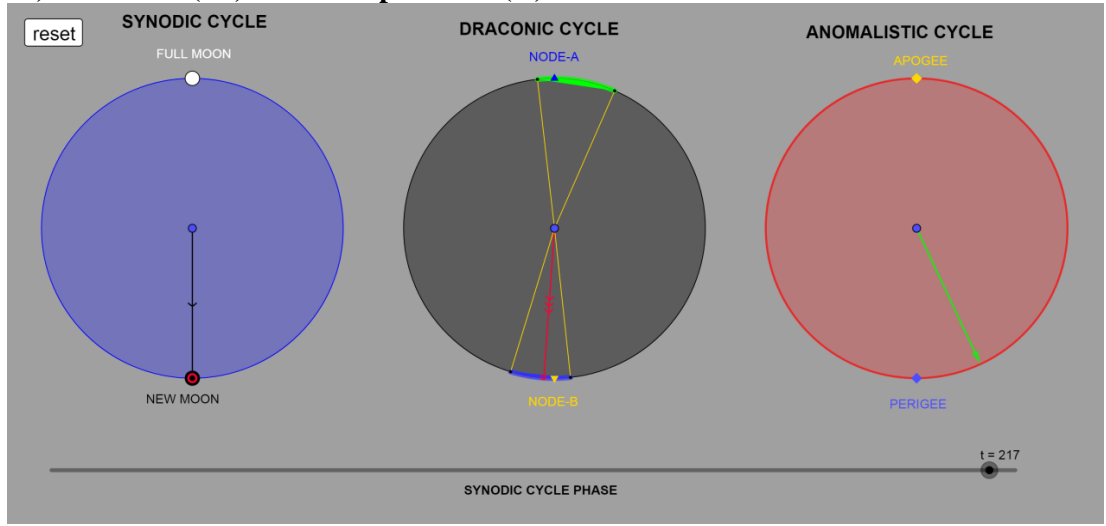

**V) Cell 01/Cell 224/A1:** *DracoNod* **predicts a Lunar eclipse event (Σ). Full Moon close to the ecliptic limit. On cell 112 (one Sar after Cell 01), is not engraved the solar eclipse event. Therefore, on Cell 01 could not be engraved the lunar eclipse event.**

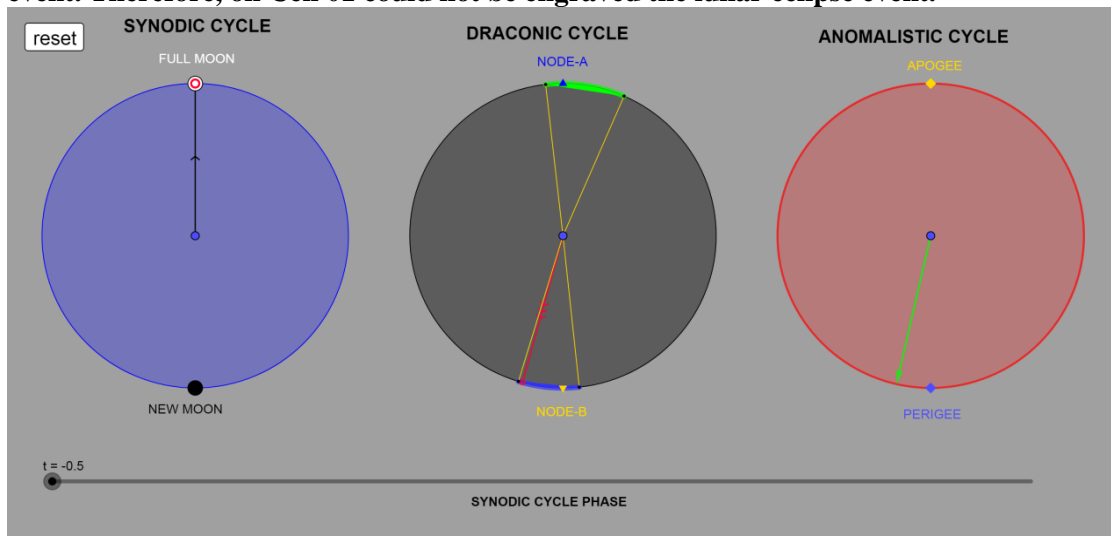

# 7. Eclipse event Statistics and Comments

Eclipse events, occurred just right on or too close to the ecliptic limit, present a high indeterminacy. Moreover, as result of the possible gear(s) eccentricities and the teeth un-uniformity, the final position of the Draconic pointer, or the Lunar pointer, or the Golden sphere pointer, may (slightly) differ than the ideal position, which was calculated via *DracoNod* program. Minor gear(s) imperfections affect and change the position of the Draconic pointer: sometimes the Draconic pointer is located out of the ecliptic window, although it should be on the ecliptic window (therefore no eclipse event occurs). Other times the Draconic pointer is located inside the ecliptic window (therefore an eclipse event occurs), although it should be out of the ecliptic limit.

| *DracoNod* eclipse events prediction ||
|---|---|
| **Solar eclipse events on cell** | **Lunar eclipse events on cell** |
| 1, 7, 12, 24, 30, 36, 42, 48, 54, 59, 71, 77, 83, 89, 95, 106, 118, 124, 130, 136, 142, 153, 159, 165, 171, 177, 183, 200, 206, 212, 218 (31 events) | 7, 13, 19, 25, 31, 42, 48, 54, 60, 66, 72, 78, 89, 95, 101, 107, 113, 119, 124, 130, 136, 142, 148, 154, 160, 171, 177, 183, 189, 195, 201, 207, 218 (33 events) |
| Cell-65 non engraved event-predicted H, Pointer at the limit (indeterminacy) | Cell-130 engraved-not predicted Pointer just out of limit |
| Cell-112 non engraved-predicted H, Pointer close to the limit (gearing error) | Cell-166 predicted Σ at limit (indeterminacy), but it should be not exist based on index numbering |
| Cell-189 non engraved-predicted H, Pointer at the limit (indeterminacy) | Cell-224/Cell-01 predicted Σ It could be exist or not exist (same as Cell-112) |

*DracoNod* predicted all the preserved solar eclipses plus three additional eclipses which are non-engraved events and present high indeterminacy.

*DracoNod* predicted all the preserved lunar eclipses except one which presents high indeterminacy. Was also predicted one non-engraved eclipse which presents high indeterminacy. The predicted lunar eclipse on cell 01 is too doubtful if finally was real engraved.

**I) Cell 65: Full Moon just right on the ecliptic limit. Based on the events' index numbering it should be no engraved event. Indeterminacy or eccentricity error.**

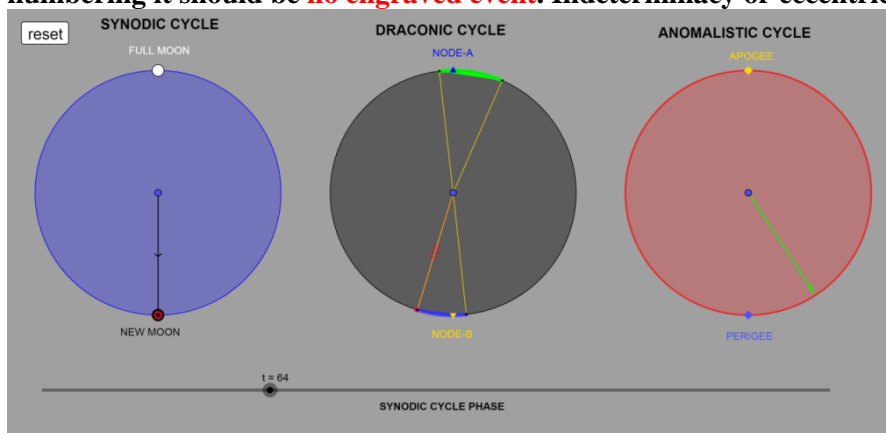

**II) Cell 112: Solar eclipse event (H). Calculated by the program but is not an engraved event,** according to the present sequence of the index letters. New Moon too close to the ecliptic limit. Indeterminacy or eccentricity error.

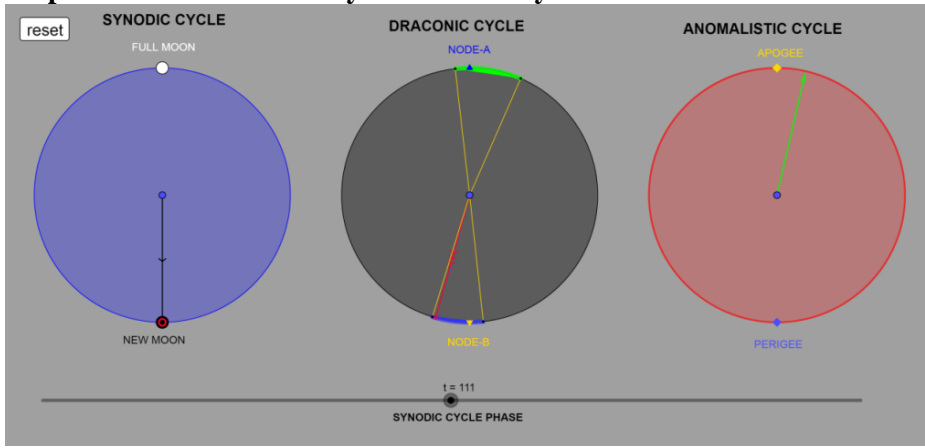

**38) Cell 130/H2: engraved event Lunar eclipse-Σ**, but according to *DracoNod* program **no eclipse event**: the Full Moon is just out of the ecliptic limit. Probably error of eccentricity.

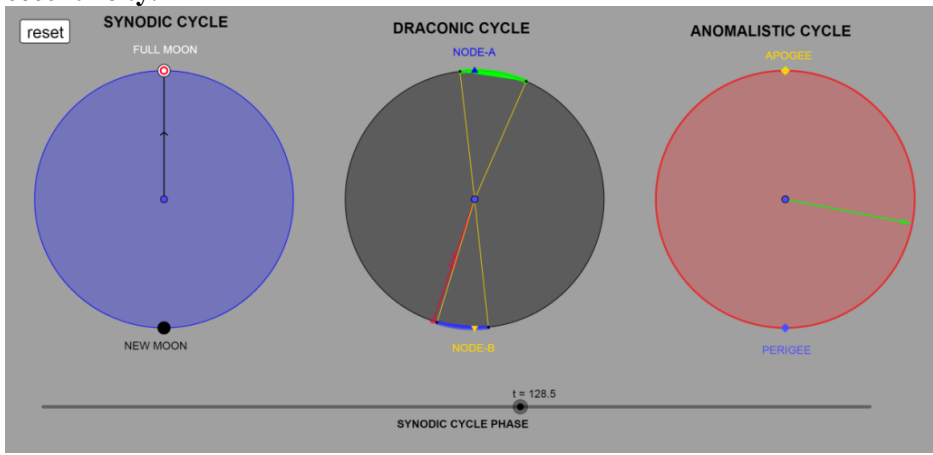

**III) Cell 166: DracoNod program predicts a Lunar eclipse. Full Moon just right on the ecliptic limit. Based on the events' index numbering, between cells 137 and 170 they should be 7 cells with events. Cell 166 is the 8$^{th}$ cell with event, and it is the only one event presenting high indeterminacy. Cell 166 is considered as cell without event.**

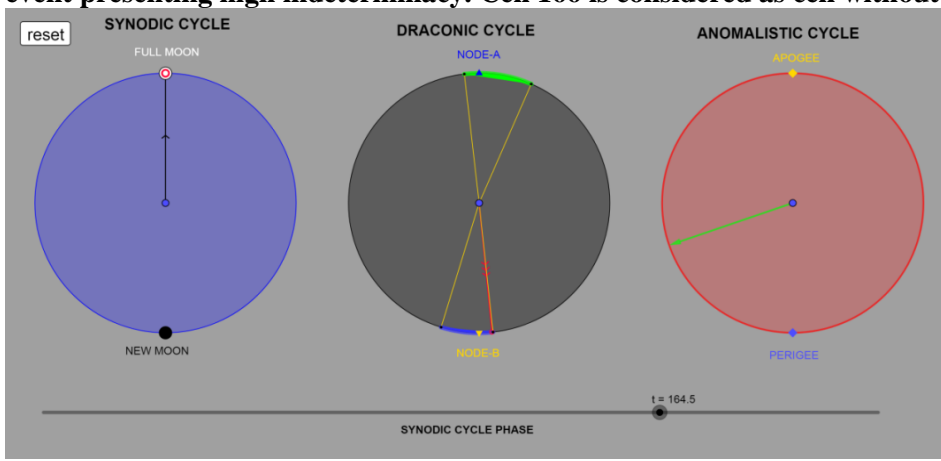

**IV) Cell 189:** <mark>No engraved event.</mark> DracoNod program predicts a second event, a Solar eclipse. New Moon just right on the ecliptic limit. Indeterminacy or gearing errors.

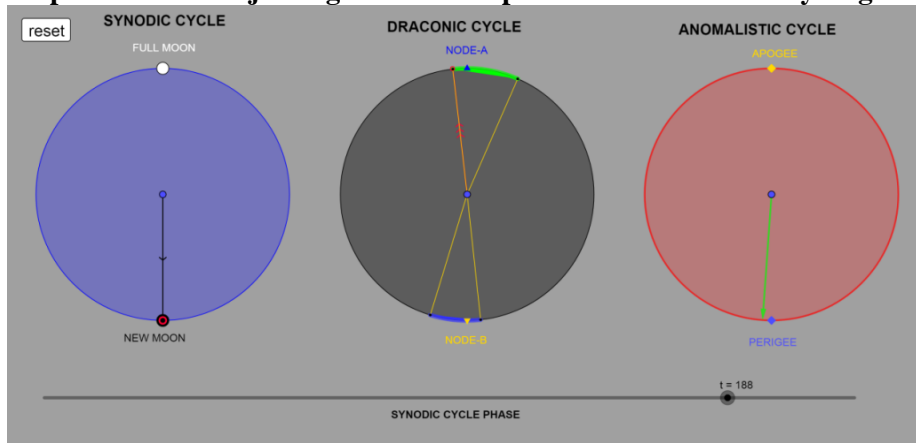

**V) Cell 01/Cell 224/A1:** *DracoNod* predicts a Lunar eclipse event (Σ). Full Moon close to the ecliptic limit. On cell 112 (one Sar after Cell 01), is not engraved the solar eclipse event. Therefore, on Cell 01 could not be engraved the lunar eclipse event.

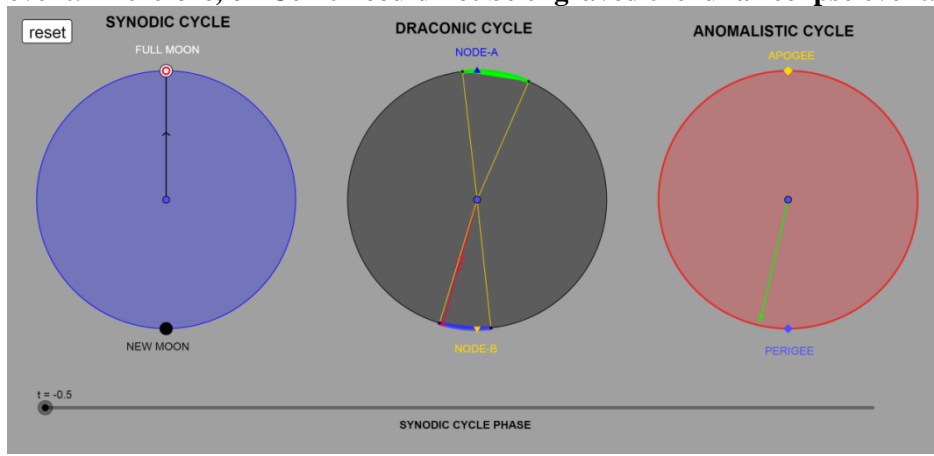

## 8. The Sar period (half Saros) of the Antikythera Mechanism

Dividing by two the equality of 223 Synodic months = 242 Draconic months = 239 Anomalistic months, the equality of the 111.5 Synodic months = 121 Draconic months = 119.5 Anomalistic months arises. Starting with the New Moon at Node-A and at Apogee, after 111.5 Synodic months, the Full Moon is located at the same Node-A and at Perigee. I.e., after a half Saros period, named Sar, the sequence of the eclipse events is repeated in inversed mode (Voulgaris et al., 2021).

Below, five pairs of eclipse events in time span of one Sar are presented (Cell 01 A1/Solar – Cell 113 Γ2/Lunar, Cell 07 B1/Lunar – Cell 118 Δ2/Solar, Cell 19 E1/Lunar – Cell 130 H2/Solar, Cell 101 Ω1/Lunar – Cell 212 ῌ/Solar, ῌ=A3). The Moon is located at the same position relative to the same Node and in a symmetrical inversed distance from the Earth.

## Cell 01 A1/Solar – Cell 113 Γ2/Lunar

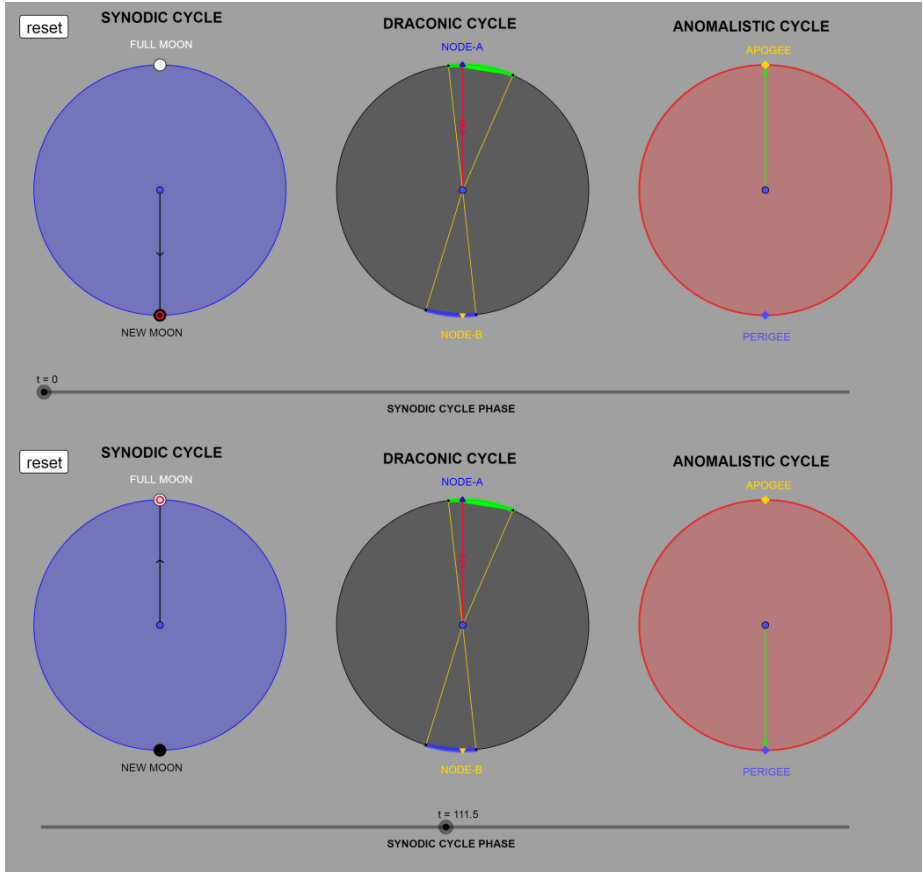

## Cell 07 B1/Lunar – Cell 118 Δ2/Solar

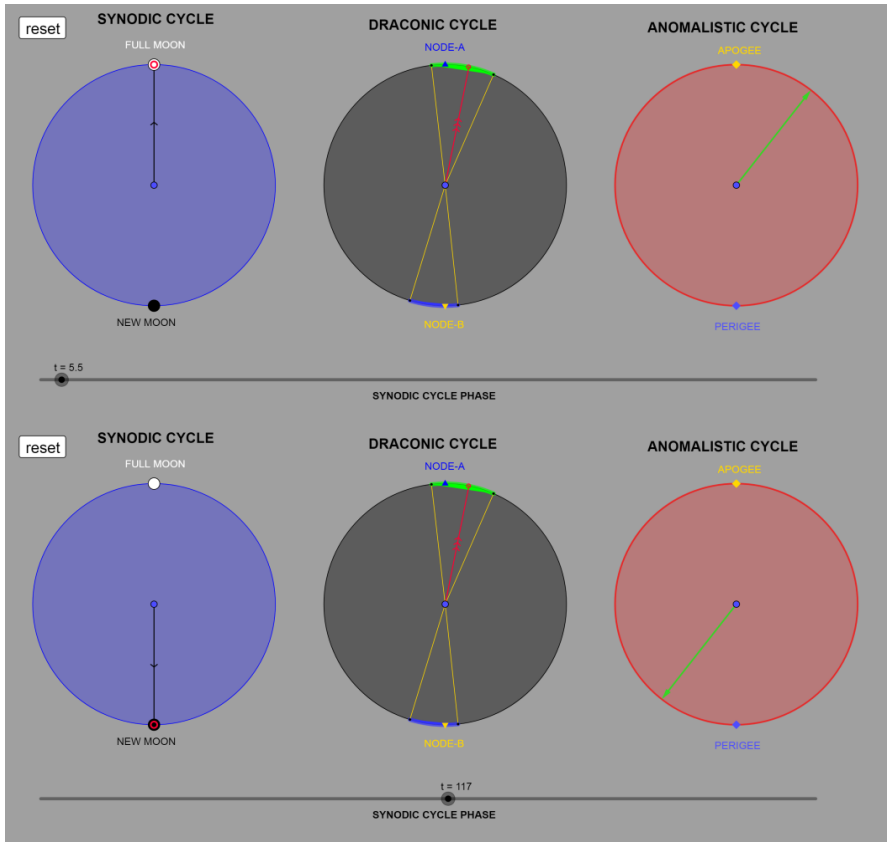

## Cell 19 E1/Lunar – Cell 130 H2/Solar

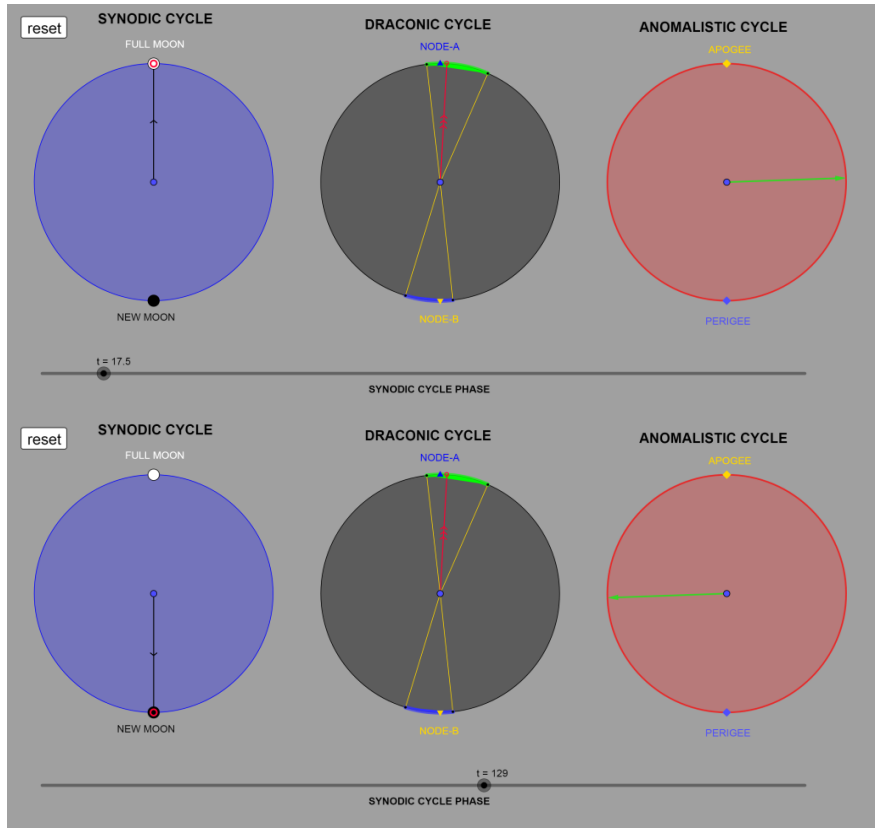

## Cell 101 Ω1/Lunar – Cell 212 ℌ/Solar, ℌ=A3

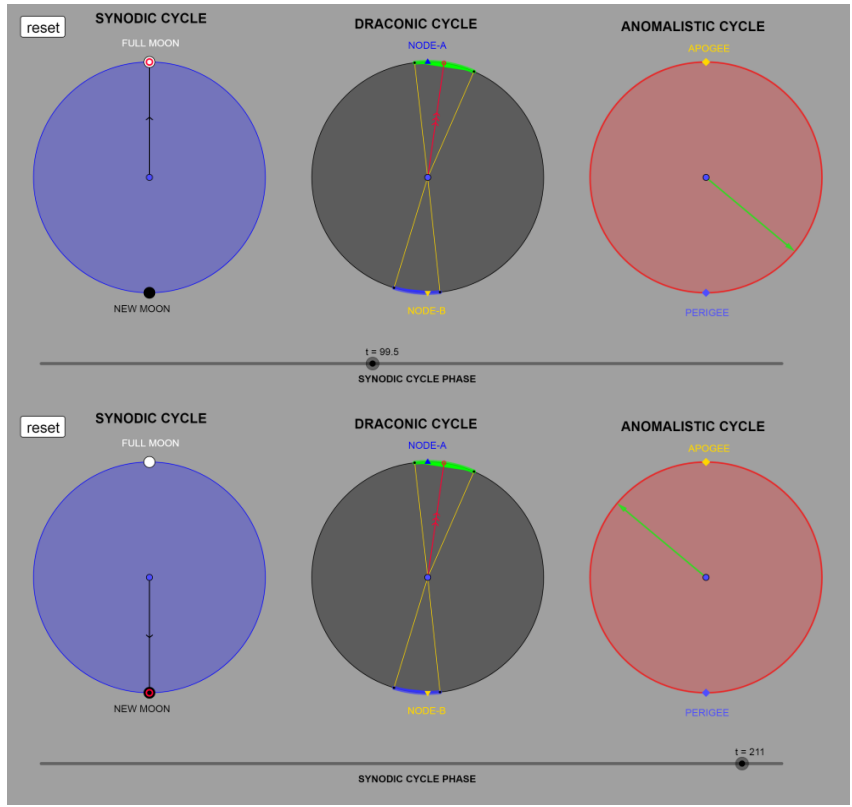

## 9. Calculating the true (symmetrical) ecliptic limits of the Draconic scale

In order to find the actual symmetrical ecliptic limits (http://www.steveholmes.net/ecliptic-limits/; Green 1985), we attempted to detect the position of the 4 points (2 points ×2 ecliptic limits) placed on a circular perimeter-Draconic scale, so that the epicenter angles defined by the points, to present central symmetry. At the same time, the line connects the two Nodes (Line of Nodes) should be a diameter of the circle and the Nodes should be points on the perimeter of the Draconic scale.
The approximate solution was found in the graphic design environment with error ±0.5°: The symmetrical ecliptic limits, probably adopted from the ancient Manufacturer, might be:
- For the Ecliptic Window-A: –19.9° and +5.7° from Node-A (measured CCW)
- For the Ecliptic Window-B: –19.4° and +6.2° from Node-B (measured CCW).
The mean values for both of the Ecliptic Windows are 19.65° and 5.95°, approaching the values 20° and 6°.

This solution (which is not the only one that occurs) places the center of the Draconic scale/axis pointer about 4mm lower to the initial center and exact on the Line of Nodes. The new center creates an eccentricity of the Draconic pointer relative to the Draconic scale. Therefore, there are two options:
i) Symmetrical ecliptic limits and Draconic pointer with eccentricity,
ii) Asymmetrical ecliptic limits and Draconic pointer without eccentricity.
On both of the above options the prediction of the eclipse events sequence is (approximately) the same, but the second seems to be more realistic.

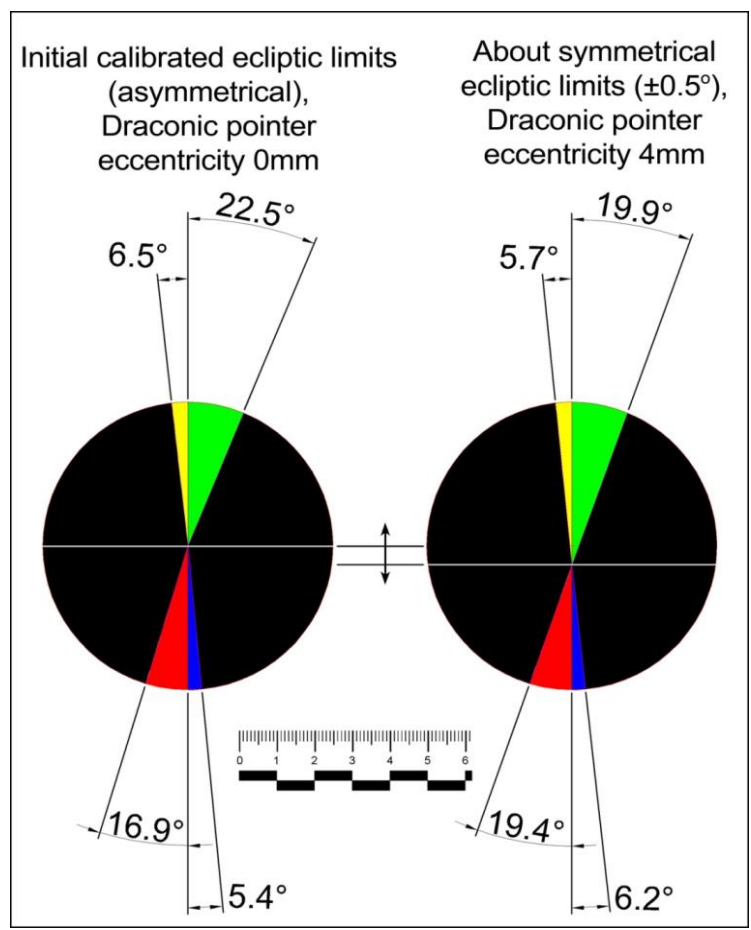

# 10. Changing *DracoNod* parameters

*DracoNod* program was designed for parameterization of several factors.
We can replace the general equality of the lunar cycles for one Saros for more precise values, using the values presented in NASA eclipse page, *Periodicity of solar eclipses, § 1.10 Secular Variations in the Saros and Inex,*
https://eclipse.gsfc.nasa.gov/SEsaros/SEperiodicity.html#section106, by Fred Espenak, NASA's GSFC.

For era 1 AD, it is valid that 223 Synodic months = 241.998703 Draconic months = 238.991950 Anomalistic months.
After running *DracoNod* program, the shifting of Node and the Apogee/Perigee positions through time is visible.

**Exeligmos/Saros 0. New moon at Node-A and at Apogee.**

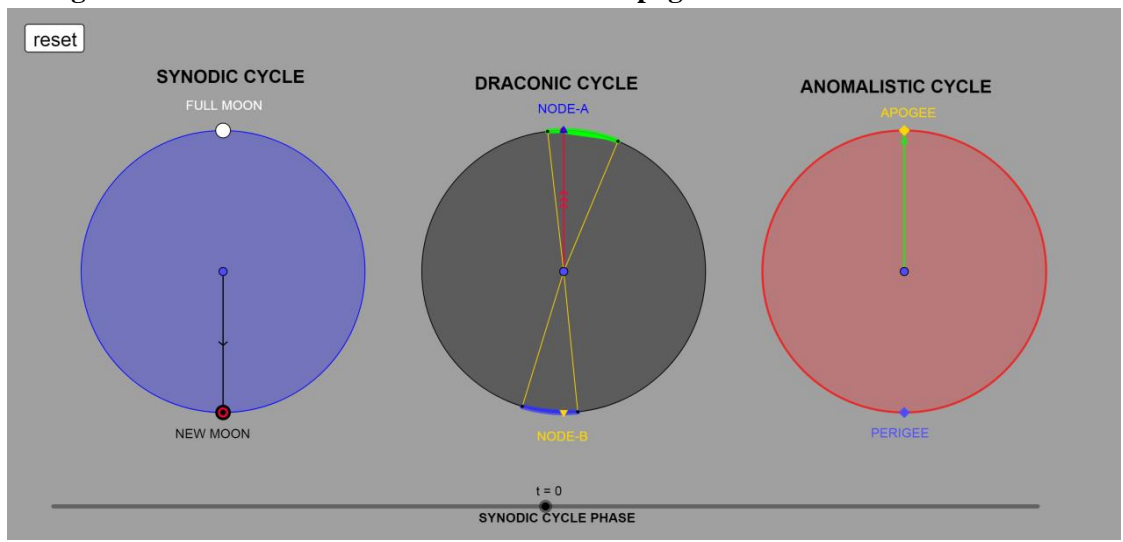

**After one Exeligmos/3 Saros.**

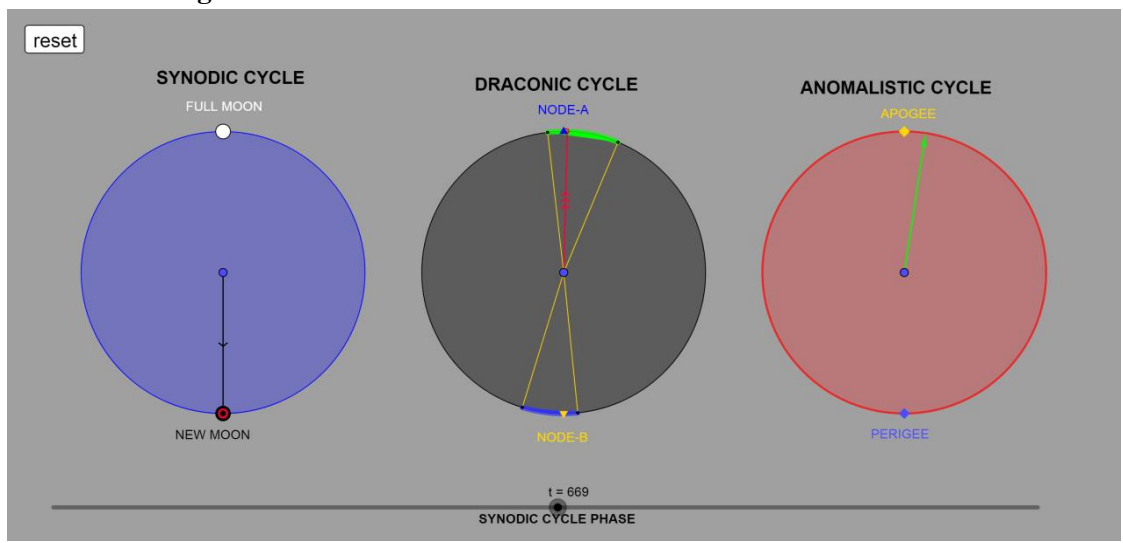

**After 3 Exeligmos/9 Saros.**

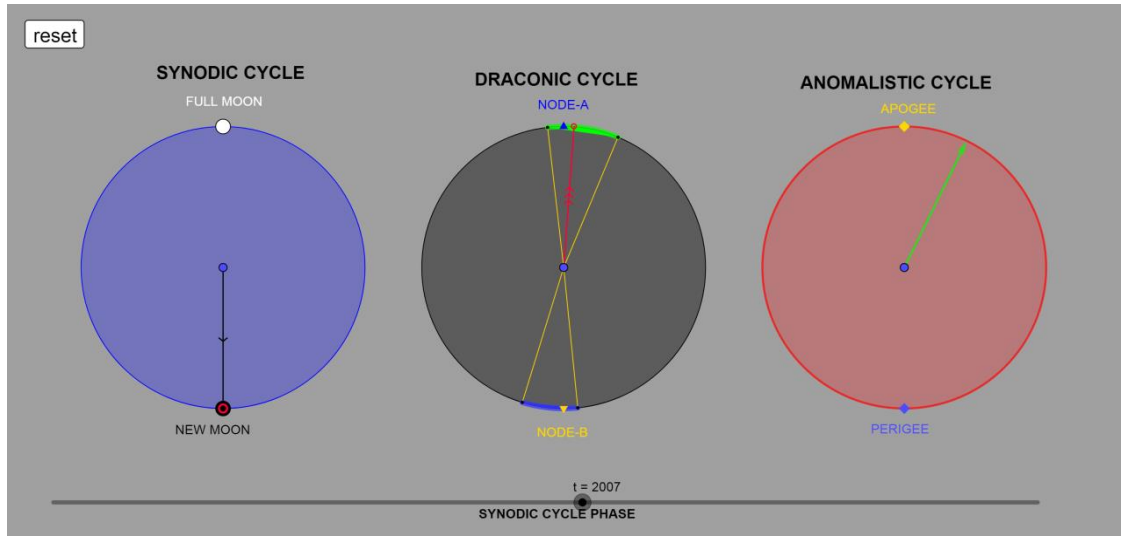

**After 6 Exeligmos/18 Saros.**

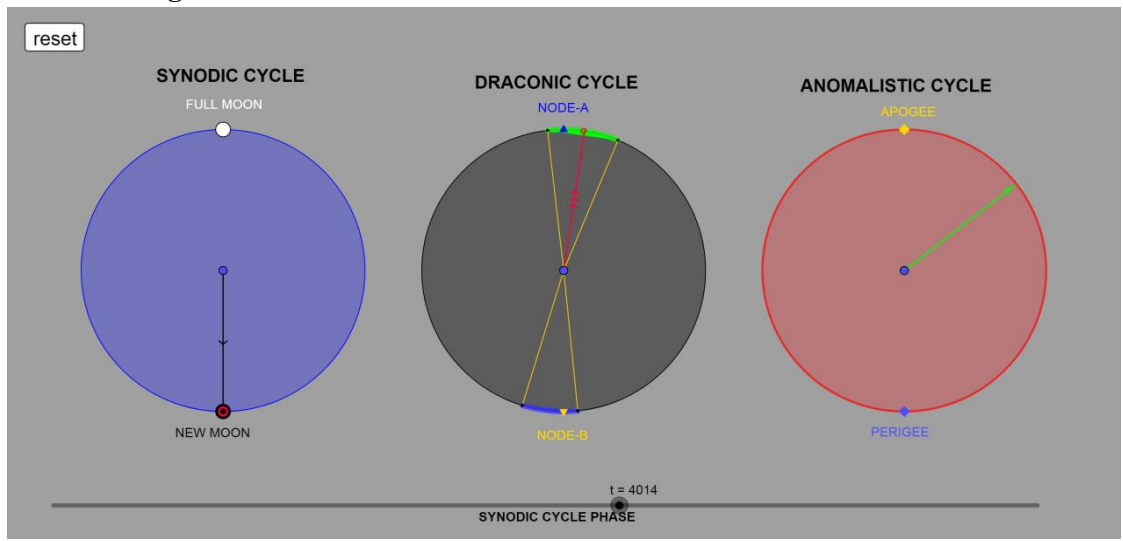

**After 9 Exeligmos/27 Saros.**

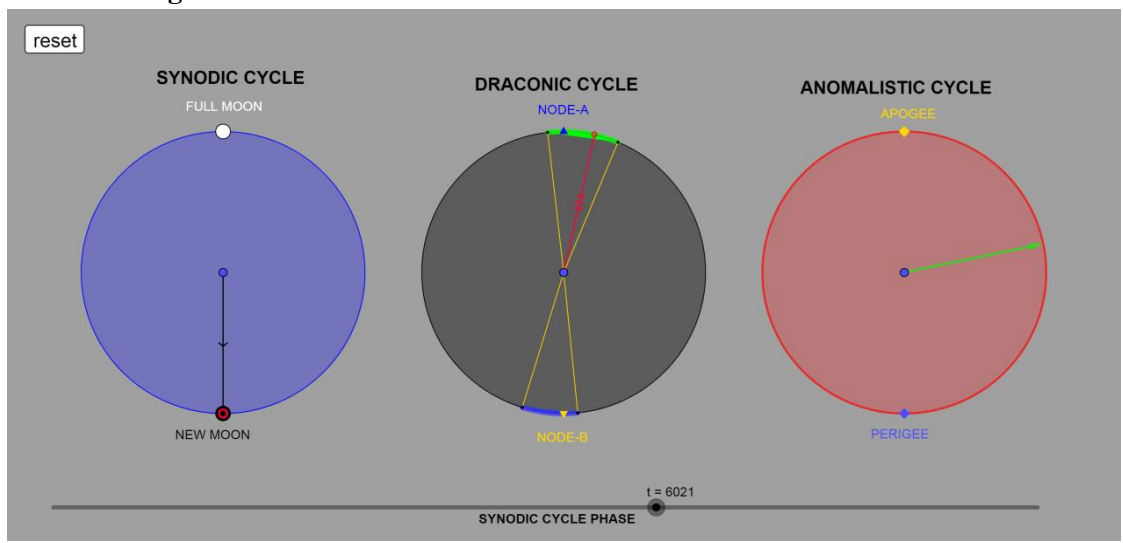

As the positions of the Nodes and the Apogee/Perigee are continuously shifting, the Saros cycle should be re-calculated during time.